\documentclass[11pt]{article}
\usepackage{verbatim,amsmath,amssymb}
\usepackage{epsfig,float,color}
\usepackage{geometry}
\usepackage{setspace}
\usepackage{natbib}
\usepackage{hyperref}
\usepackage[parfill]{parskip}
\usepackage{url}
\usepackage{xfrac}
\usepackage{relsize}
\usepackage{mathtools}
\usepackage[titletoc,title]{appendix}
\usepackage{enumitem}
\usepackage{array}
\usepackage{multirow}
\usepackage{subfig}
\usepackage{graphicx,import}
\usepackage{adjustbox}
\usepackage{mdframed}

\definecolor{darkblue}{rgb}{0,0,1}
\hypersetup{pdftex=true, colorlinks=true, breaklinks=true, linkcolor=darkblue, menucolor=darkblue, pagecolor=darkblue, citecolor=darkblue, urlcolor=darkblue}

\begin{document}


\begin{center}
\Large{\bf{A strategy to interface isogeometric analysis with Lagrangian finite elements --- application to fluid-structure interaction problems}}\\

\end{center}

\begin{center}
\large{Raheel Rasool\,$^a$, Maximilian Harmel\,$^b$ and Roger A. Sauer$\,^{b,}\!\!$
\footnote{corresponding author, email: sauer@aices.rwth-aachen.de}} \\
\vspace{4mm}

\small{\textit{$^a$Institute of Mineral Processing Machines, TU Bergakademie Freiberg, Lampadiusstrasse~4, 09599~Freiberg, Germany}}

\small{\textit{$^b$Aachen Institute for Advanced Study in Computational Engineering Science (AICES), RWTH Aachen
University, Templergraben~55, 52056~Aachen, Germany}}

\vspace{4mm}


\end{center}

\vspace{3mm}


\rule{\linewidth}{.15mm}
{\bf Abstract}		
Isogeometrically enriched finite elements~(\cite{corbett2014,rasool2016,harmel2016}) offer efficient localized isogeometric analysis (IGA) enrichment for numerical simulations involving large computational domains. This is achieved by employing surface enriched elements to interface isogeometric elements with classical Langrangian finite elements. In this paper, we explore their applicability and merits for fluid-structure interaction (FSI) analysis. The implemented approach not only offers an enrichment of the finite element space, but also offers a framework for discretizing and analyzing fluid and structure with different finite element approaches, namely, classical Lagrange finite elements and IGA. \\
In this context, a monolithic solution approach with an explicit grid update mechanism is implemented for FSI. The applicability and the impact of the isogeometric enrichment approach on the accuracy of the numerical solution is assessed by comparing the obtained results with existing reference solutions of FSI benchmark examples involving two- and three-dimensional incompressible fluid flow past hyper-elastic solids. \\

\vspace{-4mm}
\rule{\linewidth}{.15mm}

\section{Introduction}   \label{sec_intro}
	Fluid-structure interaction (FSI) belongs to a special class of multi-physics where a deformable solid structure interacts with a body of fluid. The interactions between the constituent media occur at the interface, where the two media meet, and typically involve transfer of interfacial forces from one media to the other. These forces then determine the kinematical response of the constituent media. These interactions are typically strong for many physical and engineering applications, and the absence of a coupled analysis will lead to misleading, and possibly catastrophic conclusions. Blood flow in arteries and veins, design of mechanical cardiovascular valves and pumps, flutter and fatigue of aircraft wing and turbine blades under aerodynamic loadings, and safety calculations of suspension bridges and tall buildings under strong winds, are only a few of the many examples that demand a coupled analysis approach that takes into account the cross-interactions between solid and fluid media.
	
	Mathematical models of FSI problems are often too complex to be analyzed analytically. Therefore such problems are largely investigated through experiments and numerical simulations. In the context of numerical analysis, the available modeling approaches for FSI analysis can broadly be categorized into three groups, namely field elimination, partitioned approaches and monolitic approaches. Field elimination methods (e.g. see~\cite{belvinsbook,duetsch1998,ohayon2004}) constitute the simplest of these approaches, where one or more of the associated field variables are eliminated from the governing model through reduction techniques such as substitution, integral transforms or model reductions. Such approaches, however, are restricted to the modeling of linear problems where a decoupling between the field variables can be formulated. On the other hand, available FSI solvers are generally classified as a partitioned (\cite{felippa2001}) or a monolithic (\cite{bendiksen1991}) approach, depending upon the mechanism by which the kinematic and the kinetic description is transfered between solid and fluid media at the fluid-solid interface. 
	
	Partitioned approaches make use of existing stand-alone fluid and structure solvers by invoking them in a sequential manner. The coupling interactions at the interface are enforced as forcing effects communicated through prediction, substitution and synchronization techniques. Whether these communications are performed iteratively or in a staggered manner, partitioned approaches are further categorized as strongly (\cite{tezduyar2006,dettmer2006,matthies2006,kuettler2008}) or weakly coupled schemes (\cite{piperno1995,farhat2004,wang2004,vanzuijlen2007}), respectively. In contrast, monolithic solvers cast the governing equations of the fluid and the solid model into a singular unit and attempt a solution in a single iteration with consistent time integration schemes (\cite{huebner2004, walhorn2005,bazilevs2008,kloeppel2011,mayr2015}). 
	
	Partitioned and monolithic approaches have been compared by many researchers (e.g.~see~\cite{rugonyi2001,heil2008,kuettler2010}), however the choice of a preferred approach for a particular application predominantly remains problem and resource specific. The strongest merit of partition approaches is the flexibility to incorporate validated and optimized stand-alone fluid and solid solvers in the coupled analysis framework, irrespective of the spatial discretization methodology. Finite volume method (FVM) is a popular choice for fluid flow problems, while finite element method (FEM) is used extensively in the structural mechanics community. However, the sequential invocation of the solvers leads to time-inconsistent field description, which if left untreated, can lead to numerical instabilities (e.g.~the added mass effect~\cite{causin2005}, incompressibility dilemma at the interface~\cite{kuettler2006}, etc.). Although monolithic methods lead to time-consistent solutions and subsequently to a more robust solution approach, they demand larger computational resources due to storage and solution of a larger system of equations. Moreover, due to a combined cast for the governing equations of both models, monolithic schemes tend to lead to ill-conditioned global matrices for cases where material parameters of the constituents are orders of magnitude different from one-another. For such cases, preconditioning strategies (such as~\cite{heil2004a,gee2011}) should be employed or developed before the monolithic system is solved.  
	
	In this paper, we present a monolithic FSI solution procedure in the framework of isogeometric finite elements. Initially proposed as a subset of the FEM in~\cite{hughes2005}, isogeometric analysis (IGA) offers an improvement to the classical FEM by incorporating CAD discretizations based on NURBS~(\cite{piegelbook}) and T-splines --- rather than interpolatory Lagrange polynomials --- to the trial and test function spaces of the discrete finite element formulation. In this manner, the original CAD description (and consequently the original geometry) is retained in the discrete finite element setting and a direct correspondence between CAD and finite element analysis is maintained. Although initially formulated to speed-up the design-to-analysis transition, the use of spline-based functions provides additional merits for numerical analysis over classical interpolatory polynomials due to many desirable properties of the spline basis (such as accurate geometry representation, higher continuity across elements and the variation diminishing property). As a results, IGA has lately been the focus of intense technical research with wide-range engineering and biomedical applications. For an elaborate overview, interested readers are invited to consult~\cite{cottrellbook}. 
	
	From biomedical applications~(\cite{zhang2007,bazilevs2008,bazilevs2009,chivukula2014}) to the design of wind turbines~(\cite{bazilevs2011,bazilevs2012,bazilevs2016}), IGA has been successively applied to various FSI problems. The superiority of IGA over interpolatory FEM for uncoupled problems has been investigated by many researchers, e.g. \cite{bazilevs2007vms,akkerman2008,motlagh2013,rasool2016}~for fluid dynamics and \cite{cottrell2006,hughes2008,sauer2014,morganti2015}~for structural mechanics applications. For many of these studies, IGA yields higher accuracy per degree-of-freedom ($\mathrm{dof}$) than the classical FEM. However, modeling the entire computational domain with spline-based isogeometric elements may not be desirable for every problem. Due to recursive evaluation of the basis functions (which for the case of NURBS and T-splines can be distinct over each element), IGA demands additional computational resources for discretizations with comparable $\mathrm{dof}$s. Moreover, it has also been observed that the extended continuity of the basis across element boundaries incurs additional burden on linear solvers~(see \cite{collier2012, collier2013}). For many problems, especially true for simulation of physical flow in unbounded regions, it is sufficient to have an enriched analysis only in certain localized regions of the discretized domain, while the remaining bulk can be represented with a low-order computationally efficient discretization, e.g.~\cite{rasool2016} and \cite{harmel2016}. Such applications motivates a framework where an interpolatory finite element discretization is enriched with isogeometric elements only at regions where an enhanced representation and analysis is beneficial. 
	
	The motivation to blend isogeometric elements to an interpolatory finite element discretization is certainly not novel. Perhaps, the earliest reference in this regard can be found in~\cite{bazilevs2012}, where T-spline shells for solids were interfaced with low-order Lagrangian finite elements for fluid at the fluid-solid interface. Such a treatment was achieved by weakly enforcing the coupling conditions at the interface, which also relieved the necessity of a matching discretization at the interface. However, the interfacing between isogeometric and Lagrange finite elements is only available at the fluid-solid interface. Interpolating B-spline curves with Lagrangian functions, transformation maps were developed in~\cite{lu2013} and \cite{ge2016} to construct ``Blended elements'' which can interface two- and three-dimensional NURBS-based elements with Lagrangian elements. 
	
	Another strategy developed with a similar motivation is the isogeometric zone enrichment approach. It was proposed in~\cite{rasool2016} for fluid flow problems and in~\cite{harmel2016} for thermal analysis. In this paper, we explore its applicability and benefits for FSI applications. Isogeometrically enriched surface elements~(see~\cite{corbett2014,corbett2015}) are used to interface IGA elements with Lagrange finite elements, yielding the possibility of a locally enriched analysis. The strategy additionally offers a monolithic mechanism, where one medium can be analyzed in the framework of IGA, while the other in the classical finite element setting. The potential of such an analysis is also investigated. 
	
	The remainder of this paper is organized in the following manner: The framework of the governing monolithic FSI model is discussed in Section~\ref{sec-continuum}. It includes the continuous differential model, its discretized weak form and a grid motion mechanism for incorporating moving surfaces. The isogeometric zone enrichment strategy is explained in Section~\ref{sec-enrich}, while results from several numerical experiments are presented in Section~\ref{sec-ex}. Conclusions from the analysis are discussed in Section~\ref{sec-conc}.
	
%
\section{Monolithic FSI model}		\label{sec-continuum}
The theoretical framework for the developed monolithic FSI solver is discussed in this section. It begins with the description of the governing equations that describe momentum and mass balance for the entire system. The equations are first expressed in strong form. Their variational (weak) description follows consequently. The implemented finite element formulation, in conjunction with the stabilization technique used in this study, is then presented. A note on the explicit grid motion mechanism concludes this section.
%
\subsection{Governing equations in monolithic setting}
Let us consider disjointed spatial domains $\mathcal{B}_\text{f} \subset \mathbb{R}^{d}$ and $\mathcal{B}_\text{s} \subset \mathbb{R}^{d}$ at any given time $t\in\left[0,T\right]$. Here $d$ denote the number of spatial dimensions involved, while $\mathcal{B}_\text{f}$ and $\mathcal{B}_\text{s}$ represent the solid and the fluid domain respectively. Let $\partial\mathcal{B}_\text{f} = \partial_v \mathcal{B}_\text{f} \cup \partial_t \mathcal{B}_\text{f} $ represent the boundary of the fluid domain with $\partial_v \mathcal{B}_\text{f}$ as the Dirichlet part and $\partial_t \mathcal{B}_\text{f}$ as the Neumann part of $\partial \mathcal{B}_\text{f}$. Similarly, for the solid domain we denote $\partial\mathcal{B}_\text{s} = \partial_u \mathcal{B}_\text{s} \cup \partial_t \mathcal{B}_\text{s} $ as the boundary of $\mathcal{B}_\text{f}$ with the associated Dirichlet and Neumann boundary segments, respectively. The fluid-solid interface can then be represented as $\partial\mathcal{B}_\text{f} \cap \partial \mathcal{B}_\text{s}$.
\subsubsection{Conservation laws for the fluid}
Let us consider incompressible fluid flow within $\mathcal{B}_\text{f}$. The conservation law for momentum and mass of the fluid thus reads,
\begin{align}
	\rho_\text{f} \, \boldsymbol{a}_\text{f} &=  \text{div} \, \boldsymbol{\sigma}_\text{f} + \rho_\text{f} \, \boldsymbol{b}_\text{f} \,, \text{\hspace{1cm}} \forall \, \boldsymbol{x} \in \mathcal{B}_\text{f} \, , \label{eq:fl_cons_mom_st}\\
\text{div} \, \boldsymbol{v}_\text{f} &= 0 \, , \text{\hspace{2.9cm}}  \forall \, \boldsymbol{x} \in \mathcal{B}_\text{f} \,, \label{eq:fl_cons_mass_st}
\end{align}
where $\rho_\text{f}$ is the density of the fluid at time $t$, $\boldsymbol{\sigma}_\text{f}$ denotes the Cauchy stress tensor and $\boldsymbol{b}_\text{f}$ represents external volumetric forces acting on $\mathcal{B}_\text{f}$. The variables $\boldsymbol{a}_\text{f}$ and $\boldsymbol{v}_\text{f}$ represent the acceleration and the velocity of the fluid at a given point $\boldsymbol{x}$. In order to accommodate moving surfaces, typically at the fluid-solid interface, we express the conservation laws for the fluid media in an arbitrary Lagrangian-Eulerian (ALE) framework~(for details, refer to~\cite{doneabook}). Within the ALE description of motion, the acceleration of a particle is given as
\begin{align}
	\boldsymbol{a}_\text{f} = \frac{\partial \boldsymbol{v}_\text{f}}{\partial t} \bigg|_{\boldsymbol{\chi}}  + \boldsymbol{c} \cdot \text{grad} \,{\boldsymbol{v}_\text{f}} \,,
\end{align}
where $\boldsymbol{c} \coloneqq \boldsymbol{v}_\text{f} - \hat{\boldsymbol{v}}$, with $\hat{\boldsymbol{v}}$ being the velocity of the fluid mesh. The notation $(\cdot) \big|_{\boldsymbol{\chi}}$ denotes an observation made in the referential domain described by $\boldsymbol{\chi}$, relative to which both the material $\boldsymbol{X}$ and the mesh $\boldsymbol{x}$ can deform.
\subsubsection{Conservation law for the solid}
Let us express the solid body $\mathcal{B}_\text{s}$ with a purely Lagrangian representation, which can also be inferred as a special case of ALE with coincident referential and material coordinates for all $t$. The conservation of momentum for $\mathcal{B}_\text{s}$ then follows,
\begin{align}
	\rho_\text{s} \, \boldsymbol{a}_\text{s} = \text{div} \, \boldsymbol{\sigma}_\text{s} + \rho_\text{s} \, \boldsymbol{b}_\text{s} \, ,	\text{\hspace{1.0cm}} \forall \, \boldsymbol{x} \in \mathcal{B}_\text{s} \,, \label{eq:sol_cons_mass_st}
\end{align}
where $\rho_\text{s}$ is the solid density, $\boldsymbol{\sigma}_\text{s}$ is the Cauchy stress tensor and $\boldsymbol{b}_\text{s}$ is the cumulative external body force acting on $\mathcal{B}_s$. The acceleration $\boldsymbol{a}_\text{s}$, in terms of the displacement $\boldsymbol{u}_\text{s}$ is given as
\begin{align}
	\boldsymbol{a}_\text{s} = \frac{\partial^2 \boldsymbol{u}_\text{s}}{\partial t^2} \bigg|_{\boldsymbol{X}} =: \frac{D^2 \boldsymbol{u}_\text{s}}{D t^2} \,.
\end{align} 
It is often convenient to express Eq.~(\ref{eq:sol_cons_mass_st}) over a reference configuration, for which we will particularly consider the initial configuration (distinguished as $\mathcal{B}_{\text{s}_\text{o}}$) at $t=0$ such that
\begin{align}
	\rho_{\text{s}_\text{o}}\frac{D^2 \boldsymbol{u}_\text{s}}{D t^2} = \text{Div} \left( \boldsymbol{F \, S_{\text{s}} } \right)+ \rho_{\text{s}_\text{o}}  \boldsymbol{b}_\text{s} \, ,	\text{\hspace{1.0cm}} \forall \, \boldsymbol{X} \in \mathcal{B}_{\text{s}_\text{o}} \,, \label{eq:sol_cons_mass_st_ref}
\end{align}
where $\rho_{\text{s}_\text{o}}$ is the solid density at $t=0$, $\boldsymbol{F} = \partial \boldsymbol{x}/\partial \boldsymbol{X}$ is the deformation gradient and $\boldsymbol{S}_\text{s}$ is the second Piola-Kirchhoff stress tensor.
\subsubsection{Constitutive laws}
For the fluid model, let us consider an incompressible Newtonian fluid, for which the shear~stress is modeled as a linear function of the strain-rate tensor (symmetric gradient of the fluid velocity). Air and water are typical examples of Newtonian fluids. For such fluids, the relation between stress and velocity is expressed as,
\begin{align}
	\boldsymbol{\sigma}_\text{f} = -p_\text{f} \, \textbf{I} + \mu_\text{f}  \left( \text{grad}\boldsymbol{v}_\text{f} + \text{grad}^\text{T} \boldsymbol{v}_\text{f} \right) \,,	\label{eq:fl_const_NS}
\end{align} 
where $p_\text{f}$ is the pressure, $\mu_\text{f}$ is the dynamic viscosity, $\boldsymbol{v}_\text{f}$ is the velocity of the fluid and $\textbf{I}$ is the identity tensor. For $\mathcal{B}_\text{s}$, we consider a rubber-like hyper-elastic solid capable of undergoing large deformations. In keeping consistency with the benchmark problems considered in this study, we employ the Saint Venant-Kirchhoff material model for the constitution of the solid, i.e.,
\begin{align}
	\boldsymbol{S}_\text{s} = \lambda_\text{s} \, \text{tr}(\boldsymbol{E}_\text{s}) \, \textbf{I}+ 2 \, \mu_\text{s} \, \boldsymbol{E}_\text{s} \,.		\label{eq:sol_cont_StVen}
\end{align}
Here $\lambda_\text{s}$ and $\mu_\text{s}$ represents the material parameters known as the Lam\'{e} constants, while $\boldsymbol{E}_\text{s}$ is the Green-Lagrange strain tensor. Having formulated $\boldsymbol{S}_\text{s}$, the Cauchy stress tensor can be readily obtained by employing the push-forward operation
\begin{align}
	\boldsymbol{\sigma}_\text{s} = \frac{1}{J} \, \boldsymbol{F}\, \boldsymbol{S}_\text{s}\, \boldsymbol{F}^\text{T} \, ,
\end{align}
with $J=\det(\boldsymbol{F})$. The above discussed material models are chosen specifically to suit the benchmark configurations and represent only two of the many possible constitutive models for the continuum. It should be noted that the enrichment strategy discussed in this paper is a discretization strategy and can be easily used together with alternative material constitutions.  
\subsubsection{Initial, boundary and interface conditions}
In order to close the continuum model, we need to specify a suitable combination of initial and boundary conditions for the continuum. For the fluid, we assume a velocity description $\boldsymbol{v}_{\text{f}_\text{o}}$ that essentially satisfies Eq.~(\ref{eq:fl_cons_mass_st}) at $t=0$. For the solid, we assume a stress-free initial state and refer to it as the reference configuration, i.e., $\boldsymbol{x}=\boldsymbol{X} \text{ at } t=0$. Hence the initial conditions are formulated as
\begin{align}
	\boldsymbol{v}_\text{f} &= \boldsymbol{v}_{\text{f}_\text{o}} (\boldsymbol{x})\, , \text{\hspace{1cm}} \forall \, \boldsymbol{x} \in \mathcal{B}_\text{f} \text{ at } t=0 \,, \\
	\boldsymbol{u}_\text{s} &= \boldsymbol{0} \, , \text{\hspace{1.8cm}} \forall \, \boldsymbol{x} \in \mathcal{B}_{\text{s}} \text{ at } t=0 \, ,
\end{align}
while the boundary conditions, comprising the Dirichlet and the Neumann boundaries, are
\begin{align}
	\boldsymbol{v}_\text{f} &= \bar{\boldsymbol{v}}_\text{f}(\boldsymbol{x},t) \, , \text{\hspace{1cm}} \forall \, \boldsymbol{x} \in \partial_v \mathcal{B}_\text{f} \,, \\
	\boldsymbol{u}_\text{s} &= \bar{\boldsymbol{u}}_\text{s}(\boldsymbol{x},t) \, , \text{\hspace{0.95cm}} \forall \, \boldsymbol{x} \in \partial_u \mathcal{B}_\text{s} \,, \\
	\boldsymbol{\sigma}_\text{f} \cdot \boldsymbol{n}_\text{f} &= \bar{\boldsymbol{t}}_\text{f}(\boldsymbol{x},t) \,, \text{\hspace{1.08cm}} \forall \, \boldsymbol{x} \in \partial_t \mathcal{B}_\text{f} \,, \\
	\boldsymbol{\sigma}_\text{s} \cdot \boldsymbol{n}_\text{s} &= \bar{\boldsymbol{t}}_\text{s}(\boldsymbol{x},t) \,, \text{\hspace{1.05cm}} \forall \, \boldsymbol{x} \in \partial_t \mathcal{B}_\text{s} \,.
\end{align}
Here $\boldsymbol{n}_\text{f}$ and $\boldsymbol{n}_\text{s}$ denote the outward normal vectors at the Neumann boundary surface for the fluid and solid media respectively, while the entities with an over-bar represent the imposed fields at the boundary. The interface boundary should ensure a continuous transfer of kinematic and kinetic field, which is enforced through the following conditions:
\begin{align}
	\begin{rcases}
      		\boldsymbol{x} = \boldsymbol{X} + \boldsymbol{u}_\text{s} \, , \\
      		\boldsymbol{v}_\text{f} = \boldsymbol{v}_\text{s} \, , \text{\hspace{1.35cm}} \\
      		\boldsymbol{\sigma}_\text{f} \cdot \boldsymbol{n}_\text{f} + \boldsymbol{\sigma}_\text{s} \cdot \boldsymbol{n}_\text{s} = 0 \,,
    \end{rcases}
	\text{\hspace{0.5cm}} \forall \, \boldsymbol{x} \in \, \partial \mathcal{B}_\text{s} \cap \partial \mathcal{B}_\text{f} \,.		 	\label{eq:coupling_cond}
\end{align}
The kinematic condition on the position of the grid points $\boldsymbol{x}$ demands a pure Lagrangian description at the interface. It is automatically satisfied across the entire $\mathcal{B}_\text{s}$, since conservative laws for the solid continuum are expressed in the Lagrangian frame of reference. For $\mathcal{B}_\text{f}$, this condition at the interface is accommodated through the ALE description. It should also be noted that the interfacial condition on the velocity field, together with Eq.~(\ref{eq:fl_cons_mass_st}), necessitates a divergence-free velocity of the solid interface. For a partitioned FSI solution procedure such a description is not always guaranteed and may lead to numerical artifacts~(e.g. see~\cite{kuettler2006}).
%
\subsection{Weak form and the finite element model}
The continuum model presented in the previous section is analyzed in the framework of the FEM. To obtain the discrete finite element equations, let us first express Eqs.~(\ref{eq:fl_cons_mom_st}-\ref{eq:fl_cons_mass_st}) and Eq.~(\ref{eq:sol_cons_mass_st_ref}) in its variational, i.e., weak form. We define the following functional spaces for the unknown field variables $(\boldsymbol{v}_\text{f}, p_\text{f},\boldsymbol{u}_\text{s})$ and their respective test functions $(\boldsymbol{w},q,\delta \boldsymbol{u})$:
\begin{align}
\boldsymbol{\mathcal{S}}_v &= \left\{ \boldsymbol{v}_\text{f} \,|\, \boldsymbol{v}_\text{f} \, \in \, \text{H}^{1}(\mathcal{B}_\text{f}), \hspace{0.1cm} \boldsymbol{v}_\text{f} = \bar{\boldsymbol{v}}_\text{f} \text{ on }\partial_v \mathcal{B}_\text{f}\right\} \,, \\
\boldsymbol{\mathcal{S}}_u &= \left\{ \boldsymbol{u}_\text{s} \,|\, \boldsymbol{u}_\text{s} \, \in \, \text{H}^{1}(\mathcal{B}_\text{s}), \hspace{0.1cm} \boldsymbol{u}_\text{s} = \bar{\boldsymbol{u}}_\text{s} \text{ on }\partial_u \mathcal{B}_\text{s}\right\} \,, \\
\boldsymbol{\mathcal{V}}_v &= \left\{ \boldsymbol{w} \,|\, \boldsymbol{w} \, \in \, \text{H}^{1}(\mathcal{B}_\text{f}), \hspace{0.1cm} \boldsymbol{w} = \boldsymbol{0} \text{ on }\partial_v \mathcal{B}_\text{f}\right\} \,, \\
\boldsymbol{\mathcal{V}}_u &= \left\{ \delta \boldsymbol{u} \,|\, \delta \boldsymbol{u} \, \in \, \text{H}^{1}(\mathcal{B}_\text{s}), \hspace{0.1cm} \delta\boldsymbol{u} = \boldsymbol{0} \text{ on }\partial_u \mathcal{B}_\text{s}\right\} \,, \\
\mathcal{S}_p &= \mathcal{V}_p = \left\{ q|q \, \in \text{L}^2(\mathcal{B}_\text{f}) \right\} \,,
\end{align} 
where $\text{L}^2(\cdot)$ represents a collection of functions that are square-integrable over the given domain, while $\text{H}^1(\cdot)$ is a subclass of $\text{L}^2(\cdot)$ comprising functions that additionally posses finite square-integrable first-order derivatives. The weak form can now be obtained by multiplying Eqs.~(\ref{eq:fl_cons_mom_st}-\ref{eq:fl_cons_mass_st}) with the test function $\boldsymbol{w}$ and Eq.~(\ref{eq:sol_cons_mass_st_ref}) with $\delta \boldsymbol{u}$, and integrating over the entire domain. After certain mathematical manipulations, we arrive at the following weak form: 
\begin{align}
&\text{Find } \boldsymbol{v}_\text{f} \in \boldsymbol{\mathcal{S}}_v\text{, } p _\text{f}\in \mathcal{S}_p \text{ and } \boldsymbol{u}_\text{s} \in \boldsymbol{\mathcal{S}}_u,  \text{ such that} \nonumber\\
&\int\limits_{\mathcal{B}_\text{f}} \! \rho_\text{f} \, \boldsymbol{w}\cdot \frac{\partial \boldsymbol{v}_\text{f}}{\partial t} \bigg|_\chi \text{d}v + \int\limits_{\mathcal{B}_\text{f}} \!\rho_\text{f} \,\boldsymbol{w} \cdot \boldsymbol{c} \cdot \text{grad}\, \boldsymbol{v}_\text{f} \, \text{d}v + \int\limits_{\mathcal{B}_\text{f}}\!\text{grad}\boldsymbol{w}: \boldsymbol{\sigma}_\text{f}  \, \text{d}v -  \int\limits_{\mathcal{B}_\text{f}} \!\rho_\text{f} \, \boldsymbol{w} \cdot \boldsymbol{b}_\text{f} \, \text{d}v  \nonumber \\
& \text{\hspace{9.8cm}}- \int\limits_{\partial_t \mathcal{B}_\text{f}} \!\! \boldsymbol{w} \cdot \bar{\boldsymbol{t}}_\text{f} \, \text{d}a =0 \,, \label{eq:weak_fl_mom} \\
	&\int\limits_{\mathcal{B}_\text{f}} q \,\, \text{div}\, \boldsymbol{v}_\text{f} \, \text{d}v = 0 \,,	\label{eq:weak_fl_mass} \\
&\int\limits_{\mathcal{B}_{\text{s}_\text{o}}} \rho_{\text{s}_\text{o}} \delta \boldsymbol{u} \cdot\frac{D^2 \boldsymbol{u}_\text{s}}{Dt^2} \, \text{d}V + \int\limits_{\mathcal{B}_{\text{s}_\text{o}}} \text{Grad} \, \delta \boldsymbol{u} : \left( \boldsymbol{F S_\text{s}}\right) \, \text{d}V - \int\limits_{\mathcal{B}_{\text{s}_\text{o}}}  \rho_{\text{s}_\text{o}} \,\delta \boldsymbol{u} \cdot  \boldsymbol{b}_\text{s} \, \text{d}V - \int\limits_{\partial_t \mathcal{B}_{\text{s}_\text{o}}} \delta \boldsymbol{u} \cdot \bar{\boldsymbol{t}}_\text{s}\, \text{d}A =0\,, \label{eq:weak_sol_mom} \\
&\forall \, \boldsymbol{w} \in \boldsymbol{\mathcal{V}}_v, \, q \in \mathcal{V}_p \text{ and }\delta \boldsymbol{u} \in \boldsymbol{\mathcal{V}}_u \, .  \nonumber
\end{align}
\subsubsection{Spatially discrete stabilized formulation}
Let us now discretize the spatial domains $\mathcal{B}_\text{f}$ and $\mathcal{B}_\text{s}$ with $n_{fe}$ and $n_{se}$ non-overlapping finite elements such that
\begin{align}
	\mathcal{B}^\text{h}_\text{f} = \bigcup_{e=1}^{n_{fe}} \, \Omega_\text{f}^e \text{\hspace{0.5cm}and \hspace{0.5cm}}\mathcal{B}^\text{h}_\text{s} &= \bigcup_{e=1}^{n_{se}}\, \Omega_\text{s}^e 	\label{eq:domain_fe}
\end{align}
constitute a reasonable approximation of the respective continuous domains. Here $\Omega^e$ denotes a unique finite element identified with the index $e$, while the superscript h denotes a discretized entity. We further define a discrete subset of our trial and test function spaces, denoted as $\boldsymbol{\mathcal{S}}_v^\text{h}, \mathcal{S}_p^\text{h}, \boldsymbol{\mathcal{S}}_u^\text{h}, \boldsymbol{\mathcal{V}}_v^\text{h}, \mathcal{V}_p^\text{h} \text{ and }\boldsymbol{\mathcal{V}}_u^\text{h}$, such that the chosen functions now only span in the vicinity of element $e$ and vanish elsewhere. The discrete formulation, also known as the Bubonov-Galerkin finite element formulation, is readily obtained by formulating the weak form of Eqs.~(\ref{eq:weak_fl_mom}-\ref{eq:weak_sol_mom}) over the discretized space.

The Bubonov-Galerkin formulations leads to a discretization scheme that is similar in character as the central-difference scheme. Central-difference schemes are known to be numerically unstable for problems where the advection operator dominates the diffusion term, such as convective fluid flows. These instabilities manifest themselves in the form of spurious node-to-node oscillations, primarily in the velocity field. Moreover for incompressible fluid flow, the peculiar form of Eq.~(\ref{eq:fl_cons_mass_st}), which is devoid of the pressure, leads to a saddle-point problem where the pressure merely acts as a Lagrange multiplier to satisfy the incompressibility constraint. Such problems are constrained by a solvability condition, known as the LBB (Ladyzhenskay-Babu\v{s}ka-Brezzi) condition, the violation of which leads to spurious oscillations in the pressure field.

Several solution strategies were developed to circumvent the numerical instabilities associated with the Bubonov-Galerkin form for problems involving incompressible fluid flow. For this study, we employ the consistently stabilized approach, the streamline upwind Petrov-Galerkin (SUPG)~(\cite{brooks1982}) to overcome the difficulties associated with the convective operator, while the pressure stabilizing Petrov-Galerkin (PSPG)~(\cite{hughes1986b}) approach is used to circumvent the LBB constraint. Additionally, the least-square incompressibility stabilization (LSIC)~(\cite{hansbo1990}) approach is utilized for ensuring mass conservation at the discrete level. The discrete stabilized formulation of the coupled problem is given as:
\begin{align}
&\text{Find } \boldsymbol{v}^\text{h}_\text{f} \in \boldsymbol{\mathcal{S}}^\text{h}_v\text{, } p^\text{h}_\text{f}\in \mathcal{S}^\text{h}_p \text{ and } \boldsymbol{u}^\text{h}_\text{s} \in \boldsymbol{\mathcal{S}}^\text{h}_u,  \text{ such that} \nonumber\\
 &\int\limits_{\mathcal{B}_\text{f}^\text{h}} \rho_\text{f} \, \boldsymbol{w}^\text{h}\cdot \frac{\partial \boldsymbol{v}_\text{f}^\text{h}}{\partial t} \bigg|_\chi \text{d}v + \int\limits_{\mathcal{B}_\text{f}^\text{h}} \!\rho_\text{f} \,\boldsymbol{w}^\text{h} \cdot \boldsymbol{c}^\text{h} \cdot \text{grad} \, \boldsymbol{v}^\text{h}_\text{f} \, \text{d}v + \int\limits_{\mathcal{B}^\text{h}_\text{f}}\!\text{grad} \,\boldsymbol{w}^\text{h}: \boldsymbol{\sigma}_\text{f}^\text{h}  \, \text{d}v   \nonumber \\ 		
 &\hspace{1.0cm} - \int\limits_{\mathcal{B}_\text{f}^\text{h}} \!\rho_\text{f} \, \boldsymbol{w}^\text{h} \cdot \boldsymbol{b}_\text{f} \,\, \text{d}v  - \int\limits_{\partial_t \mathcal{B}_\text{f}^\text{h}} \! \boldsymbol{w}^\text{h} \cdot \bar{\boldsymbol{t}}_\text{f} \,\, \text{d}a + \sum_{e=1}^{n_{fe}} \, \underbrace{\int\limits_{\Omega_e} \! \tau_m \, \boldsymbol{c}^\text{h} \cdot \text{grad}\, \boldsymbol{w}^\text{h} \cdot \boldsymbol{\mathcal{R}}_m(\boldsymbol{v}_\text{f}^\text{h},p_\text{f}^\text{h})\, \text{d}v }_\text{SUPG stabilization}  \nonumber \\   
 &\hspace{8.0cm} + \sum_{e=1}^{n_{fe}} \,  \underbrace{\int\limits_{\Omega_e} \tau_c \,  (\,\text{div} \, \boldsymbol{w}^\text{h}\,) \, \mathcal{R}_c(\boldsymbol{v}_\text{f}^\text{h})\, \text{d}v}_\text{LSIC stabilization}  = 0 \,, \label{w_st_mom} \\		
	&\int\limits_{\mathcal{B}_\text{f}^\text{h}} q^\text{h} \, \text{div}\, \boldsymbol{v}_\text{f}^\text{h} \, \text{d}v + \sum_{e=1}^{n_{fe}} \, \underbrace{\int\limits_{\Omega_e}  \tau_m \,\, \text{grad}\,q^\text{h} \cdot \boldsymbol{\mathcal{R}}_m(\boldsymbol{v}_\text{f}^\text{h},p_\text{f}^\text{h})\,\, \text{d}v}_\text{PSPG stabilization}= 0 \,,\label{w_st_mass} 	\\ 
	&\int\limits_{\mathcal{B}_{\text{s}_\text{o}}^\text{h}} \!\rho_{\text{s}_\text{o}} \delta \boldsymbol{u}^\text{h} \cdot\frac{D^2 \boldsymbol{u}^\text{h}_\text{s}}{Dt^2} \, \text{d}V + \! \int\limits_{\mathcal{B}_{\text{s}_\text{o}}^\text{h}} \! \text{Grad} \, \delta \boldsymbol{u}^\text{h} : \left( \boldsymbol{F S_\text{s}^\text{h}}\right) \, \text{d}V - \int\limits_{\mathcal{B}_{\text{s}_\text{o}}^\text{h}} \!\!  \rho_{\text{s}_\text{o}} \,\delta \boldsymbol{u}^\text{h} \cdot  \boldsymbol{b}_\text{s} \, \text{d}V - \! \int\limits_{\partial_t \mathcal{B}_{\text{s}_\text{o}}^\text{h}} \!\! \delta \boldsymbol{u}^\text{h} \cdot \bar{\boldsymbol{t}}_\text{s}\, \text{d}A =0\,, \label{eq:w_st_solid} \\
	&\forall \, \boldsymbol{w}^\text{h} \in \boldsymbol{\mathcal{V}}^\text{h}_v, q^\text{h} \in \mathcal{V}^\text{h}_p \text{ and }\delta \boldsymbol{u}^\text{h} \in \boldsymbol{\mathcal{V}}^\text{h}_u \, .  \nonumber
	\end{align}
	Here $\boldsymbol{\mathcal{R}}_m(\boldsymbol{v}_\text{f}^\text{h},p_\text{f}^\text{h})$ and $\mathcal{R}_c(\boldsymbol{v}_\text{f}^\text{h})$ represent the residuals obtained from Eqs.~(\ref{eq:fl_cons_mom_st}-\ref{eq:fl_cons_mass_st}), such that
\begin{align}
	\boldsymbol{\mathcal{R}}_m(\boldsymbol{v}_\text{f}^\text{h},p_\text{f}^\text{h}) &\coloneqq \rho_\text{f} \, \left(  \frac{\partial \boldsymbol{v}_\text{f}^\text{h}}{\partial t} \bigg|_{\boldsymbol{\chi}}  + \boldsymbol{c}^\text{h} \cdot \text{grad} \,{\boldsymbol{v}_\text{f}^\text{h}} \right) -  \text{div} \, \boldsymbol{\sigma}_\text{f}^\text{h} - \rho_\text{f} \, \boldsymbol{b}_\text{f} \,, \label{eq:discrete_residual} \\
	\mathcal{R}_c(\boldsymbol{v}_\text{f}^\text{h}) &\coloneqq \text{div} \, \boldsymbol{v}_\text{f}^\text{h} \, , \label{eq:discrete_residual_cont}
\end{align}
while $\tau_m$ and $\tau_c$ are stabilization parameters which are also known as the intrinsic time scales. The optimal definition of the stabilization parameters has been an area of intense research and over the years several alternate definitions have been proposed in the context of stabilized finite elements for fluid dynamics (e.g. see~\cite{brooks1982,hughes1986b,codina2000,franca2000,tezduyar2000,bazilevs2007vms}). Here, we employ the definition presented in~\cite{bazilevs2007vms} in the context of the variational multiscale approach. Thus, for $\tau_m$, we have
\begin{align}
	\tau_m = \left[ \, \frac{4}{\Delta t^2} + \boldsymbol{c}^\text{h} \cdot \boldsymbol{G} \, \boldsymbol{c}^\text{h} + \frac{12}{\text{m}^e} \, \nu_\text{f}^2 \, \boldsymbol{G}: \boldsymbol{G} \, \right]^{-\frac{1}{2}} \,,
\end{align}
with the tensor $\boldsymbol{G}$ defined by
\begin{align}
	G_{ij} = \sum\limits_{k=1}^{d_\text{f}} \frac{\partial \xi_k}{\partial x_i} \, \frac{\partial \xi_k}{\partial x_j} \,.
\end{align}
Here, $\partial \boldsymbol{\xi}/\partial \boldsymbol{x}$ can be construed as the inverse Jacobian matrix of the mapping between the parametric and physical element, $\Delta t$ is the discrete time step size and  $\nu_\text{f}=\mu_\text{f}/\rho_\text{f}$ is the kinematic viscosity of the fluid. Additionally, $\tau_m$ also depends on the constant $\text{m}^e$, where
\begin{align}
	\text{m}^e = \min \left\{ \frac{1}{3}, \text{C}_\text{inv} \right\} \,.
\end{align}
The value of $\text{C}_\text{inv}$ is chosen such that the stabilized weak form is coercive for a given discretization. Estimates for $\text{C}_\text{inv}$ for triangular and quadrilateral Lagrangian elements were formulated in~\cite{harari1992}, which were later shown to be reasonable estimates for NURBS-based finite elements as well in~\cite{gamnitzer2010b}. For $\tau_c$, we have
\begin{align}
	\tau_c = \left( \, \tau_m \, \boldsymbol{g} \cdot \boldsymbol{g} \, \right)^{-1} \,,
\end{align}
where 
\begin{align}
	g_i = \sum\limits_{j=1}^{d_\text{f}} \frac{\partial \xi_j}{\partial x_i} \,.
\end{align}

Within a unique finite element $e$, the discrete value of the involved variables are obtained through a linear combination of basis functions chosen from the discrete functional space. We retain the isoparameteric concept and represent all variables with similar basis functions. Hence, the value of the test functions and the trial solutions within the element $e$ are estimated as
\begin{align}
	&\boldsymbol{w} \approx\boldsymbol{w}^\text{h} = \textbf{N} \, \textbf{w}^e \,,\text{\hspace{0.5cm}}	q \approx q^\text{h} = \tilde{\textbf{N}} \, \textbf{q}^e \,, \text{\hspace{0.5cm}} \delta \boldsymbol{u} \approx\delta \boldsymbol{u}^\text{h} = \textbf{N} \, \delta\textbf{u}^e\,, \label{eq:disc_delu} \\
	&\boldsymbol{v}_\text{f} \approx \boldsymbol{v}_\text{f}^\text{h} = \textbf{N} \,  \textbf{v}_\text{f}^e\,,  \text{\hspace{0.6cm}} p_\text{f} \approx p_\text{f}^\text{h} =\tilde{\textbf{N}} \, \textbf{p}_\text{f}^e \,,\text{\hspace{0.5cm}}\boldsymbol{u}_\text{s} \approx \boldsymbol{u}_\text{s}^\text{h} = \textbf{N} \, \textbf{u}_\text{s}^e\,,	\label{eq:disc_p}
\end{align}   
respectively, and a similar representation of the geometry is employed, i.e.,
\begin{align}
\boldsymbol{x} \approx \boldsymbol{x}^\text{h} = \textbf{N} \, \textbf{x}^e \text{\hspace{0.6cm}and\hspace{0.6cm}} \boldsymbol{X} \approx \boldsymbol{X}^\text{h} = \textbf{N} \, \textbf{X}^e \,. \label{eq:disc_x}
\end{align}
The basis functions for the element $e$ are collected in the vectors $\textbf{N}$ and $\tilde{\textbf{N}}$, such that
\begin{align}
	\textbf{N} &= \left[ \, N_1 \, \textbf{I}, \, N_2 \, \textbf{I}, \, \cdots , N_{n_n} \textbf{I} \, \right] \,, \\
	\tilde{\textbf{N}} &=  \left[ \, N_1, \, N_2 , \, \cdots , N_{n_n} \, \right] \,,
\end{align}
where $n_n$ denotes the total number of nodes and control points of element $e$. The definition of the individual basis functions $N_\text{A}$ is discussed in Section~\ref{sec-enrich}. The scalar and the vector variables are collected as
\begin{align}
	\left(\textbf{w}^e \right)^\text{T} &= \left[ \,  \boldsymbol{w}_1^\text{T},\,  \boldsymbol{w}_2^\text{T}, \, \cdots, \, \boldsymbol{w}_{n_n}^\text{T}\, \right] \,, \\
	\left(\textbf{q}^e\right)^\text{T} &= \left[ \, q_1, \, q_2,\, \cdots, \, q_{n_n}\,\right] \,,
\end{align}
and similar arrangements for $\delta \textbf{u}^e, \, \textbf{v}_\text{f}^e, \, \textbf{p}_\text{f}^e, \, \textbf{u}_\text{s}^e, \, \textbf{x}^e$ and $\textbf{X}^e$. The approximations of Eqs.~(\ref{eq:disc_delu}-\ref{eq:disc_x}) are substituted into Eqs.~(\ref{w_st_mom}-\ref{eq:w_st_solid}) and a finite element assembly operation is performed to obtain the discrete balance equations at the global level. Considering that the weak form is valid for all admissible choices of $(\boldsymbol{w},\,q,\,\delta \boldsymbol{u})$, we obtain the nodal force balances at the non-Dirichlet nodes of the FSI model:
\begin{align}
	\textbf{M}_\text{f}\, \frac{\partial \textbf{v}_\text{f}}{\partial t} \bigg|_\chi + \textbf{f}_\text{adv}(\textbf{v}_\text{f}) +\textbf{f}_\text{fint}(\textbf{v}_\text{f},\textbf{p}_\text{f}) + \textbf{f}_{\text{SUPG}}(\textbf{v}_\text{f},\textbf{p}_\text{f})  + \textbf{f}_{\text{LSIC}}(\textbf{v}_\text{f}) -  \textbf{f}_\text{fext} &= \textbf{0} =:  \textbf{R}_{\text{m}}\, , \label{eq:fe_fl_mom_stab}\\
\textbf{f}_\text{con} (\textbf{v}_\text{f})  + \textbf{f}_\text{PSPG}(\textbf{v}_\text{f},\textbf{p}_\text{f})&= \textbf{0} =: \textbf{R}_{\text{c}}  \,, \label{eq:fe_fl_mass_stab} \\
\textbf{M}_\text{s}\, \frac{D^2 \textbf{u}_\text{s}}{Dt^2} + \textbf{f}_{\text{sint}}(\textbf{u}) - \textbf{f}_\text{sext} &= \boldsymbol{0}  =:\textbf{R}_{\text{s}}\, . \label{eq:fe_s_mom_stab}
\end{align}
The terms $\textbf{f}_\text{con}$ and $\textbf{f}_\text{sint}$ (and subsequently $\textbf{f}_\text{SUPG}$ and $\textbf{f}_\text{PSPG}$) are inherently nonlinear in nature due to the solid material model and the Eulerian description of the fluid, respectively. A Newton-Raphson method based predictor-multicorrector algorithm is used to linearize and obtain the solution of this system. A detailed derivation and the definition of the individual terms involved in Eqs.~(\ref{eq:fe_fl_mom_stab}-\ref{eq:fe_s_mom_stab}) is presented in Appendix~\ref{appx:fe_eq}.
\subsubsection{Temporal discretization and linearization}
Eqs.~(\ref{eq:fe_fl_mom_stab}-\ref{eq:fe_s_mom_stab}) represent a set of first- and second-order ODEs in time, respectively. In order to integrate these equations in time, we employ the generalized-$\alpha$ method, which is an extension of the Newmark method. It was first presented in~\cite{chung1993} in the context of second-order systems (solid mechanics) and was later extended to first-order Navier-Stokes in~\cite{jansen1999}. It has since been successively applied to many FSI problems (e.g., see~\cite{kuhl2003,bazilevs2010,eken2016}).

Let us proceed by first expressing the nodal force balance of Eqs.~(\ref{eq:fe_fl_mom_stab}-\ref{eq:fe_s_mom_stab}) in a unified residual vector such that,
\begin{align}
	\textbf{R}\left(\dot{\textbf{v}}_\text{f}, \textbf{v}_\text{f},\textbf{p}_\text{f},\ddot{\textbf{u}}_\text{s},\textbf{u}_\text{s}\right) := \left\{ \,
		\begin{array}{c}
			\textbf{R}_\text{m}(\dot{\textbf{v}}_\text{f}, \textbf{v}_\text{f},\textbf{p}_\text{f}) \\
			\textbf{R}_\text{c} (\dot{\textbf{v}}_\text{f}, \textbf{v}_\text{f},\textbf{p}_\text{f}) \\
			\hspace{-0.3cm}\textbf{R}_\text{s}(\ddot{\textbf{u}}_\text{s},\textbf{u}_\text{s})
		\end{array} \,
	\right\}  = \boldsymbol{0} \,,
\end{align}
where an over-dot denotes a time-derivative. The generalized-$\alpha$ method requires the evaluation of the residual at intermediate time-levels $n+\alpha_m$ and $n+\alpha_f$ to advance the system from time-level $n$ to $n+1$, i.e., 
\begin{align}
	\textbf{R}^{n+1} = \textbf{R}\left(\dot{\textbf{v}}_\text{f}^{n+\alpha_m}, \textbf{v}_\text{f}^{n+\alpha_f},\textbf{p}_\text{f}^{n+1},\ddot{\textbf{u}}_\text{s}^{n+\alpha_m},\textbf{u}_\text{s}^{n+\alpha_f}\right) \,, \label{eq:alp_residual}
\end{align}
where the intermediate time-levels are given as
\begin{align}
	t^{n+\alpha_m} &= t^n + \alpha_m \, \Delta t = t^n + \alpha_m \,(\,t^{n+1}-t^n\,) \,, \\
	t^{n+\alpha_f} &= t^n + \alpha_f \, \Delta t = t^n + \alpha_f\,(\,t^{n+1}-t^n\,)
\end{align}
and consequently the intermediate description of the field variables is obtained as
\begin{align}
	\dot{\textbf{v}}_\text{f}^{n+\alpha_m} &= \dot{\textbf{v}}_\text{f}^{n} + \alpha_m \, ( \, \dot{\textbf{v}}_\text{f}^{n+1} - \dot{\textbf{v}}_\text{f}^{n} \,) \, , \\
	\textbf{v}_\text{f}^{n+\alpha_f} &= \textbf{v}_\text{f}^{n} + \alpha_f \, ( \, \textbf{v}_\text{f}^{n+1} - \textbf{v}_\text{f}^{n} \, ) \,, \\
	\ddot{\textbf{u}}_\text{s}^{n+\alpha_m} &= \ddot{\textbf{u}}_\text{s}^{n} + \alpha_m \, ( \, \ddot{\textbf{u}}_\text{s}^{n+1} - \ddot{\textbf{u}}_\text{s}^{n} \,) \, , \\
	\textbf{u}_\text{s}^{n+\alpha_f} &= \textbf{u}_\text{s}^{n} + \alpha_f \, ( \, \textbf{u}_\text{s}^{n+1} - \textbf{u}_\text{s}^{n} \, ) \,.
\end{align}
Additionally, the discrete Newmark formulae are used to estimate the field description at time level $n+1$, i.e.,
\begin{align}
	\textbf{v}_\text{f}^{n+1} &= \textbf{v}_\text{f}^n + \Delta t \, \left[ \,  \left(1-\gamma \right) \dot{\textbf{v}}^n + \gamma \, \dot{\textbf{v}}_\text{f}^{n+1} \right] \,, \\
	\textbf{u}_\text{s}^{n+1} &= \textbf{u}_\text{s}^n + \Delta t \, \dot{\textbf{u}}_\text{s}^n + \frac{\Delta t^2}{2} \left[ \, (1-2\,\beta) \, \ddot{\textbf{u}}_\text{s}^n + 2\,\beta\, \ddot{\textbf{u}}_\text{s}^{n+1}\,\right] \, .
\end{align}
Here, $\alpha_m, \, \alpha_f, \, \gamma$ and $ \beta$ are real-valued parameters that determine the character of the time-integration scheme. A subsequent Taylor expansion of a two-step time-integration scheme obtained through the generalized-$\alpha$ method reveals a second-order accurate method if
\begin{align}
	\gamma = \frac{1}{2} + \alpha_m - \alpha_f \, \text{\hspace{0.5cm}and\hspace{0.5cm}} \beta = \frac{1}{4} \left( 1+\alpha_m - \alpha_f\right)^2 \,,
\end{align}
and a subsequent evaluation of the eigenvalues associated with the amplification matrix of the scheme guarantees unconditional stability if
\begin{align}
	\alpha_m \geq \alpha_f \geq \frac{1}{2} \,.
\end{align}
Using the spectral radius $\rho_\infty$ of the amplification matrix of the two-step method, a family of methods were proposed in~\cite{chung1993} to obtained second-order, unconditionally stable schemes for second-order system. It was suggested to have
\begin{align}
	\alpha_m \big|_2 = \frac{2-\rho_\infty}{1+\rho_\infty} \text{\hspace{0.5cm}and\hspace{0.5cm}} \alpha_f \big|_2 = \frac{1}{1+\rho_\infty} \,,   \label{eq:alp_2nd}
\end{align}
for second-order systems, while a similar analysis in~\cite{jansen1999} recommends
\begin{align}
	\alpha_m \big|_1 = \frac{1}{2} \left( \frac{3-\rho_\infty}{1+\rho_\infty} \right) \text{\hspace{0.5cm}and\hspace{0.5cm}} \alpha_f \big|_1 = \frac{1}{1+\rho_\infty} \,,  \label{eq:alp_1st}
\end{align} 
for the first-order Navier-Stokes system. A value of $\rho_\infty=1$ results in a non-dissipative scheme, where all frequencies are retained in the analysis, essentially requiring a very fine time discretization. While $\rho_\infty=0$ corresponds to the asymptotic annihilation case, where high-frequency components are annihilated in a single solution step. It is evident from Eqs.~(\ref{eq:alp_2nd}-\ref{eq:alp_1st}), that in the context of FSI analysis, there is a mismatch between $\alpha_m \big|_2$ and $\alpha_m \big|_1$. Such an arrangement has the tendency to lead to asynchronous evaluation of time-derivatives, which can subsequently lead to the loss of accuracy. We employ the definition given by Eq.~(\ref{eq:alp_1st}) for both the fluid and the solid system, as suggested in~\cite{bazilevs2008}. Such an arrangement guarantees consistent evaluation of all the time-derivatives and provides optimal numerical dissipation for the fluid problem, which is stiffer in nature. Moreover, with this choice it is ensured that the spectral radius for the second-order system remains well bounded and always stays below 1, hence a stable scheme. For all subsequent examples we consider $\rho_\infty=1/2$.

As mentioned earlier, the semi-discrete system of Eqs.~(\ref{eq:fe_fl_mom_stab}-\ref{eq:fe_s_mom_stab}) is inherently nonlinear by nature. The generalized-$\alpha$ method is invoked in the framework of a Newton-Raphson method to linearize and subsequently advance the simulation in time. Using an iterative predictor-multicorrector approach, we expand the residual of Eq.~(\ref{eq:alp_residual}) about a previously $k^{th}$-iterate solution for the residual at $(k+1)^{th}$-iterate, which should ideally be zero, i.e., 
\begin{align}
	\textbf{R}^{n+1,k+1} = \textbf{R}^{n+1,k} + \frac{\partial \textbf{R}^{n+1,k}}{\partial \bar{\textbf{a}}^{n+1}} \, \Delta \bar{\textbf{a}}^{n+1,k}= \boldsymbol{0} \,, \label{eq:newt_raph}
\end{align} 
where $\bar{\textbf{a}} = \left[\, \dot{\textbf{v}}_\text{f}^\text{T}, \, \textbf{p}_\text{f}^\text{T}, \, \ddot{\textbf{u}}_\text{s}^\text{T} \right]^\text{T}$ is a vector composed of the nodal accelerations and fluid pressure $\mathrm{dofs}$. In the predictor-multicorrector framework, the solution procedure is implemented as follows:
\paragraph*{Predictor:}
The counter $k$ is set to zero and a reasonable prediction is made for the field variables. We employ a constant velocity predictor such that:
\begin{align}
	\textbf{v}_\text{f}^{n+1,0} &=\textbf{v}_\text{f}^n \, , \\					
	\textbf{p}_\text{f}^{n+1,0} &= \textbf{p}_\text{f}^n \, ,	\\	
	\dot{\textbf{u}}_\text{s}^{n+1,0} &= \dot{\textbf{u}}_\text{s}^n \, , \\	
	\dot{\textbf{v}}_\text{f}^{n+1,0} &=  \left( \frac{\gamma-1}{\gamma} \right)\, \dot{\textbf{v}}_\text{f}^n + \frac{\textbf{v}_\text{f}^{n+1,0} - \textbf{v}_\text{f}^{n}}{\gamma \, \Delta t} \, , \\ 
	\ddot{\textbf{u}}_\text{s}^{n+1,0} &= \left( \frac{\gamma-1}{\gamma} \right)\, \ddot{\textbf{u}}_\text{s}^n + \frac{\dot{\textbf{u}}_\text{s}^{n+1,0} - \dot{\textbf{u}}_\text{s}^{n}}{\gamma \, \Delta t} \, ,\\		
	\textbf{u}_\text{s}^{n+1,0} &= \textbf{u}_\text{s}^n + \Delta t \, \dot{\textbf{u}}^n_\text{s} + \frac{\Delta t^2}{2}\left[ \, (1-2\, \beta)\, \ddot{\textbf{u}}_\text{s}^n + 2\,\beta\, \ddot{\textbf{u}}_\text{s}^{n+1,0}\,\right] \,. 
\end{align}
\paragraph*{Corrector:} 
After the predictor step, the execution of the solver moves into an iterative corrector step for $k=0,1,2,\dots,k_{max}$ or until an acceptable convergence is achieved. The multi-corrector step conducts the following sequence recursively:  
\begin{enumerate}[label= \arabic*:]
	\item Collocation of field entities at the intermediate time-level:
\begin{align}
\dot{\textbf{v}}_\text{f}^{n+\alpha_m,k} &= \dot{\textbf{v}}_\text{f}^n + \alpha_m \left( \dot{\textbf{v}}_\text{f}^{n+1,k} - \dot{\textbf{v}}_\text{f}^n \right) \, ,\\		
\textbf{v}_\text{f}^{n+\alpha_f,k} &= \textbf{v}_\text{f}^n + \alpha_f \left( \textbf{v}_\text{f}^{n+1,k} - \textbf{v}_\text{f}^n \right) \, ,\\		
\textbf{p}_\text{f}^{n+1,k} &= \textbf{p}_\text{f}^{n+1,k} \, , \\   
\ddot{\textbf{u}}_\text{s}^{n+\alpha_m,k} &= \ddot{\textbf{u}}_\text{s}^n + \alpha_m \left( \ddot{\textbf{u}}_\text{s}^{n+1,k} - \ddot{\textbf{u}}_\text{s}^n \right) \, ,\\		
\dot{\textbf{u}}_\text{s}^{n+\alpha_f,k} &= \dot{\textbf{u}}_\text{s}^n + \alpha_f \left( \dot{\textbf{u}}_\text{s}^{n+1,k} - \dot{\textbf{u}}_\text{s}^n \right) \, ,\\		
\textbf{u}_\text{s}^{n+\alpha_f,k} &= \textbf{u}_\text{s}^n + \alpha_f \left( \textbf{u}_\text{s}^{n+1,k} - \textbf{u}_\text{s}^n \right) \, .		
\end{align}
	\item Setup of the linear system of Eq.~(\ref{eq:newt_raph}) using the intermediate field entities and the solution for the incremental improvements $\Delta \bar{\textbf{a}}$ for the $k^{th}$-iteration:
\begin{align}
	\frac{\partial \textbf{R}^{n+1,k}}{\partial \bar{\textbf{a}}^{n+1}} \, \Delta \bar{\textbf{a}}^{n+1,k} = - \, \textbf{R}^{n+1,k}  \,.		\label{eq:residual_eq}
\end{align}
	\item Update of the field variables using the incremental vector $\Delta \bar{\textbf{a}}^{n+1,k}$:
	\begin{align}
		\dot{\textbf{v}}_\text{f}^{n+1,k+1} &= \dot{\textbf{v}}_\text{f}^{n+1,k} + \Delta \dot{\textbf{v}}_\text{f}^{n+1,k} \, , \\
		\textbf{v}_\text{f}^{n+1,k+1} &= \textbf{v}_\text{f}^{n+1,k} + \gamma \, \Delta t  \, \Delta \dot{\textbf{v}}_\text{f}^{n+1,k} \, , \\
		\textbf{p}_\text{f}^{n+1,k+1} &= \textbf{p}_\text{f}^{n+1,k} + \Delta \textbf{p}_\text{f}^{n+1,k} \, , \\
		\ddot{\textbf{u}}_\text{s}^{n+1,k+1} &= \ddot{\textbf{u}}_\text{s}^{n+1,k} + \Delta \ddot{\textbf{u}}_\text{s}^{n+1,k} \, , \\
		\dot{\textbf{u}}_\text{s}^{n+1,k+1} &= \dot{\textbf{u}}_\text{s}^{n+1,k} + \gamma \, \Delta t \,\Delta \ddot{\textbf{u}}_\text{s}^{n+1,k} \, , \\
		\textbf{u}_\text{s}^{n+1,k+1} &= \textbf{u}_\text{s}^{n+1,k} + \beta\, (\Delta t)^2 \Delta \ddot{\textbf{u}}_\text{s}^{n+1,k} \, .
	\end{align}
The convergence of the algorithm for the iteration $k$ is tested by formulating a relevant norm which is a function of the incremental vector $\Delta \bar{\boldsymbol{a}}^{n+1,k}$ and the residual vector $\textbf{R}^{n+1,k}$. If this value is less than an acceptable convergence criteria, the time step $n+1$ is considered converged, otherwise the flow of the program is transfered back to corrector-step 1 for the $(k+1)^{th}$ corrector iteration. 
\end{enumerate}
\subsubsection{The linearized monolithic coupled system}
The construction and the solution of the linearized system of Eq.~(\ref{eq:residual_eq}) constitute the most computational intensive part of the predictor-multicorrector algorithm. Once all the element level contributions have been calculated, the entries are moved to a global system through a finite element assembly operation. For a monolithic solution procedure, the global system prior to the enforcement of the coupling conditions is of the form:
\begin{align}
	\left[
	\begin{array}{c c}
		\textbf{K}_\text{f}	& \textbf{0} \\
		\textbf{0}			& \textbf{K}_\text{s} 
	\end{array}
	\right]	\, \left\{
	\begin{array}{c}
		\Delta \bar{\textbf{a}}_\text{f}	 \\
		\Delta \bar{\textbf{a}}_\text{s}		 
	\end{array}
	\right\} = -\left\{
	\begin{array}{c}
		\textbf{f}_\text{f}	 \\
		\textbf{f}_\text{s}		 
	\end{array}
	\right\} \,,		\label{eq:mon_fsi_orig}
\end{align} 
where $\textbf{K}_\text{f}$ and $\textbf{f}_\text{f}$ are the tangent and the residual force vector associated with the fluid media, while $\textbf{K}_\text{s}$ and $\textbf{f}_\text{s}$ are corresponding entities for the solid media. Detail derivations for these term are given in Appendix~\ref{appx:fe_eq}. It should be noted that \mbox{$\Delta \bar{\textbf{a}}_\text{f}= \left[ \, \Delta \dot{\textbf{v}}_\text{f}^\text{T}, \, \Delta \textbf{p}_\text{f}^\text{T}	\, \right]^\text{T}$}, while $\Delta \bar{\textbf{a}}_\text{s} = \Delta \ddot{\textbf{u}}_\text{s}$. In order to enforce the interfacial conditions at the fluid-solid interface in a finite element setting, let us isolate the interfacial $\mathrm{dofs}$. Let $\mathrm{I}$ be the set of interfacial $\mathrm{dofs}$ for the respective media on which it operates. Consequently, we define A as a set that holds all non-interfacial $\mathrm{dofs}$ for the fluid, while the set B is comprised of all non-interfacial $\mathrm{dofs}$ for the solid media. The coupling conditions of Eq.~(\ref{eq:coupling_cond}) in a monolithic finite element setting are enforced as
\begin{align} 
\left[
	\arraycolsep=3pt\def\arraystretch{1.3}
	\begin{array}{c c c}
		\textbf{K}_\text{f}^\text{AA}	& \textbf{K}_\text{f}^\text{AI}	& \textbf{0} \\
		\textbf{K}_\text{f}^\text{IA}	& \left(\textbf{K}_\text{f}^\text{II}+\textbf{K}_\text{s}^\text{II}\right)		& \textbf{K}_\text{s}^\text{IB} \\
		\textbf{0}						& \textbf{K}_\text{s}^\text{BI}	& \textbf{K}_\text{s}^\text{BB} 
	\end{array}
	\right]	\, \left\{
	\arraycolsep=3pt\def\arraystretch{1.3}
	\begin{array}{c}
		\Delta \bar{\textbf{a}}_\text{f}^\text{A}	 \\
		\Delta \bar{\textbf{a}}^\text{I}	\\
		\Delta \bar{\textbf{a}}_\text{s}^\text{B}		 
	\end{array}
	\right\} = -\left\{
	\arraycolsep=3pt\def\arraystretch{1.3}
	\begin{array}{c}
		\textbf{f}_\text{f}^\text{\,A}	 \\
		\textbf{f}_\text{f}^\text{\,I} + \textbf{f}_\text{s}^\text{\,I}	 \\
		\textbf{f}_\text{s}^\text{\,B}		 
	\end{array}
	\right\} \,.		\label{eq:coupled_fe_sys}
\end{align} 
Eq.~(\ref{eq:coupled_fe_sys}) constitute the fully-coupled linearized FSI system. Dirichlet boundary conditions can now be imposed on the coupled system and a solution to Eq.~(\ref{eq:coupled_fe_sys}) yields the incremental improvements $\Delta \bar{\textbf{a}}$ for the iterate $k$.
%
\subsection{Grid motion}
For FSI simulations, the motion of the fluid-solid interface boundary needs to be accommodated smoothly within the fluid mesh $\mathcal{B}^\text{h}_\text{f}$. At the discrete level, the ALE representation of Eq.~(\ref{eq:fl_cons_mom_st}) needs to be augmented with a mesh update algorithm such that the motion of the interface boundary can be adjusted in $\mathcal{B}^\text{h}_\text{f}$ without excessive degradation and distortion of the computational mesh. Several mesh update algorithms exist in this regard. These include algebraic schemes~(e.g. \cite{yigit2008,ahn2010}), Laplace smoothers~(e.g. \cite{loehner1996}), spring/elastic medium analogy~(e.g. \cite{farhat1998,johnson1994}) and mesh regeneration schemes~(e.g. \cite{johnson1999}), among others.  

\begin{figure}[!ht]
  	\centering
  	\includegraphics[width=1\textwidth]{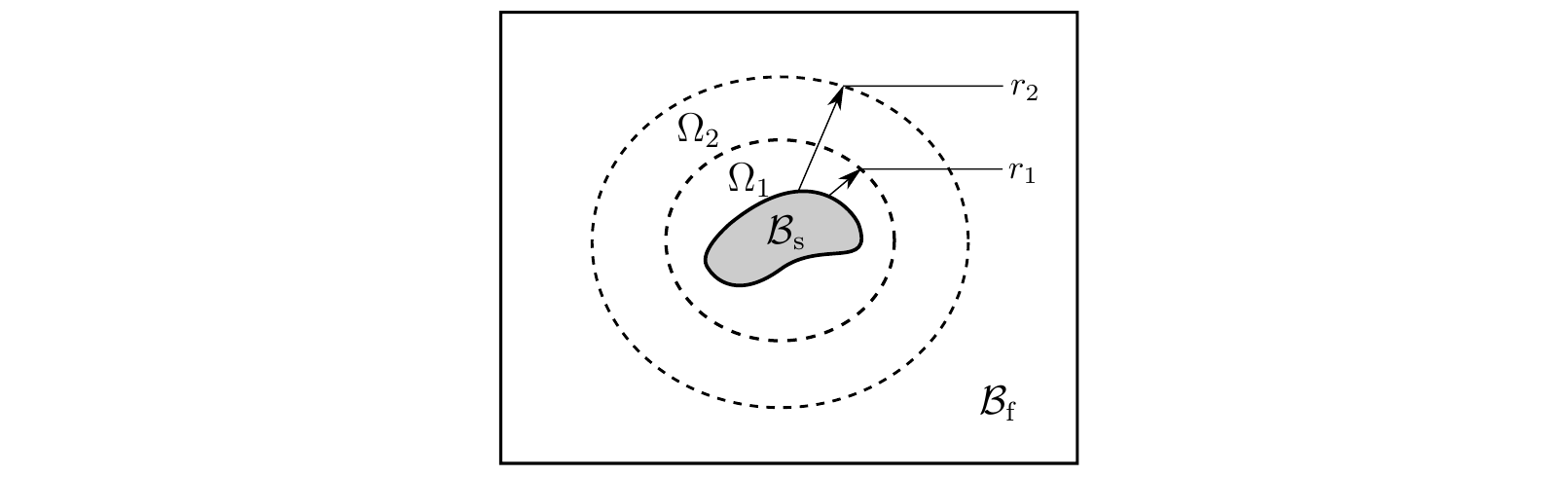}
  	\caption[Region decomposition for the mesh motion technique.]{Region decomposition for the mesh motion technique. The rigidly moving region in the immediate vicinity of $\mathcal{B}_\text{s}$ is denoted as $\Omega_1$, while $(\mathcal{B}_\text{f} \backslash \Omega_2)$ is the region farthest away from $\mathcal{B}_\text{s}$ and remains static. $(\mathcal{B}_\text{f} \backslash \Omega_1)$ is the buffer region where the displacements are absorbed.}
	\label{fig:mesh_motion}
\end{figure}

For this study, we employ a simplistic algebraic mesh motion algorithm, similar to the strategy described in~\cite{ahn2010}. With reference to Figure~\ref{fig:mesh_motion}, let us consider a rigidly moving solid body $\mathcal{B}_\text{s}$. We identify three distinct regions in $\mathcal{B}_\text{f}$. The region $\Omega_1$ contains all those points of $\mathcal{B}_\text{f}$ whose radius from the center of $\mathcal{B}_\text{f}$ doesnot exceed $r_1$. Similarly, $\Omega_2$ contains points of $\mathcal{B}_\text{f}$ whose radius is smaller than $r_2$. Note that $\Omega_1$ and $\Omega_2$ are overlapping. The remaining outer region of $\mathcal{B}_\text{f}$ can then be identified as $\mathcal{B}_\text{f} \backslash \Omega_2$. A purely Lagrangian description is imposed on the points included in $\Omega_1$. This is achieved by moving the points with the same velocity as the rigidly moving solid body $\mathcal{B}_\text{s}$. At the same time, points in the $\mathcal{B}_\text{f} \backslash \Omega_2$ are kept static. The region between these two domains $(\text{i.e.},~\Omega_2 \backslash \Omega_1)$ then constitutes the buffer region, where points are moved so as to gradually absorb the largest displacements occuring at the periphery of $\Omega_1$ to zero displacement of points in $\mathcal{B}_\text{f} \backslash \Omega_2$. Several weighting functions $\vartheta$ can be formulated to interpolate the displacement of the points in the buffer region, the simplest of which is of the following linear form:
\begin{align}
	\vartheta(r) = 
	\begin{cases}
	1 \,, \text{\hspace{1.9cm}} \forall \, \boldsymbol{x} \in \Omega_1\,,\\
	\dfrac{r_2 - r}{ r_2 - r_1 } \,,   \text{\hspace{0.85cm}}\forall \, \boldsymbol{x} \in \left( \Omega_2 \backslash \Omega_1\, \right) , \\
	0 \,, \text{\hspace{1.9cm}} \forall \, \boldsymbol{x} \in \left( \mathcal{B}_\text{f} \backslash \Omega_2 \right).
	\end{cases}
\end{align}
Once the weighting associated with a particular point in the fluid domain has been determined, the new position and the velocity of the point are given as
\begin{align}
	\boldsymbol{x} &= \boldsymbol{X} + \vartheta(r) \, \boldsymbol{u}_\text{s}\ ,  \label{eq:mesh_updt_disp} \\ 
	\hat{\boldsymbol{v}} &= \vartheta(r) \, \boldsymbol{v}_\text{s} \, .		\label{eq:mesh_updt_vel}		
\end{align}
The scheme is easily extended to problems involving solid bodies undergoing deformations. This is achieved by coupling discrete points in $\mathcal{B}_\text{f}$ to the closest point on the fluid-solid interface. The updated configuration of the candidate point can then be obtained as a weighted multiple of the coupled interface point, as done in Eqs.~(\ref{eq:mesh_updt_disp}-\ref{eq:mesh_updt_vel}). For the numerical examples considered in this study, we employ an explicit representation of the mesh update scheme, i.e., the mesh configuration of the domain $\mathcal{B}_\text{f}^\text{h}$ for the time step $(n+1)$ is determined from the interface description of the state at the time step $n$. Such a treatment allows mesh updates without the need of solving additional equations and is observed to be reliable for the time step sizes considered in this study. 
%
\section{Isogeometric enrichment of the discretized domain}	\label{sec-enrich} 
Isogeometric analysis has the potential to increase the accuracy of finite element simulations, but it simultaneously increases the required computational effort (see Section~\ref{sec_intro} and Section~\ref{sec-ex}). Moreover, pure isogeometric discretization of volumetric domains is a challenging task that is still not solved for arbitrary geometries. Isogeometric enrichment mixes isogeometric elements (spline interpolation) with standard finite elements (Lagrange interpolation) in order to overcome these deficiencies. Within this framework, two general isogeometric enrichment strategies are described in Section~\ref{sec_strategy}, while the construction of the enriched elements for interfacing standard elements with isogeometric regions is presented in Section~\ref{sec_enrich2}. 


\subsection{Enrichment strategies}\label{sec_strategy}
The discretization of an entire domain with standard finite elements (see Figure~\ref{fig:enrich}a)) leads to an approximative representation of the geometry and to low-order continuity throughout the whole domain. Such drawbacks are addressed by isogeometric enrichment. The key idea is to identify and discretize regions that are most crucial for the accuracy of the physical problem with accurate and smooth isogeometric elements. These crucial regions of interest can be identified from a-priori knowledge of the physical problem being modeled. The remaining domain is discretized with efficient and low-order standard elements. The following two cases can be distinguished:

\begin{figure}[ht]
	\centering
  	\includegraphics[width=1.0\textwidth]{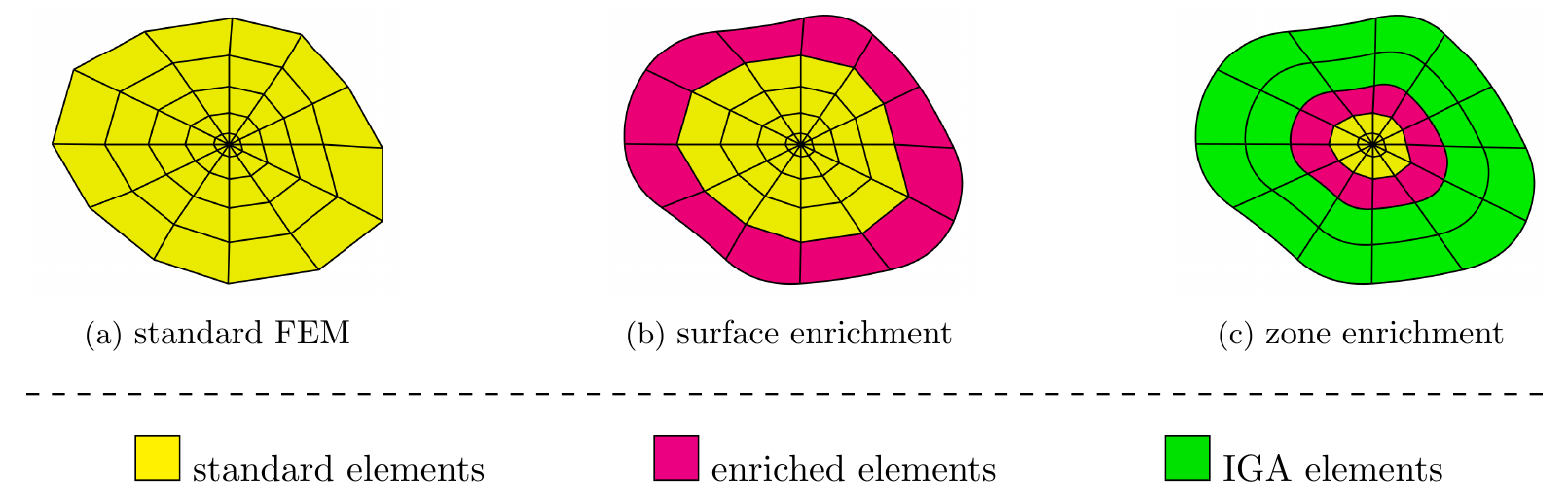} 
	\caption{Discretization of an exemplary two-dimensional domain with and without isogeometric enrichment. The enrichment strategies can be applied analogously in three-dimensions.} \label{fig:enrich}
\end{figure}

\subsubsection{Surface enrichment}
Isogeometric surface enrichment of the standard FEM (see Figure~\ref{fig:enrich}b) results in the exact representation of the underlying surface geometry. Moreover, such an enrichment provides a representation, where the field and the solution variables at the surface are continuously differentiable. In order to obtain a surface enriched discretization, a layer of isogemeotrically enriched elements (see Section~\ref{sec_enrich2}) is created at the surface. These special elements interface smooth spline representation at the surface with standard finite elements in the bulk volume. Surface enrichment yields significantly more accurate results and is particularly beneficial for surface-dominated engineering problems such as contact, friction and adhesion (see \cite{roger2011,roger2013} for two-dimensional and~\cite{corbett2014, corbett2015} for three-dimensional analysis). At the same time, the number of $\mathrm{dof}$s is only slightly increased compared to pure standard finite element discretizations. For coupled problems, isogeometric surface enrichment additionally offers the possibility of employing formulations that demand continuous surface representations without performing IGA for the entire coupled domain. An example is a wind turbine discretized with rotation-free shell elements~(e.g. formulated in~\cite{kiendl2009} and \cite{duong2016}) in a discretized fluid domain that is composed largely of standard finite elements. However, for problems where solution gradients are dominant within the volume, surface enrichment offers only marginal accuracy gains as demonstrated in~\cite{rasool2016} and \cite{harmel2016}.

\subsubsection{Zone enrichment}
Zone enrichment (see Figure~\ref{fig:enrich}c) denotes isogeometric discretization of certain volumetric regions of the domain, while the remaining domain is discretized with standard elements. A layer of isogeometrically enriched elements (see Section~\ref{sec_enrich2}) interfaces these regions. In physical problems, large gradients of the solution variables typically occur only in certain regions of the domain, such as boundary layers in flow problems or stress-concentrations in structural problems. Accurate representation of these potentially nonlinear gradients with smooth spline basis functions increases the accuracy of the numerical solution significantly compared to standard finite element discretizations, while the computational effort is substantially less than for pure isogeometric discretizations. Such an enrichment promises significant potential for many engineering applications such as fluid flow~(see~\cite{rasool2016}) and heat transfer~(see~\cite{harmel2016}).

\subsection{Isogeometrically enriched elements} \label{sec_enrich2}
The isogeometrically enriched finite element is crucial to constructing surface enriched and zone enriched discretizations. It has a spline-based representation at one of its surfaces where it can interface with a pure IGA representation, while the opposite surface has a Lagrange-based representation where it connects with standard finite elements. For simplicity, the construction of two-dimensional elements are presented first, while the extension of the concept to the three-dimensional setting is discussed afterwards.
\begin{figure}[ht]
\centering
 \includegraphics[width=1\textwidth]{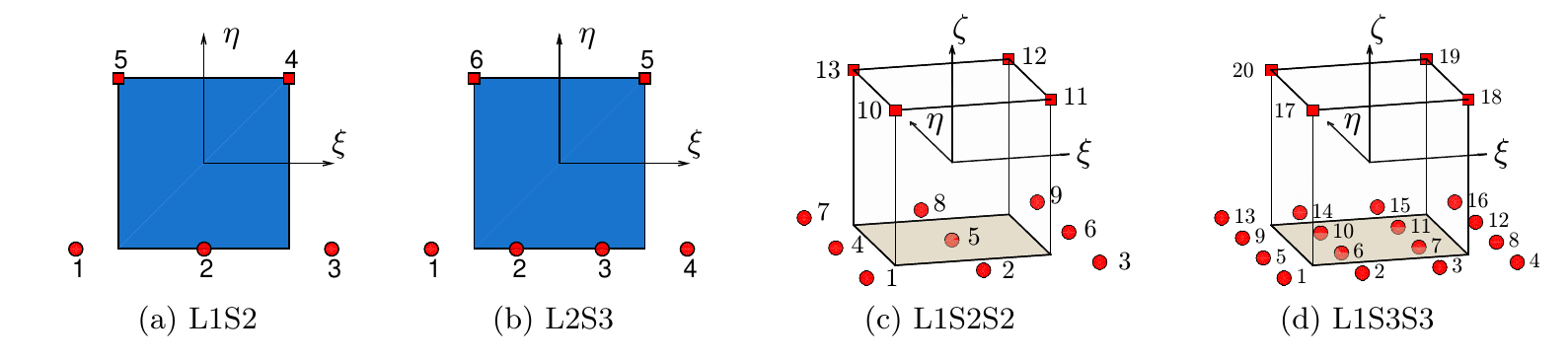}
\caption{Examples for two- and three-dimensional isogeometrically enriched master elements. Control points of the spline curve are depicted as circles, while standard nodes are represented with squares.} \label{enrich_elem}
\end{figure}

An isogeometrically enriched quadrilateral element can be obtained by replacing an edge of a standard quadrilateral element (i.e., Lagrange basis functions of order $p$) with a spline curve of order $q$. Without loss of generality, let us locate the spline curve at $\eta=-1$ (see Figure~\ref{enrich_elem}a and Figure~\ref{enrich_elem}b). We denote the respective element type as $\mathrm{L}p\mathrm{S}q$, where $q$ denotes the order of the spline curve (B-Spline, NURBS, T-Splines etc.) used to enrich the element at $\eta=-1$. An enriched element with linear Lagrange basis functions (i.e., $\mathrm{L}1\mathrm{S}q$) will possess two standard finite element nodes and $q+1$ control points of the spline curve. The basis functions of such an element are

\begin{align}
N_1&= \text{S}_1^q(\xi) \, \text{L}^1_1(\eta) = S_1^q(\xi)\, \frac{1}{2}(1-\eta) \,,\nonumber\\
 & \text{\hspace{0.2cm}}\vdots \nonumber\\
N_{q+1}&= \text{S}_{q+1}^q(\xi) \, \text{L}^1_1(\eta)  = \text{S}_{q+1}^{q}(\xi)\,\frac{1}{2}(1-\eta) \,,\label{eq:enrich_basis} \\
N_{q+2}&= \text{L}^1_1(\xi) \, \text{L}^1_2(\eta)  = \frac{1}{4} (1+\xi) (1+\eta) \,,\nonumber\\
N_{q+3}&=\text{L}^1_2(\xi) \, \text{L}^1_2(\eta) = \frac{1}{4} (1-\xi)(1+\eta) \,.\nonumber 
\end{align}
Here $\text{L}_i^p(\cdot)$ denotes the $i^{th}$~Lagrangian basis function of order $p$, while $\text{S}_j^q(\cdot)$ represents the $j^{th}$~spline basis function of order $q$. For the elements in Figure~\ref{enrich_elem}a and Figure~\ref{enrich_elem}b, $i=[1,2]$ and $j=[1,2,\dots,q+1]$.
 
A three-dimensional enriched element can be obtained by replacing one face of a standard hexahedral element (e.g. the surface at $\zeta=-1$ surface) with a spline surface (see Figure~\ref{enrich_elem}c and Figure~\ref{enrich_elem}d). The shape functions corresponding to the control points of the enriched surface will be a tensor product of the spline basis (in $\xi$- and $\eta$-direction) and Lagrangian basis ($\zeta$-direction). Extending the earlier defined nomenclature, three-dimensional isogeometrically enriched elements are denoted as $\mathrm{L}p\mathrm{S}q\mathrm{S}r$ where $q$ and $r$ are the order of the polynomial defining the enriching spline surface. A three-dimensional element with linear Lagrangian basis functions (i.e., $\mathrm{L}1\mathrm{S}q\mathrm{S}r$) will have $(q+1)(r+1)$ control points and four interpolatory nodes.

Isogeometric enrichment of standard elements leads to continuous representation of the field and the solution variables at the enrichment interface. The enriched basis functions retain the partition of unity over the entire element. Moreover, at the enriched surface, the enriched element type fulfills the variation diminishing property and its basis functions are always pointwise non-negative, similar to spline basis functions.
%
%
\section{Numerical examples}		\label{sec-ex}
The monolithic FSI model and the proposed IGA enrichment strategy, discussed in the previous sections, is validated and assessed in this section for different FSI benchmark problems. These include two-dimensional fluid-flow in a box cavity with a flexible base, the propagation of a pressure pulse in a deformable pipe and channel flow past a circular cylinder with a flexible tail. The consistency of the implemented model and the influence of the zone enrichment strategy is assessed through convergence analysis of relevant error norms for successive spatial refinements. Due to the non-availability of an exact solution and dearth of robust mesh convergence reference studies, relative error measures are formulated in relation to the obtained numerical solution of the finest discretization. Only for the third example, the obtained numerical solution is benchmarked with an available reference solution of very fine discretization. The response from the finest discretizations are also provided for future benchmark studies.      
%
\subsection{Lid driven cavity with a deformable base}			\label{sec:ex_cavity}
Internal fluid flow within a square cavity, due to the motion of the top surface, is a popular benchmark problem for fluid flow solvers. A modification to the pure fluid flow setup was proposed in~\cite{wall1999} to account for a base that could deform under the influence of fluid forces, essentially setting up an FSI example. Since then the problem has been used as a demonstration example for many numerical FSI studies~(\cite{mok2001b,gerbeau2003,foerster2007,kassiotis2011,mayr2015}). 

\begin{figure}[!ht]
	\centering
  	\includegraphics[width=1\textwidth]{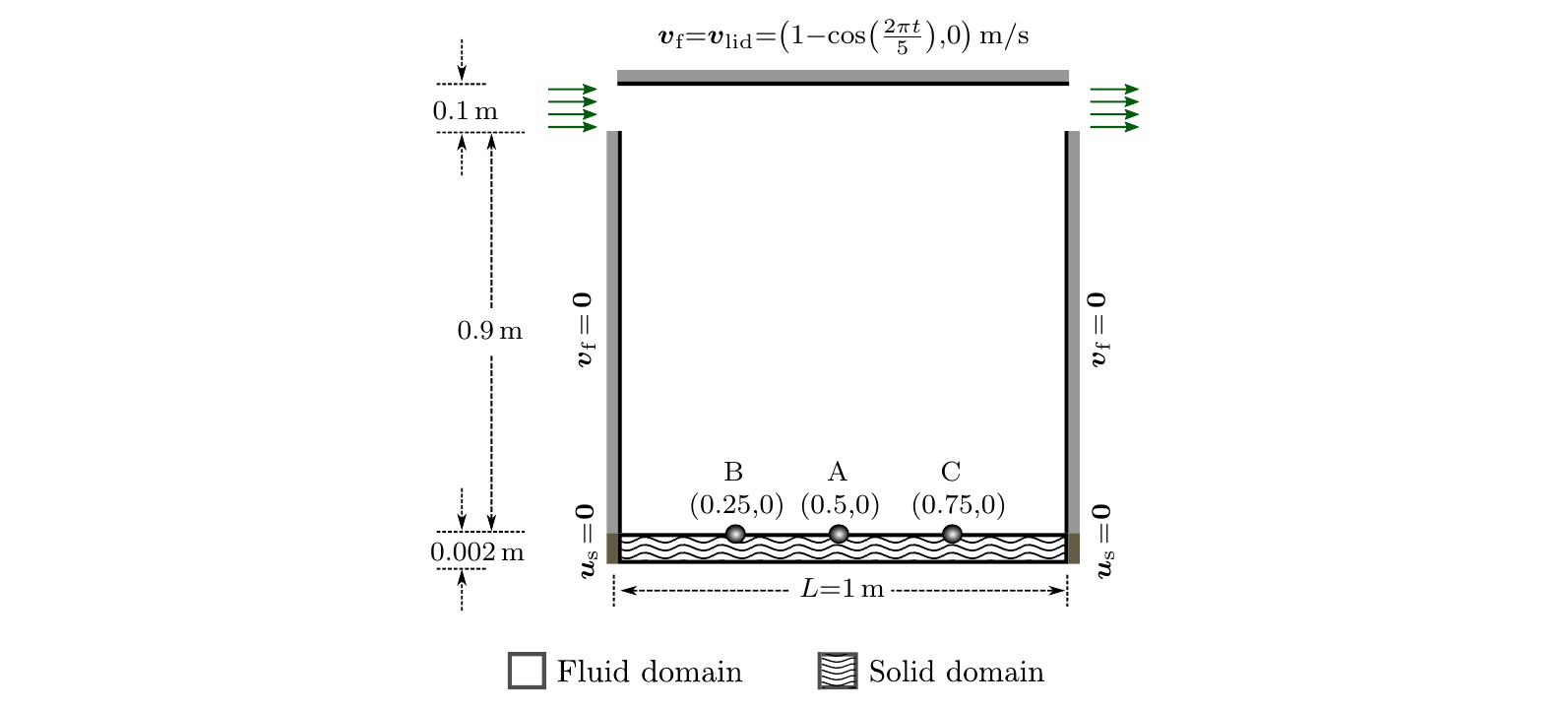}
	\caption{Lid driven cavity with a deformable base: Problem description with associated dimensions and boundary conditions.}
	\label{fig:cavity_fsi_config}
\end{figure}

Let us consider a square cavity where the fluid domain is enclosed in a box of ${\left[ 0,1\right] \times \left[ 0,1\right]}\,\text{m}^2$. The boundaries of the fluid domain are modeled as no-slip walls, where the lateral faces of the cavity are kept stationary, while the top surface (i.e., the lid) moves with an oscillating velocity $\boldsymbol{v}_{\mathrm{lid}}$ resulting in a flow with Reynolds number $0\leq Re_{\mathrm{lid}} \leq 200$. The base of the cavity is modeled as a thin hyper-elastic solid of dimensions ${\left[ 0,1\right] \times \left[ -0.002,0\right]}\,\text{m}^2$, fixed permanently at its lateral boundaries. The material parameters for the fluid and the solid are listed in Table~\ref{tab:cavity_prop}. To avoid numerical singularities in the pressure field associated with typical cavity flow benchmark studies (e.g. see~\cite{doneabook}), small segments close to the lid are modeled as traction free boundaries on both lateral faces, consequently allowing fluid inflow and outflow at these segments. Figure~\ref{fig:cavity_fsi_config} illustrates this setup in its entirety.
\begin{table}[!ht]
\begin{center}
\begin{tabular}{| >{\centering\arraybackslash}m{1.2in} | >{\centering\arraybackslash}m{1.1in} | >{\centering\arraybackslash}m{0.7in} |}
\hline
\textbf{parameters}		& \textbf{values}		& \textbf{units}							\parbox{0pt}{\rule{0pt}{2ex+\baselineskip}}\\
\hline
$\rho_{\text{s}_o}$  	& $500$		 			& $\frac{\text{kg}}{\text{m}^3}$		\parbox{0pt}{\rule{0pt}{2ex+\baselineskip}}\\
\hline
$ E_\text{s}$   			& $250$					& $\frac{\text{N}}{\text{m}^2}$			\parbox{0pt}{\rule{0pt}{2ex+\baselineskip}}\\
\hline
$\nu_\text{s}$ 			& $0.0$					& - 										\parbox{0pt}{\rule{0pt}{2ex+\baselineskip}}\\
\hline
$\rho_\text{f}$ 			& $1.0$ 					& $\frac{\text{kg}}{\text{m}^3}$		\parbox{0pt}{\rule{0pt}{2ex+\baselineskip}}\\
\hline
$\mu_\text{f}$			& $0.01$ 				& $\frac{\text{kg}}{\text{m}\, \text{s}}$	\parbox{0pt}{\rule{0pt}{2ex+\baselineskip}} \\
\hline
$Re_\text{lid} = \frac{\rho_\text{f} \, ||\boldsymbol{v}_\text{lid}|| \, L}{\mu_\text{f}}$ & $0\leq Re_\text{lid} \leq 200 $& - \parbox{0pt}{\rule{0pt}{2.0ex+\baselineskip}}\\
\hline
\end{tabular}
\caption{Lid driven cavity with a deformable base: Material parameters for the fluid and the solid continua.}
\label{tab:cavity_prop}
\end{center}
\end{table}

\begin{figure}[!ht]
	\centering
  	\includegraphics[width=1\textwidth]{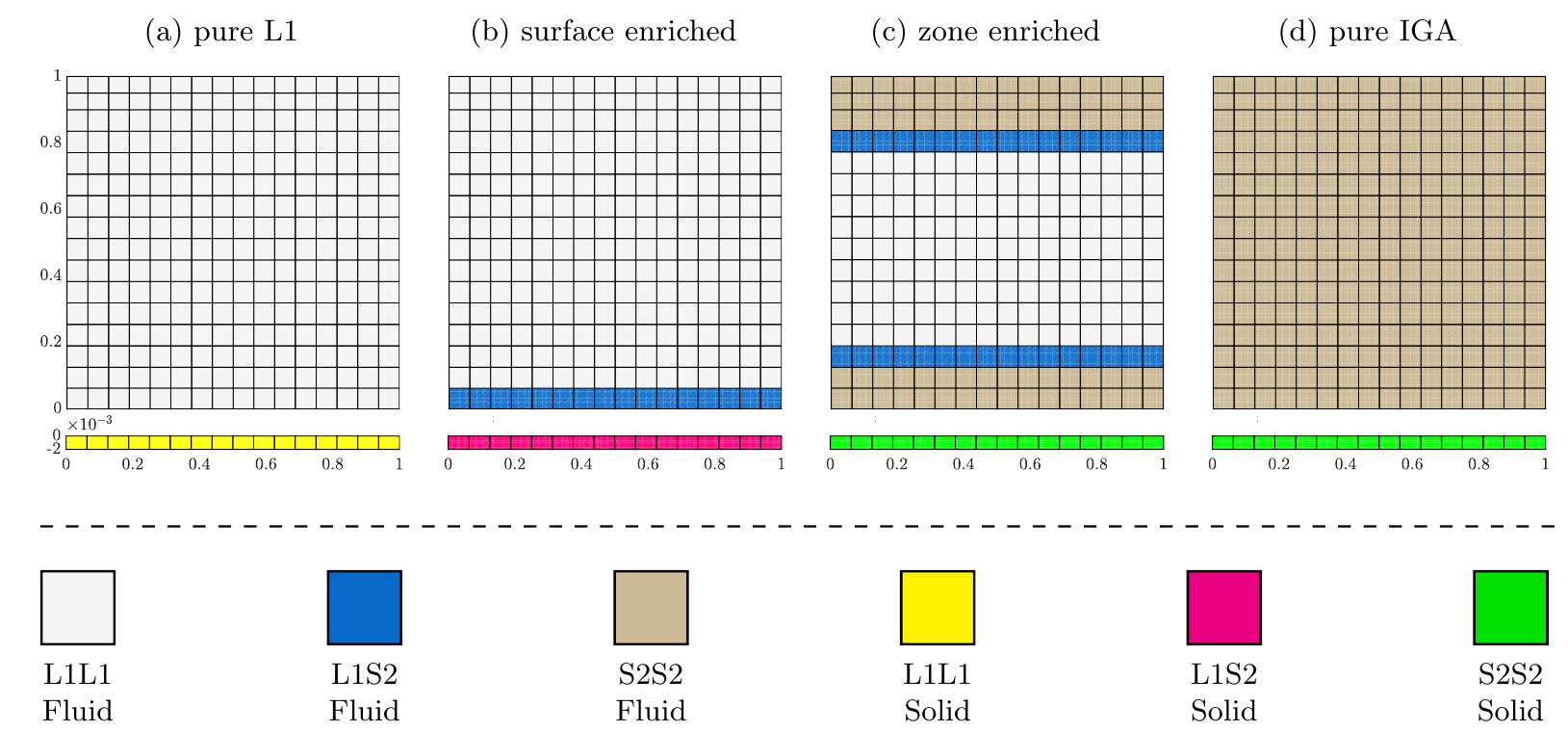}
	\caption{Lid driven cavity with a deformable base: The coarsest mesh (i.e., $m=0$) for different enrichment cases. The solid domain $\mathcal{B}^\text{h}_\text{s}$ is shown disconnected from $\mathcal{B}_\text{f}^\text{h}$ and stretched along the vertical axis only for representation purposes.}
	\label{fig:cavity_meshes_mx1}
\end{figure}

For the discretization, four enrichment cases are considered. A discretization composed entirely of second-order isogeometric spline elements $\mathrm{S}2\mathrm{S}2$ (i.e., with $p=q=2$) for both $\mathcal{B}_\text{f}^\text{h}$ and $\mathcal{B}_\text{s}^\text{h}$ corresponds to the pure IGA case. This case can be considered as the maximum IGA enrichment level. A subsequently reduced enrichment level is the zone enriched case, where only the fluid domain near the interface $\mathcal{B}_\text{f}^\text{h} \cap \mathcal{B}_\text{s}^\text{h}$ is represented with $\mathrm{S}2\mathrm{S}2$ elements, while most of $\mathcal{B}_\text{f}^\text{h}$ is represented by first-order Lagrange elements $\mathrm{L}1\mathrm{L}1$ (i.e., with $p=q=1$). The solid domain is continually discretized with $\mathrm{S}2\mathrm{S}2$ IGA elements. A further special case of the zone enriched case is the surface enriched discretization, where only the interface is provided with NURBS representation using $\mathrm{L}1\mathrm{S}2$ elements, while the remaining bulk is discretized with $\mathrm{L}1\mathrm{L}1$ elements. A discretization with no IGA enrichment is termed as the pure $\mathrm{L}1$ case and is composed entirely of $\mathrm{L}1\mathrm{L}1$ Lagrange elements. Figure~\ref{fig:cavity_meshes_mx1} shows the different enriched discretization at the coarsest level, while the respective element and $\mathrm{dofs}$ population data for each considered mesh are tabulated in Table~\ref{tab:cavity_fsi_elem}.  

\begin{table}[!ht]
\begin{center}
\begin{adjustbox}{max width=\textwidth}
\begin{tabular}{| c | c | c | c | c | c | c | c | c | c | c | c | c | c |}
  \hline
\multicolumn{3}{|c|}{} 						& \multicolumn{2}{ c |}{} 					& \multicolumn{3}{ c |}{}						& \multicolumn{4}{ c |}{} 					& \multicolumn{2}{ c |}{} \\
\multicolumn{3}{|c|}{\textbf{Refinement}} 	& \multicolumn{2}{ c |}{\textbf{pure L1}} 	& \multicolumn{3}{ c |}{\textbf{surface enrich}} 	& \multicolumn{4}{ c |}{\textbf{zone enrich}} & \multicolumn{2}{ c |}{\textbf{pure IGA}} \\ 
\multicolumn{3}{|c|}{} 				& \multicolumn{2}{ c |}{} 					& \multicolumn{3}{ c |}{}						& \multicolumn{4}{ c |}{} 					& \multicolumn{2}{ c |}{} \\ \cline{4-14}
\multicolumn{3}{|c|}{}						& Elements	& $\mathrm{dof}$s				& \multicolumn{2}{ c |}{Elements} & $\mathrm{dof}$s		& \multicolumn{3}{ c |}{Elements} & $\mathrm{dof}$s & Elements & $\mathrm{dof}$s \\ \cline{1-4} \cline{6-7} \cline{9-11} \cline{13-13}
 $m$ 						& $\bar{\mathrm{h}}^e (\text{m})$	&	& $\mathrm{L}1$ 		& 			& $\mathrm{L}1$ 		&$\mathrm{L}1\mathrm{S}2$ 	& 			& $\mathrm{L}1$ 		& $\mathrm{L}1\mathrm{S}2$ 	& $\mathrm{S}2\mathrm{S}2$ 		&   			& $\mathrm{S}2\mathrm{S}2$ 		& \parbox{0pt}{\rule{0pt}{1.2ex+\baselineskip}}\\ 
\hline
\multirow{2}{*}{0}		& \multirow{2}{*}{0.06250}	& Fluid		& 256		& 578		& 240		& 16		& 580		& 144 	& 32	& 80    	& 700	& 256 		& 720	\\
						& 			& Solid		& 16		& 68		& -			& 16		& 70		& - 		& - 		& 16  	& 108	& 16		& 108 \\
\hline
\multirow{2}{*}{1}		& \multirow{2}{*}{0.03125}	& Fluid		& 1,024		& 2,178		& 992		& 32		& 2,180		& 640 	& 64	& 320    	& 2,406	& 1,024		& 2,448 \\
						&			& Solid		& 64		& 198		& 32		& 32		& 200		& - 		& - 		& 64  	& 272 	& 64		& 272 \\
\hline
\multirow{2}{*}{$2$}	& \multirow{2}{*}{0.01563}	& Fluid		& 4,096	& 8,450	& 4,032	& 64		& 8,452	& 2,688 	& 128 		& 1,280    	& 8,890	& 4,096	& 8,976\\
						&			& Solid		& 256		& 650		& 192		& 64		& 652		& - 		& - 		& 256 		& 792 		& 256		& 792 \\
\hline
\multirow{2}{*}{$3$}	& \multirow{2}{*}{0.00781}	& Fluid		& 16,384	& 33,282	& 16,256	& 128		& 33,284	& 11,008 	& 256 		& 5,120    	& 34,146	& 16,384	& 34,320\\
						&			& Solid		& 1,024	& 2,322	& 896		& 128		& 2,324	& - 		& - 		& 1,024	& 2,600 	&1,024		& 2,600\\
\hline
\multirow{2}{*}{$4$}	& \multirow{2}{*}{0.00390}	& Fluid		& 65,536	& 132,098	& 65,280	& 256		& 132,100	& 44,544 	& 512 		& 20,480    	& 133,810	& 65,536	& 134,160\\
						&			& Solid		& 4,096	& 8,738	& 3,840	& 256		& 8,740	& - 		& - 		& 4,096	& 9,288 	& 4,096	& 9,288 \\
\hline
\multirow{2}{*}{$5$}	& \multirow{2}{*}{0.00195}	& Fluid		& 262,144	& 526,338	& 261,632	& 512		& 526,340	& 179,200	& 1,024 		& 81,920  	& 529,746 & 262,144	& 530,448\\
						&			& Solid		& 16,384	& 33,858	& 15,872	& 512		& 33,860	& - 		& - 		& 16,384	& 34,952 	& 16,384	& 34,952\\
\hline
\end{tabular}	
\end{adjustbox}
\end{center}
\caption{Lid driven cavity with a deformable base: Element and $\mathrm{dof}$ population data for successive spatial mesh refinements.}
\label{tab:cavity_fsi_elem}
\end{table}

\begin{figure}[!ht]
	\centering
  	\includegraphics[width=1\textwidth]{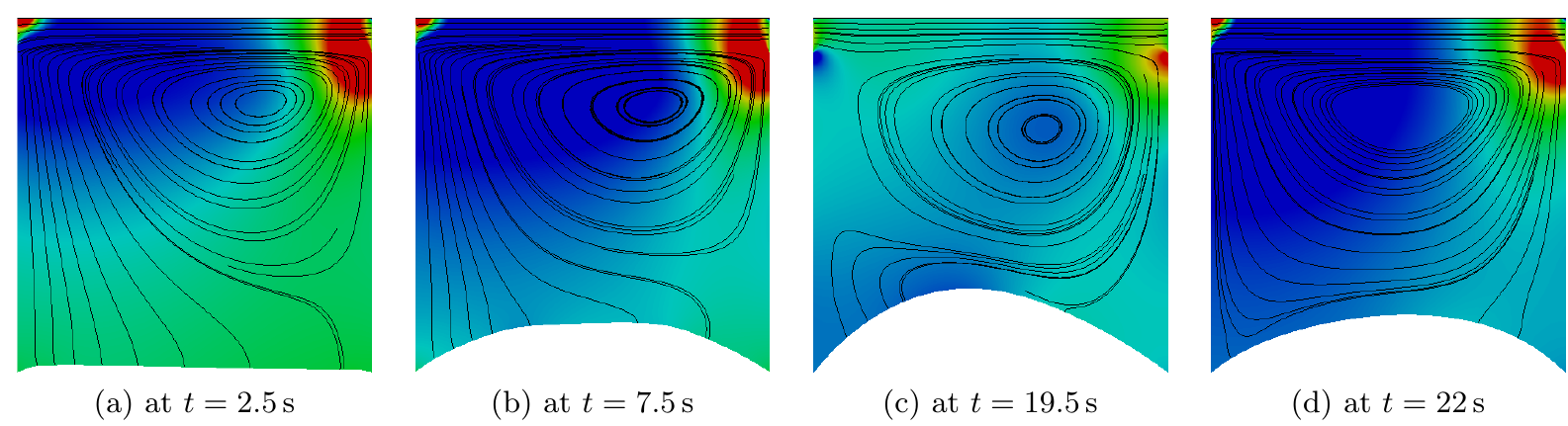}
	\caption{Lid driven cavity with a deformable base: Contours of fluid pressure $p_\text{f}$ together with flow streamlines at different time steps for mesh $m=5$}.
	\label{fig:cavity_fsi_press}
\end{figure}

\begin{figure}[h!]
	\centering
  	\includegraphics[width=1\textwidth]{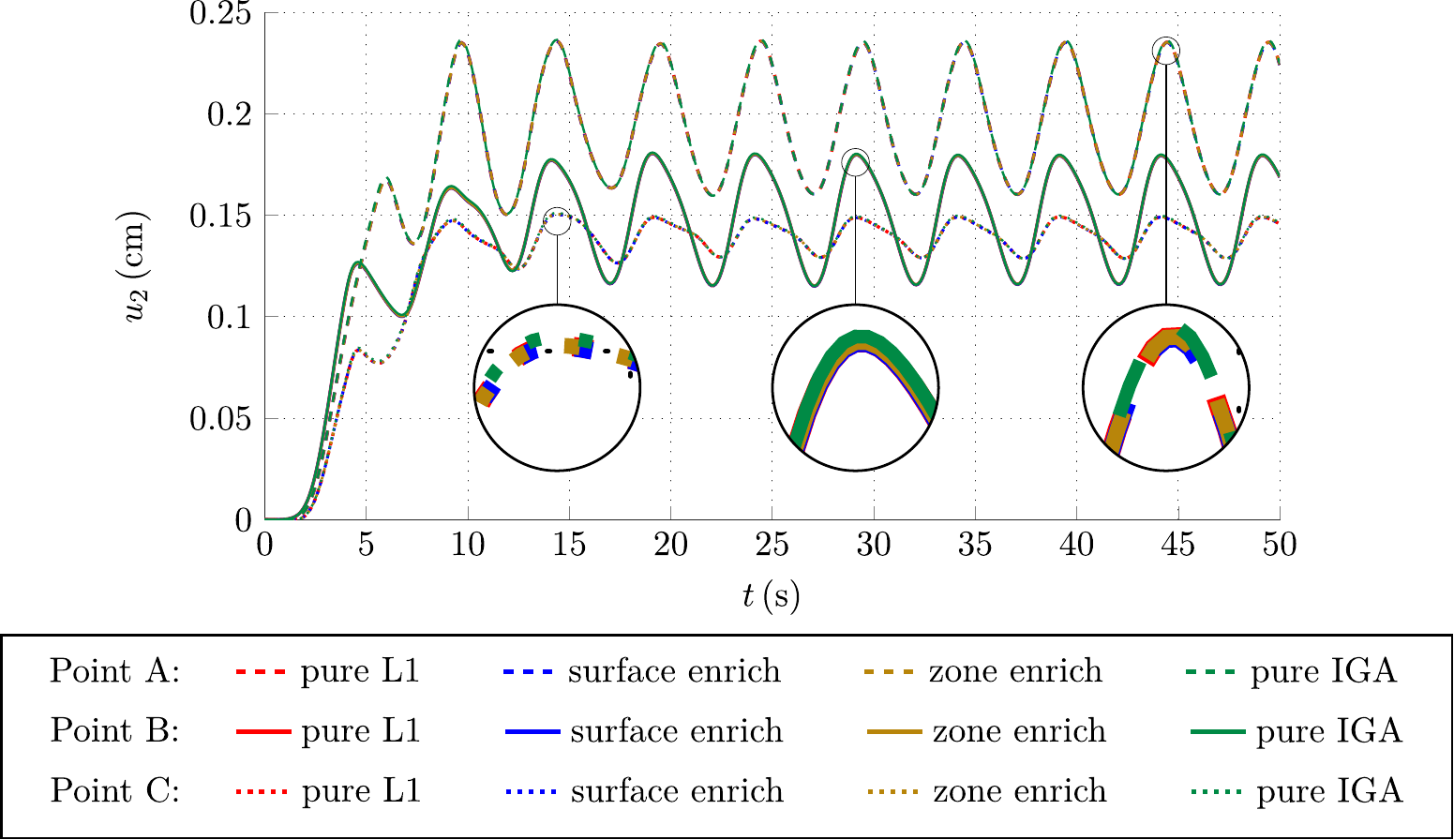}
	\caption{Lid driven cavity with a deformable base: Evolution of the vertical displacement $u_2$ of reference points A, B and C on the fluid-solid interface $\partial \mathcal{B}_\text{f} \cap \partial \mathcal{B}_\text{s}$ for mesh $m=5$.}
	\label{fig:cavity_fsi_u2_ptABC} 
\end{figure}

Numerical simulations are performed over the interval $t\in[0,50\,\text{s}]$ with a time step size of $0.1\,$s. Figure~\ref{fig:cavity_fsi_press} depicts the fluid pressure field at different stages of the simulation. Additionally, the vertical displacement $u_2$ of three reference points on the fluid-solid interface, points A, B and C as shown in Figure~\ref{fig:cavity_fsi_config}, are recorded over the course of the entire simulation. The obtained response for the evolution of the vertical displacement of these reference points over the entire time interval is shown in Figure~\ref{fig:cavity_fsi_u2_ptABC}. This response is similar in character to those reported in~\cite{mok2001b} and~\cite{gerbeau2003}. For subsequent convergence analysis, the numerical solution corresponding to the finest pure IGA discretization (i.e., $m=5$) is considered as the reference solution.

In order to assess the convergence behavior of the obtained dynamic numerical response, we formulate a Fourier approximation of the resulting discrete data and compare the relative differences in the approximated Fourier coefficients. Such a treatment allows us to encapsulate relative differences in the amplitude, the frequency and the phase-shift between the reference and the obtained solution in a single measure. The convergence of this measure is then studied for successive mesh refinements. A Fourier representation of an evolutionary function is given as
\begin{align}
	f(t) = a_0 + \sum\limits_{k=1}^\infty \left[ a_k \, \cos \left( \frac{k \pi}{L} t \right) + b_k \, \sin \left(\frac{k \pi }{L} t \right) \right]		\label{eq:fourier_formula}
\end{align}
where
\begin{align}
	a_0 &= \frac{1}{2\,L} \int\limits_{-L}^L  f(x) \, \text{d}x \,, \\
	a_k &= \frac{1}{L} \int\limits_{-L}^L f(x) \, \cos\left(\frac{k \pi}{L} x \right) \, \text{d}x\,, \\
	b_k &= \frac{1}{L} \int\limits_{-L}^L f(x) \, \sin\left(\frac{k \pi}{L} x \right) \, \text{d}x \,, 
\end{align}
with $2L$ being the period of the function $f(t)$. For a finite summation of Fourier coefficients $a_k$ and $b_k$ (i.e., $k=1, \, \dots ,n$), only an approximation of $f(t)$ is obtained. The value of $n$ is reflective of the number of distinct frequencies inherent in $f(t)$. Using the discrete data and a predetermined value of $n$, a Fourier approximation comprising the Fourier coefficients can be obtained using a curve fitting algorithm (e.g., see~\cite{batesbook}). For the analysis performed in this study, we employ the nonlinear least square method to fit the obtained numerical data to a Fourier approximation. Once the Fourier coefficients have been approximated, the relative difference between the reference and the obtained solution is assessed using the following error measure:
\begin{align}
	E_{(\cdot)} := \sqrt{ \left(\frac{L^\mathrm{ref} - L^\text{h}}{L^\mathrm{ref}}\right)^2 + \frac{\left(a_0^\mathrm{ref} - a_0^\text{h} \right)^2 + \sum\limits_{k=1}^n \left[ \left( a_k^\mathrm{ref} - a_k^\text{h} \right)^2 + \left( b_k^\mathrm{ref} - b_k^\text{h} \right)^2\right] }{\left(a_0^\mathrm{ref}\right)^2 + \sum\limits_{k=1}^n \left[\left(a_k^\mathrm{ref}\right)^2 + \left(b_k^\mathrm{ref}\right)^2 \right] } } \,,    \label{eq:fourier_error_defn}
\end{align}
where the superscript ``$\mathrm{ref}$'' denotes Fourier approximation using the reference solution, while the superscript $\text{h}$ denotes Fourier approximation using the obtained solution.

\begin{figure}[h!]
	\centering
  	\includegraphics[width=0.55\textwidth]{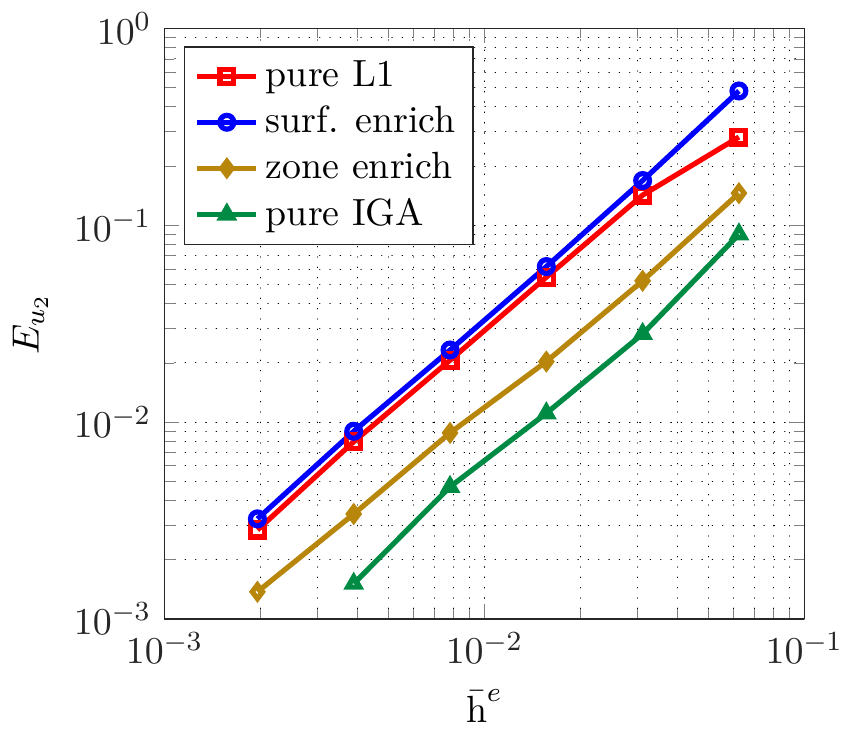}
	\caption{Lid driven cavity with a deformable base: Convergence of the error norm $E_{u_2}$, i.e., the norm formulating the relative error in the Fourier approximation of the vertical displacement ($u_2$) of the point A with $n=1$.}
	\label{fig:cavity_fsi_fourier} 
\end{figure}

Figure~\ref{fig:cavity_fsi_fourier} shows the behavior of the error measure $E_{u_2}$ for the response in the vertical displacement of the reference point A. Here, $\bar{\text{h}}^e$ denotes the average length of the elements in the domain $\mathcal{B}_\text{f}^\text{h}$. The last four periods of the response depicted in Figure~\ref{fig:cavity_fsi_u2_ptABC} were used to construct a Fourier approximation with $n=1$. A subsequent increase in the value of $n$ essentially results in plots identical to Figure~\ref{fig:cavity_fsi_fourier}. It is evident from Figure~\ref{fig:cavity_fsi_fourier} that the relative error decreases consistently for all enrichment cases, reflecting a consistent behavior. The use of S2S2 elements in the enriched zones results in a significantly improved solution for the zone enriched case, as compared to the pure $\mathrm{L1}$ and the surface enriched cases. The region near the fluid-solid interface and the fluid domain near the lid of the cavity experiences rapid movements of the boundary surface and therefore inherently retains strong velocity gradients. An IGA zone enrichment only in these areas improves the analysis considerably as Figure~\ref{fig:cavity_fsi_fourier} shows.

\subsection{Pressure pulse in a deformable tube}			\label{sec:ex_tube}
For the second benchmark example, we consider the case of incompressible fluid flow in a circular tube with a traveling pressure wave generated at the inlet. Although initially proposed as a simplified model for a human cardiovascular system in~\cite{formaggia2001}, the problem setup has since been widely employed with different geometrical and boundary condition implementations in many FSI numerical studies (e.g., see~\cite{gerbeau2003, kuettler2008, gee2011, malan2013,mayr2015,eken2016,eken2017}).

\begin{table}[!ht]
\begin{center}
\begin{tabular}{| >{\centering\arraybackslash}m{1.4in} | >{\centering\arraybackslash}m{0.6in} | >{\centering\arraybackslash}m{0.6in} |}
\hline
\textbf{parameters}		& \textbf{values}		& \textbf{units}	\parbox{0pt}{\rule{0pt}{3ex+\baselineskip}}\\
\hline
$\rho_{\text{s}_o}$  	& $1.2$		 			& $10^3 \,\frac{\text{kg}}{\text{m}^3}$ 		\parbox{0pt}{\rule{0pt}{2ex+\baselineskip}}\\
\hline
$ E_\text{s}$   			& $3$					& $10^{5}\frac{\text{N}}{\text{m}^2}$ 	\parbox{0pt}{\rule{0pt}{2ex+\baselineskip}}\\
\hline
$\nu_\text{s}$  			& 0.3					& -	\parbox{0pt}{\rule{0pt}{2ex+\baselineskip}}\\
\hline
$\rho_\text{f}$ 			& $1$ 					& $10^3 \,\frac{\text{kg}}{\text{m}^3}$ 		\parbox{0pt}{\rule{0pt}{2ex+\baselineskip}}\\
\hline
$\mu_\text{f}$			& $3$					& $10^{-3} \,\frac{\text{kg}}{\text{m}\, \text{s}}$ 	\parbox{0pt}{\rule{0pt}{2ex+\baselineskip}} \\
\hline
$Re_\mathrm{pulse} = \frac{d_\mathrm{p} \, \sqrt{\rho_\mathrm{f} \, p_\mathrm{pulse} } }{\mu_\mathrm{f}}$ &3,800& - \parbox{0pt}{\rule{0pt}{2ex+\baselineskip}}\\
\hline
\end{tabular}
\caption{Pressure pulse in a deformable tube: Material properties for fluid and solid.}
\label{tab:tube_prop}
\end{center}
\end{table}

\begin{figure}[h!]
	\centering
  	\includegraphics[width=1\textwidth]{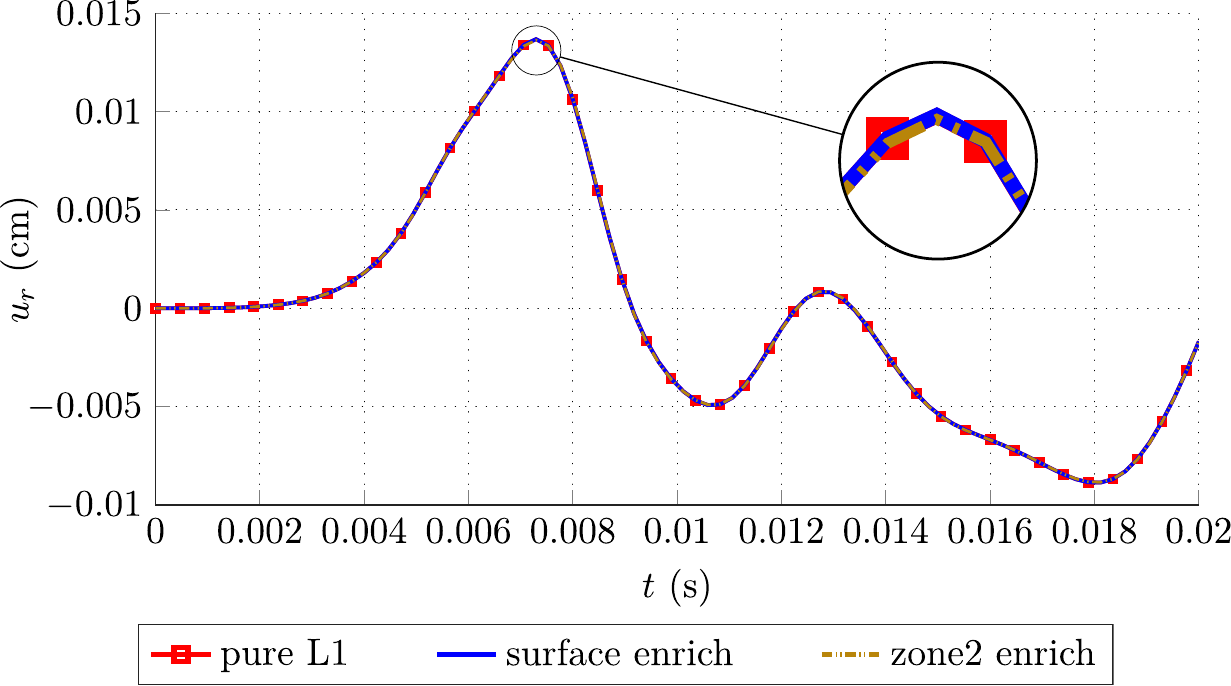}
	\caption{Pressure pulse in a deformable tube: Evolution of the radial $u_r$ displacement of a point on the tube's inner surface at halfway length of the pipe $(m=3)$.}
	\label{fig:tube_disp_typ} 
\end{figure}

The problem setup comprises a $0.05\,$m long tube of circular cross-section. The inner region of the pipe constitutes the fluid domain $\mathcal{B}_\text{f}$ and has a diameter $(d_\mathrm{p})$ of $0.01\,$m. The wall of the tube, which is $0.001\,$m thick, constitutes the solid domain $\mathcal{B}_\text{s}$. Material properties of the fluid and the solid are summarized in Table~\ref{tab:tube_prop}. The tube wall at the inlet and the outlet are held clamped over the entire simulation. The inlet and the outlet for the fluid domain are modeled as traction free boundaries, where $\boldsymbol{\sigma}_\text{f}\cdot \boldsymbol{n}_\text{f} = \boldsymbol{0}$. For a small time interval, i.e.~for $t<0.003\,$s, a fluid pressure $(p_\mathrm{pulse})$ of $1300\,$N/m$^2$ is applied at the inlet. This impulse generates a pressure pulse that continues to travel along the length of the tube, while interacting with the deformable tube. When the pulse reaches the outlet boundary, part of it gets reflected back towards the inlet due to the clamped nature of the outlet boundary. 
\begin{table}[!ht]
\begin{center}
\begin{adjustbox}{max width=\textwidth}
\begin{tabular}{| c | c | c | c | c | c | c | c | c | c | c | c | c | c | c | c |}
  \hline
\multicolumn{3}{|c|}{} 		& \multicolumn{2}{ c |}{}	& \multicolumn{3}{ c |}{}	& \multicolumn{4}{ c |}{} 	& \multicolumn{4}{ c |}{} \\
\multicolumn{3}{|c|}{\textbf{Refinement}} 		& \multicolumn{2}{ c |}{\textbf{pure L1}}	& \multicolumn{3}{ c |}{\textbf{surface enrich}}& \multicolumn{4}{ c |}{\textbf{zone1 enrich}}	& \multicolumn{4}{ c |}{\textbf{zone2 enrich}} \\ 
\multicolumn{3}{|c|}{}	& \multicolumn{2}{ c |}{}	& \multicolumn{3}{ c |}{}& \multicolumn{4}{ c |}{}	& \multicolumn{4}{ c |}{} \\  \cline{4-16}
\multicolumn{3}{|c|}{} 	& \textbf{Elements} & $\mathrm{dofs}$ & \multicolumn{2}{c|}{\textbf{Elements}} & $\mathrm{dofs}$ & \multicolumn{3}{c|}{\textbf{Elements}} & $\mathrm{dofs}$ & \multicolumn{3}{c|}{\textbf{Elements}} & $\mathrm{dofs}$ \\ \cline{1-4} \cline{6-7} \cline{9-11} \cline{13-15}	
 $m$ 				& $\bar{\mathrm{h}}^e (\text{m})$	&		& $\mathrm{L}1\mathrm{L}1\mathrm{L}1$ & & $\mathrm{L}1\mathrm{L}1\mathrm{L}1$ & $\mathrm{L}1\mathrm{S}2\mathrm{S}2$ & & $\mathrm{L}1\mathrm{L}1\mathrm{L}1$ & $\mathrm{L}1\mathrm{S}2\mathrm{S}2$ & $\mathrm{S}2\mathrm{S}2\mathrm{S}2$ & & $\mathrm{L}1\mathrm{L}1\mathrm{L}1$ & $\mathrm{L}1\mathrm{S}2\mathrm{S}2$ & $\mathrm{S}2\mathrm{S}2\mathrm{S}2$ &  \parbox{0pt}{\rule{0pt}{1.2ex+\baselineskip}} \\
\hline
\multirow{2}{*}{$0$}	& \multirow{2}{*}{$0.00156$}	& Fluid		& 640		& 3,828 		& 512		& 128		& 3,984 		& 512 		& 128 		& - 		& 3,984 		& 384		& 128		& 128 		& 4,956  \\
						&								& Solid		& 256		& 1,485 		& - 	 	& 256		& 1,719		& - 		& - 		& 256		& 2,448		& - 		& - 		& 256 		& 2,448 \\
\hline
\multirow{2}{*}{$1$}	& \multirow{2}{*}{$0.00078$}	& Fluid		& 5,120 	& 25,220 		& 4,608 	& 512 		& 25,520		& 4,608	& 512 		& - 		& 25,520		& 3,584 	& 512		& 1,024	& 28,760\\
						&								& Solid		& 2,048 	& 8,775		& 1,024 	& 1,024	& 9,225		&- 			& - 		& 2,048 	& 11,880		& -			& - 		& 2,048 	&  11,880 \\
\hline
\multirow{2}{*}{$2$}	& \multirow{2}{*}{$0.00039$}	& Fluid		& 40,960	& 182,148		& 38,912	& 2,048	& 182,736		& 38,912	& 2,048 	& - 		& 182,736		& 30,720	& 2,048	& 8,192   	& 194,448\\
						&								& Solid		& 16,384	& 59,211		& 12,288	& 4,096	& 60,093		& - 		& - 		& 16,384	& 70,200		& - 		& - 		& 16,384 	&  70,200 \\
\hline
\multirow{2}{*}{$3$}	&\multirow{2}{*}{$0.00020$} 	& Fluid		& 327,680	& 1,382,660	& 319,488	& 8,192	& 1,383,824	& 319,488	& 8,192 	& - 		& 1,383,824	& 253,952	& 8,192	& 65,536  	& 1,428,224\\
						&								& Solid		& 131,072	& 393,216		& 114,688	& 16,384	& 434,277		& - 		& - 		& 131,072	& 473,688		& - 		& - 		& 131,072 & 473,688\\
\hline
\end{tabular}	
\end{adjustbox}
\end{center}
\caption{Pressure pulse in a deformable tube: Mesh statistics for successive spatial mesh refinements.}
\label{tab:tube_mesh}
\end{table}
\begin{figure}[!ht]
	\centering
  	\includegraphics[width=1\textwidth]{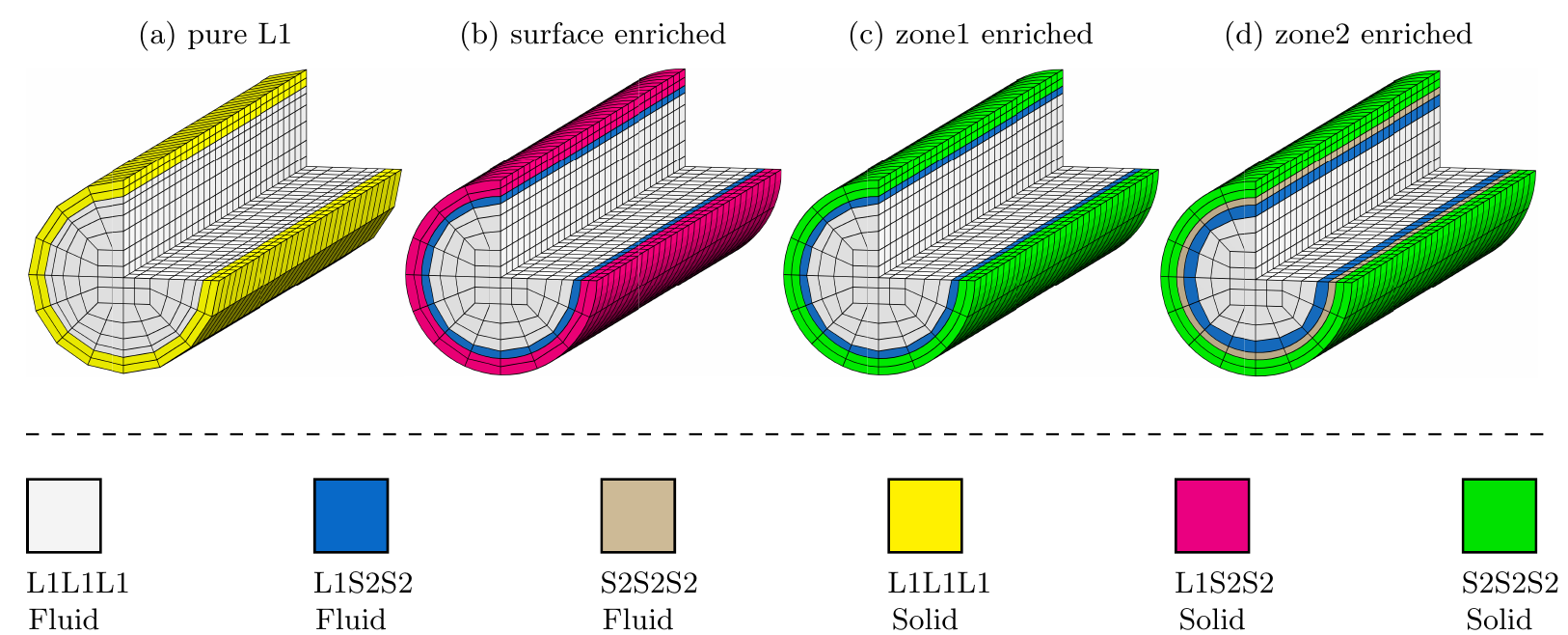}
	\caption{Pressure pulse in a deformable tube: Types of enrichment cases investigated for tube example at the coarsest refinement level (i.e., $m=0$). The figure shows a cutout view of the domain. For the computational analysis, only a single quadrant of the domain is utilized.}
	\label{fig:tube_meshes_mx1}
\end{figure}\\
Due to the geometry and the applied boundary conditions, the problem is expected to yield a symmetric solution within the laminar flow regime. Therefore only a single quadrant of the problem domain is simulated to save computational cost, while retaining a three-dimensional analysis configuration at the same time. Four IGA enrichment cases, as shown in Figure~\ref{fig:tube_meshes_mx1}, are considered for this benchmark example. The surface~enriched case comprises a discretization, where the tube's outer surface and the fluid-solid interface are discretized with $\mathrm{L}1\mathrm{S}2\mathrm{S}2$ elements to have a NURBS representation of the curved surfaces, while the remaining bulk is composed of linear Lagrange elements ($\mathrm{L}1\mathrm{L}1\mathrm{L}1$). The zone1~case is an enrichment of the surface enriched case where the solid domain $\mathcal{B}_\text{s}^\text{h}$ is discretized entirely with second-order IGA hexahedral elements ($\mathrm{S}2\mathrm{S}2\mathrm{S}2$). For the zone2~enriched case, only a thin layer of IGA hexahedral elements are added in the boundary layer region of the fluid domain, while the major fluid bulk is discretized with linear Lagrange elements. IGA hexahedral elements constitutes the discretization of $\mathcal{B}_{s}^\text{h}$ for zone2 as well. Figure~\ref{fig:tube_meshes_mx1} shows the four enrichment cases at the coarsest level, while mesh statistics are tabulated in Table~\ref{tab:tube_mesh}.
\begin{figure}[!ht]
	\centering
  	\includegraphics[width=1\textwidth]{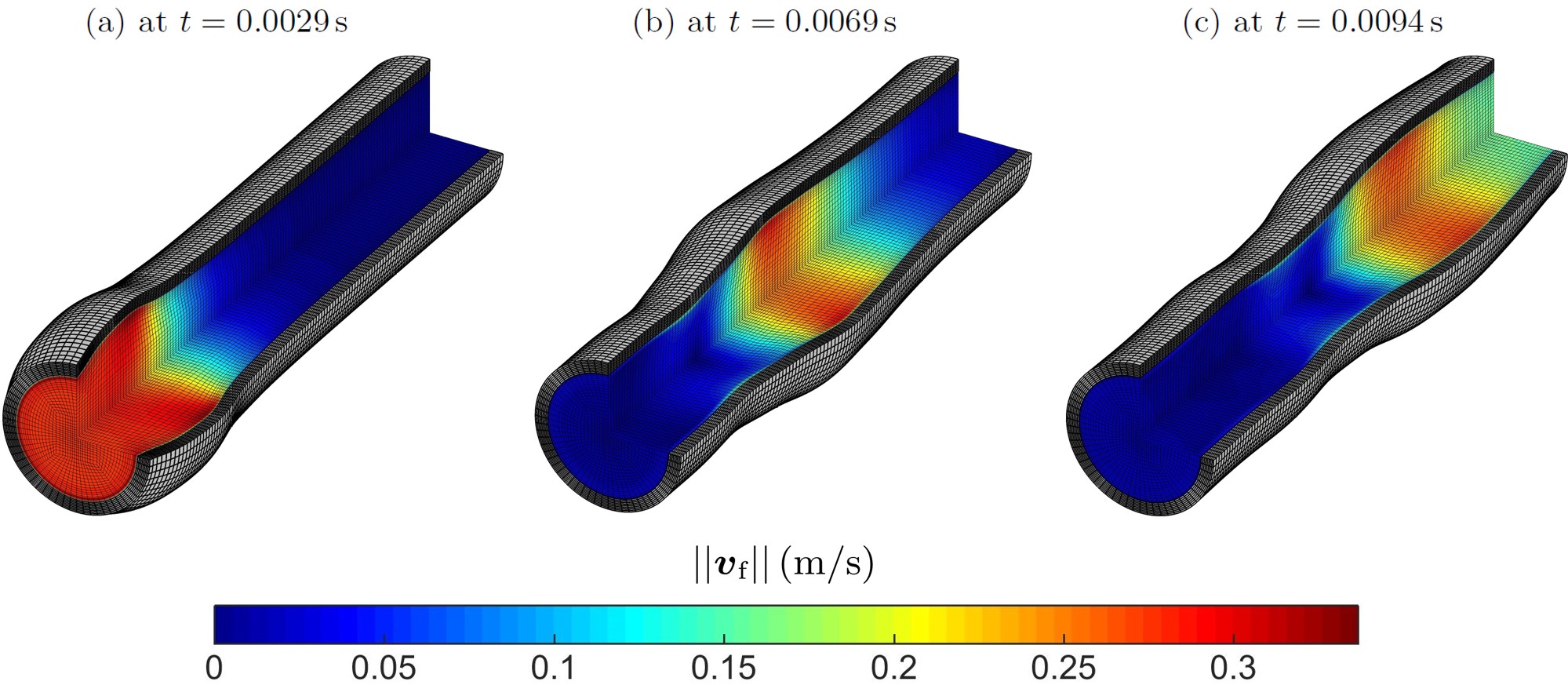}
	\caption{Pressure pulse in a deformable tube: Contours of fluid velocity magnitude $||{\boldsymbol{v}_\text{f}|| = \sqrt{ v_1^2 + v_2^2 + v_3^2}}$ at different stages for the zone2~enriched discretization $(m=2)$. The displacement of the solid is increased by a factor of 10 for illustration purposes.}
	\label{fig:tube_vel}
\end{figure}
\begin{figure}[!ht]
	\centering
  	\includegraphics[width=1\textwidth]{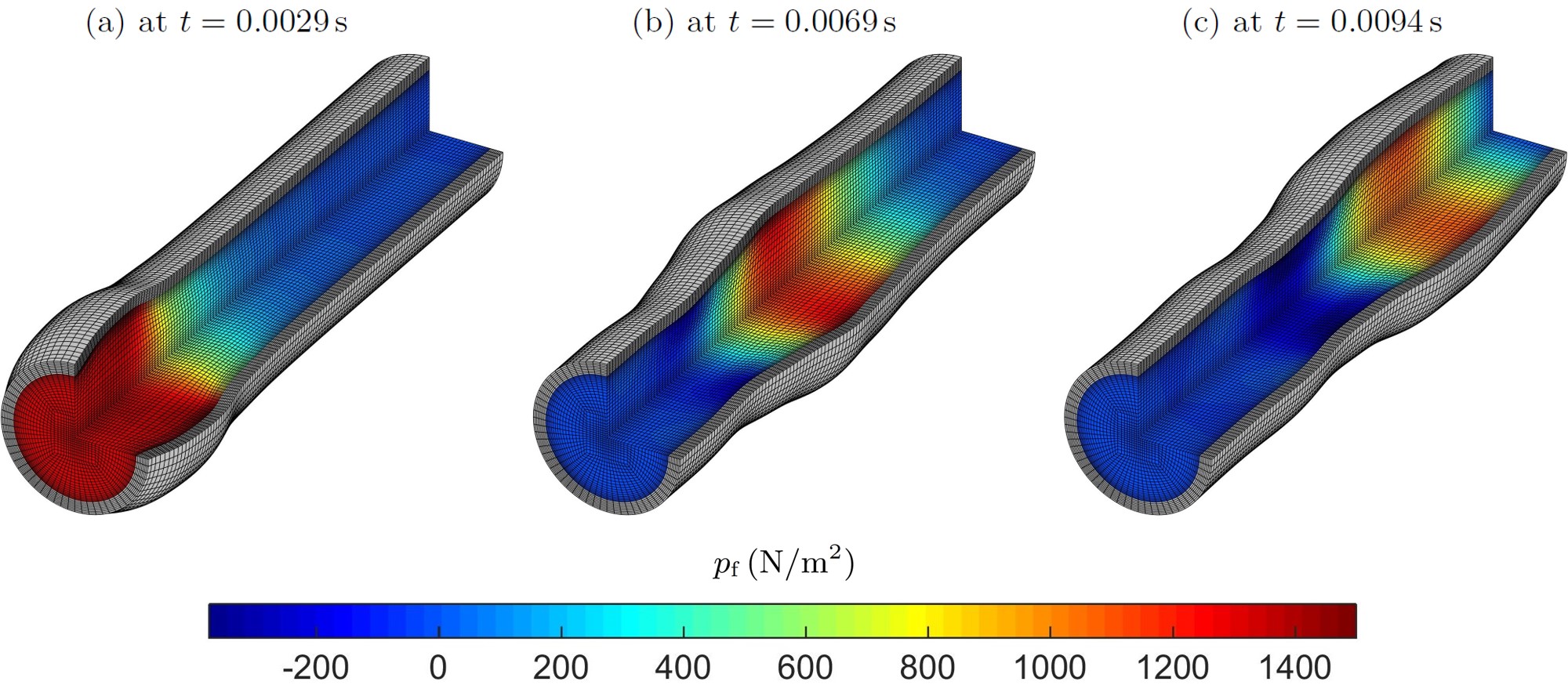}
	\caption{Pressure pulse in a deformable tube: Contours of pressure at different stages for the zone2 enriched discretization $(m=2)$. The displacement of the solid is increased by a factor of 10 for illustration purposes.}
	\label{fig:tube_press}
\end{figure}
\begin{figure}[!ht]
	\centering
  	\includegraphics[width=.9\textwidth]{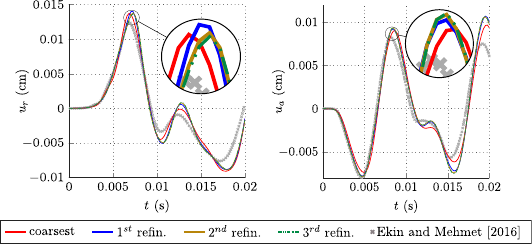}
	\caption{Pressure pulse in a deformable tube: Evolution of the radial $u_r$ and the axial $u_a$ displacement of a point on the tube's inner surface at halfway length of the pipe (pure $\mathrm{L}1$ case). }
	\label{fig:tube_disp} 
\end{figure}
\\Numerical simulations are performed over the interval $t \in [0,0.02\,\text{s}]$ with 85 time steps. During this interval, a complete reflection of the pulse from the outlet to the inlet is observed. Figure~\ref{fig:tube_vel} illustrates the fluid velocity field in the tube's interior together with the deformations of the solid domain, as the pulse travels from the inlet to the outlet. Pressure contours are shown in Figure~\ref{fig:tube_press}. In order to assess the convergence behavior of the enrichment cases, we measure the radial and the axial displacement of a point ($u_r$ and $u_a$, respectively) located on the tube's inner surface at halfway length. The evolution of these entities are shown in Figure~\ref{fig:tube_disp} for the pure~$\mathrm{L1}$ discretizations, in comparison to the findings from~\cite{eken2016}. It is evident from the figure that the numerical solution will converge to a unique finite element solution as the mesh is refined. The difference between the obtained results and the findings from~\cite{eken2016} can be attributed to the different boundary conditions at the outlet. Zero fluid pressure was imposed at the outlet in~\cite{eken2016}, whereas we impose a traction free outlet surface. It is also observed that all the considered enrichment cases tend to converge to identical numerical solutions with spatial refinements. This is evident from the response history of $u_r$ for different enrichment cases at the finest level, as shown in Figure~\ref{fig:tube_disp_typ}.

\begin{figure}[h!]
	\centering
  	\includegraphics[width=1\textwidth]{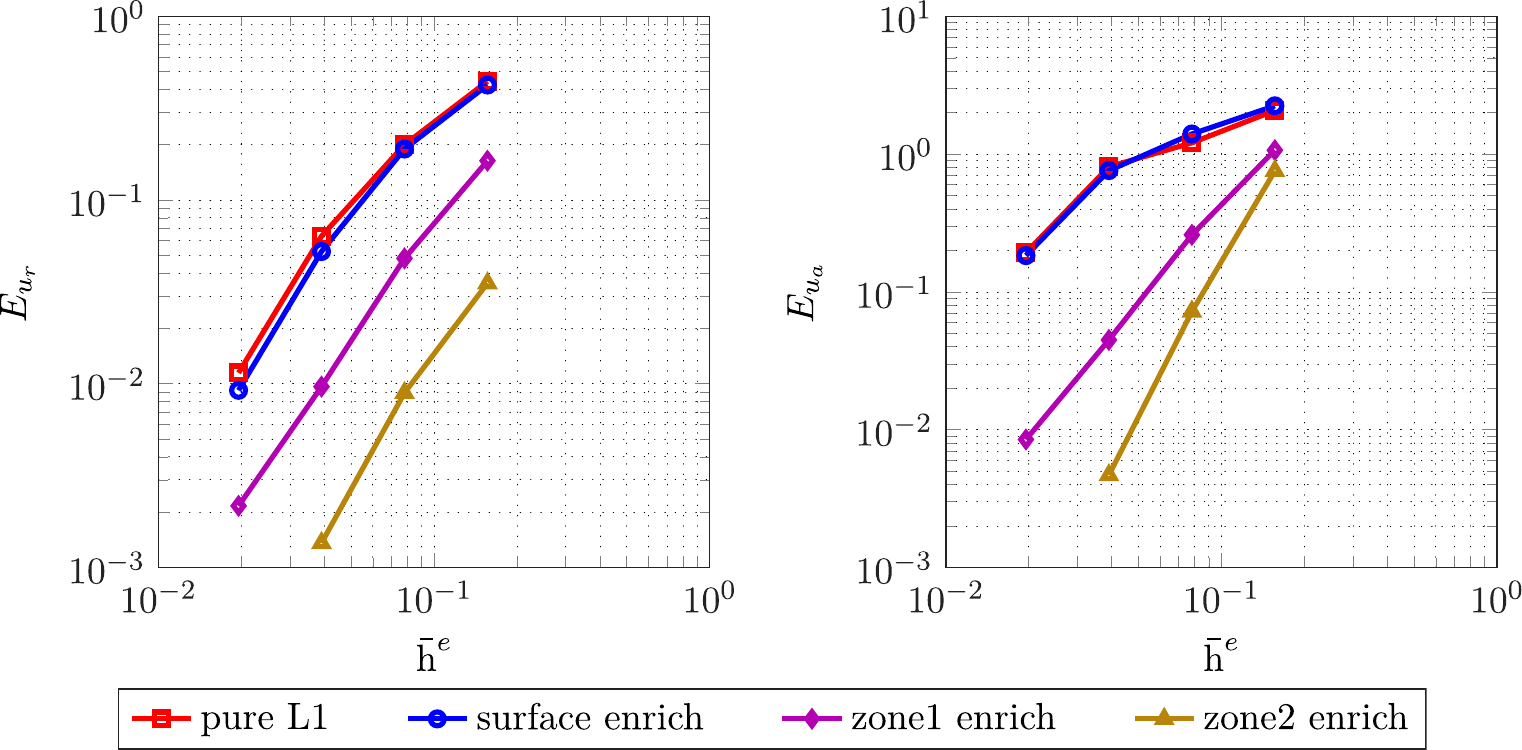}
	\caption{Pressure pulse in a deformable tube: Mesh convergence behavior of the error norm in the Fourier coefficients, i.e., $E_{u_r} \text{ and } E_{u_a}$.}
	\label{fig:tube_err_fourier} 
\end{figure}
To perform a comparative analysis for the different IGA enrichment strategies considered for this example, we use the error measure of Eq.~(\ref{eq:fourier_error_defn}) to judge the relative difference between the obtained numerical solutions. For this purpose, we consider the response history of $u_r$ and $u_a$ obtained from the finest discretization (i.e.,~$m=3$) for the zone enriched case as the reference response. The numerical solutions are fitted to a Fourier series with $n=7$, where the first acceptable approximation of the reference solution is obtained. Mesh convergence analysis for the parameters $E_{u_r}$ and $E_{u_a}$ are shown in Figure~\ref{fig:tube_err_fourier}. It is evident, that the solution obtained from the zone1 enriched discretization, with $\mathrm{S}2\mathrm{S}2\mathrm{S}2$ IGA elements only in $\mathcal{B}_\text{s}^\text{h}$, is much better than the ones obtained from the surface enriched and the pure $\mathrm{L}1$ discretizations. However, the most important observation is regarding the performance of zone2 enriched discretization, where the obtained results are significantly more accurate than all other enrichment cases. This clearly advocates the benefit of incorporating a thin layer of $\mathrm{S}2\mathrm{S}2\mathrm{S}2$ IGA elements in the boundary layer region of $\mathcal{B}_\text{f}^\text{h}$.

%
\subsection{Flow past a circular cylinder with flexible tail}			\label{sec:ex_turek}
Laminar flow past a circular cylinder with a thin flexible tail is a popular two-dimensional numerical benchmark for FSI solvers. The benchmark configuration was first proposed in~\cite{turek2006}, and has since been adopted by many researchers to validate different FSI solvers (e.g.,~\cite{heil2008,kollmannsberger2009,bhardwaj2012,lee2013} among others). Three different test cases --- based on the fluid velocity at the inlet and the material parameters of the deformable tail --- were proposed in the original paper of~\cite{turek2006}. Here, we investigate the more challenging unsteady test cases FSI-2 and FSI-3 of the original proposal. 
\begin{table}[!ht]
\begin{center}
\begin{tabular}{| >{\centering\arraybackslash}m{1.0in} | >{\centering\arraybackslash}m{0.7in} | >{\centering\arraybackslash}m{0.7in} | >{\centering\arraybackslash}m{0.7in} |}
\hline
\textbf{parameters}		& \textbf{FSI-2}		& \textbf{FSI-3} 		& \textbf{units} 								\parbox{0pt}{\rule{0pt}{5ex+\baselineskip}}\\
\hline
$\rho_{\text{s}_o}$  	& 10.0				& 1.0 				& $10^3 \, \frac{\text{kg}}{\text{m}^3}$		\parbox{0pt}{\rule{0pt}{1.5ex+\baselineskip}}\\
\hline
$\mu_\text{s}$  			& 0.5				& 2.0 				& $10^6 \, \frac{\text{kg}}{\text{m s}^2}$ 	\parbox{0pt}{\rule{0pt}{1.5ex+\baselineskip}}\\
\hline
$\nu_\text{s}$  			& 0.4				& 0.4				& - 											\parbox{0pt}{\rule{0pt}{1.5ex+\baselineskip}}\\
\hline
$\rho_\text{f}$ 			& 1.0				& 1.0 				& $10^3 \, \frac{\text{kg}}{\text{m}^3}$		\parbox{0pt}{\rule{0pt}{1.5ex+\baselineskip}} \\
\hline
$\mu_\text{f}$  			& 1.0 				& 1.0 				& $\frac{\text{kg}}{\text{m s}}$				\parbox{0pt}{\rule{0pt}{1.5ex+\baselineskip}} \\
\hline
$v_\text{in}$ 			& 1.0				& 2.0				& $\frac{\text{m}}{\text{s}}$				\parbox{0pt}{\rule{0pt}{1.5ex+\baselineskip}} \\
\hline
$Re_\mathrm{in} = \frac{\rho_\mathrm{f}\, v_\mathrm{in} \, d_\mathrm{c}}{\mu_\mathrm{f}} $ & 100 & 200 &	- \parbox{0pt}{\rule{0pt}{2ex+\baselineskip}} \\
\hline
\end{tabular}
\caption{Flow past a circular cylinder with flexible tail: Material paramteres for the FSI-2 and FSI-3 benchmark cases.}
\label{tab:turek_fsi}
\end{center}
\end{table}

\begin{figure}[!ht]
	\centering
  	\includegraphics[width=1\textwidth]{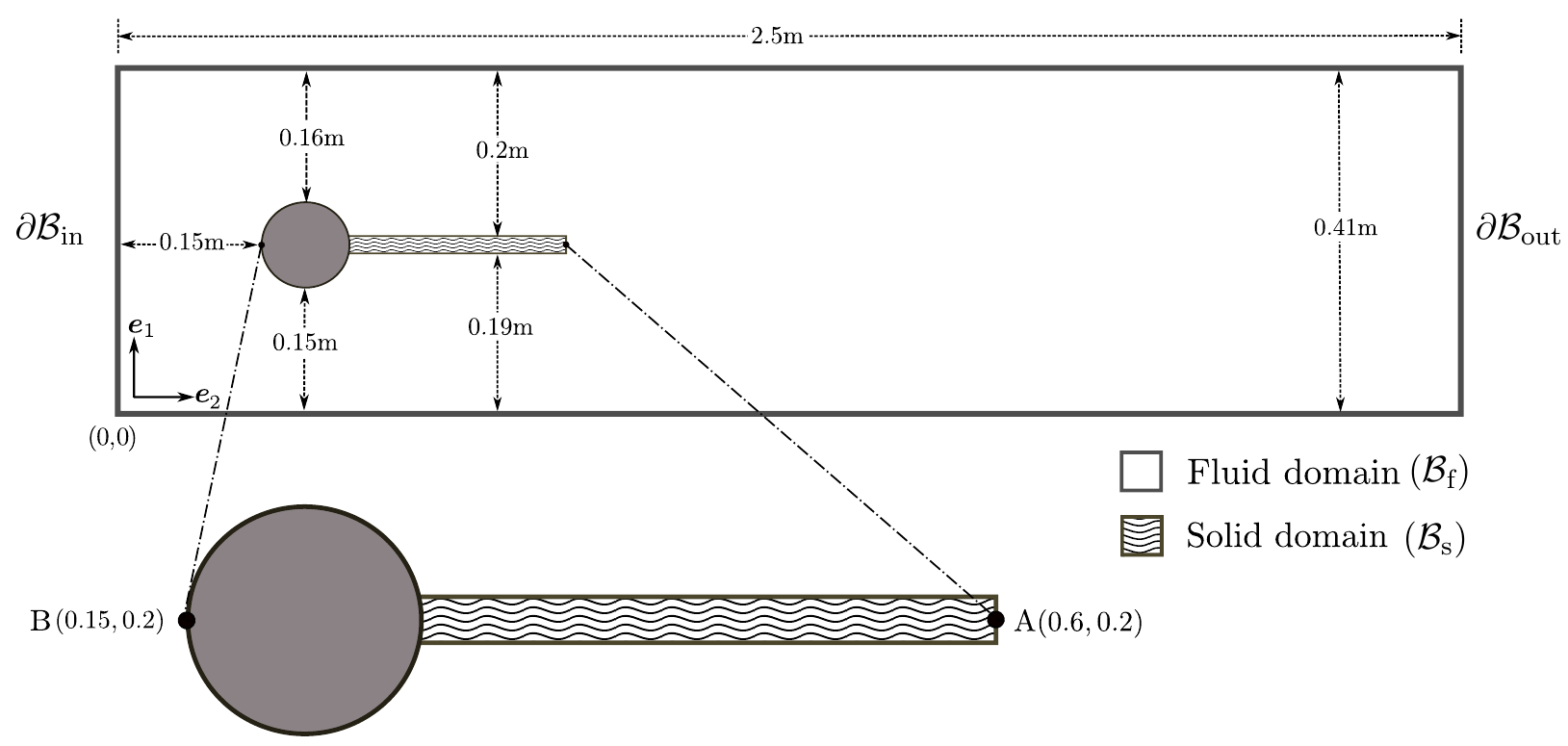}
	\caption{Flow past a circular cylinder with flexible tail: The problem description together with the associated dimensions of the analysis domain.}
	\label{fig:turek_fsi_config}
\end{figure}

Table~\ref{tab:turek_fsi} summarizes the material parameters for the fluid and the solid domain, while Figure~\ref{fig:turek_fsi_config} provides the geometrical details of the benchmark problem. The problem domain is the union of the fluid domain $\mathcal{B}_\text{f}$ and the solid domain $\mathcal{B}_\text{s}$. Let us denote the left-most face of the solid body with $\partial_u \mathcal{B}_\text{s}$. Then, the boundary conditions associated with the problem are expressed as
\begin{align}
	\boldsymbol{v}_\text{f} &= (v_\text{in},0) \,,	 \text{\hspace{0.75cm}}\forall \, \boldsymbol{x} \in \partial \mathcal{B}_\text{in} \,, \\
		\boldsymbol{v}_\text{f} &= \boldsymbol{0} \,,	\text{\hspace{1.7cm}} \forall \, \boldsymbol{x} \in \partial \mathcal{B}_\text{f} \backslash \left[\, \partial \mathcal{B}_\text{in} \cup  \partial \mathcal{B}_\text{out} \cup \left( \partial \mathcal{B}_\text{f} \cap \partial \mathcal{B}_\text{s} \right) \,\right] \,,\\
		\boldsymbol{\sigma}_\text{f} \cdot \boldsymbol{n}_\text{f} &= \boldsymbol{0} \,,	\text{\hspace{1.7cm}}\forall \, \boldsymbol{x} \in \partial \mathcal{B}_\text{out} \,, \\
		\boldsymbol{u}_\text{s} &= \boldsymbol{0} \,, \text{\hspace{1.7cm}} \forall \, \boldsymbol{x} \in \partial_u \mathcal{B}_\text{s} \,,
\end{align}
while at the fluid-solid interface 
\begin{align}
	\boldsymbol{v}_\text{f} &= \boldsymbol{v}_\text{s}\,, \text{\hspace{1.0cm}} \forall \, \boldsymbol{x} \in \partial \mathcal{B}_\text{f} \cap \partial \mathcal{B}_\text{s} \,,\\
	\boldsymbol{\sigma}_\text{f} \cdot \boldsymbol{n}_\text{f} + \boldsymbol{\sigma}_\text{s} \cdot \boldsymbol{n}_\text{s} &= \boldsymbol{0} \,, \text{\hspace{1.2cm}}\forall \, \boldsymbol{x} \in \partial \mathcal{B}_\text{f} \cap \partial \mathcal{B}_\text{s} \,.
\end{align}

The nature of the fluid flow within the domain is essentially controlled through the fluid velocity at the inlet i.e., with $v_\text{in}$, while the behavior of the solid is regulated through the material parameters $\rho_{\text{s}_o}$ and $\mu_\text{s}$.

\begin{figure}[!ht]
	\centering
  	\includegraphics[width=1\textwidth]{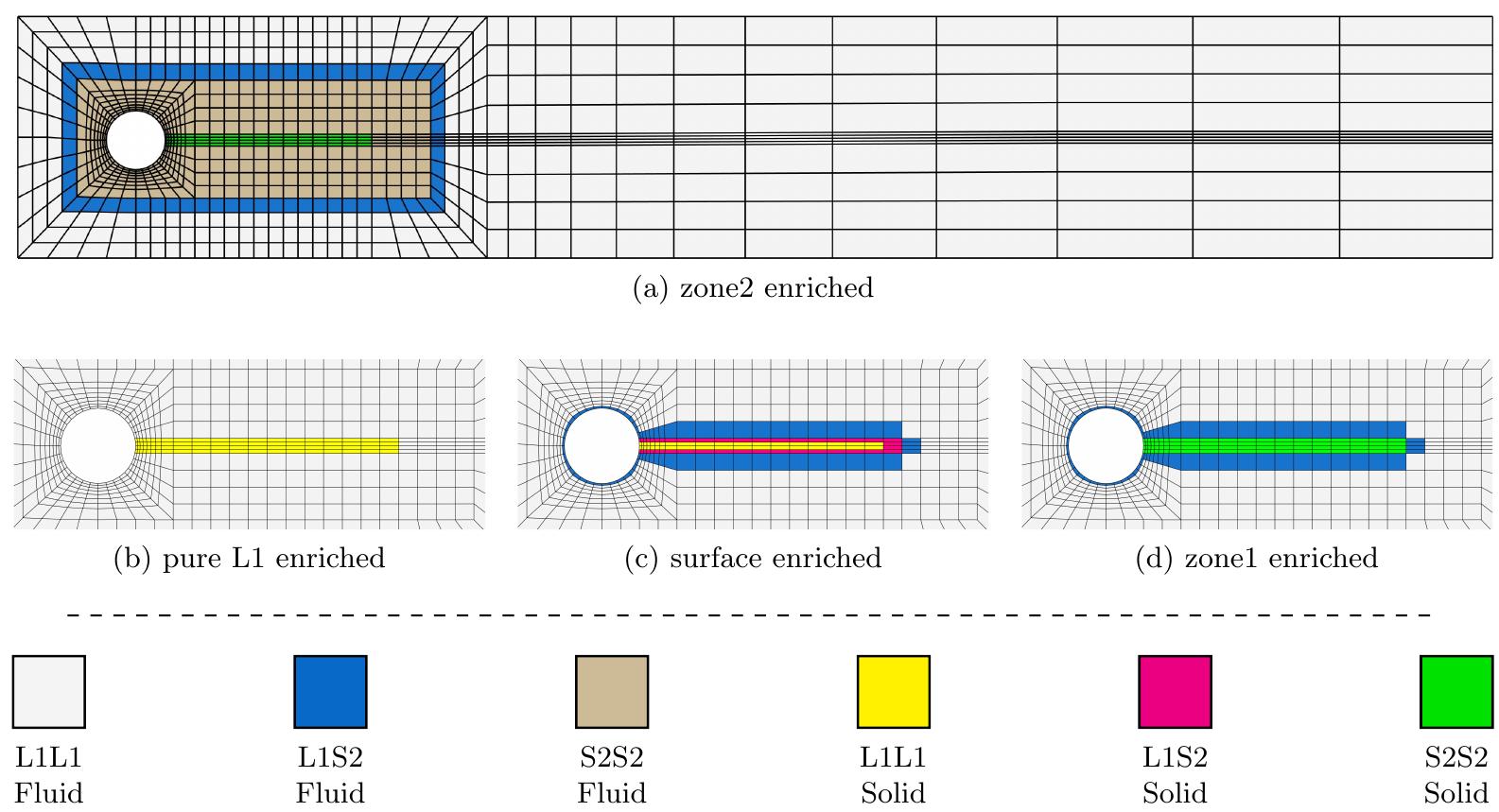}
	\caption{Flow past a circular cylinder with flexible tail: Different enriched discretizations at the coarsest mesh level (i.e., $m=0$). While (a) depicts the entire spatial domain, the subplots of (b), (c) and (d) provide exploded views of the discretization in the vicinity of the cylinder and the tail.}
	\label{fig:turek_zoneMesh_mx1}
\end{figure}

For this example we consider four IGA enrichment levels, which are similar in construction as those of the deformable tube example of Section~\ref{sec:ex_tube}. These include the pure~L1 case (with only L1L1 elements), the surface~enriched case (with L1S2 elements only at $\mathcal{B}_\text{f}\cap\mathcal{B}_\text{s}$), the zone1~enriched case (with S2S2 elements only for $\mathcal{B}_\text{s}$) and the zone2~enriched case (with zones of S2S2 elements for $\mathcal{B}_\text{f}$ and $\mathcal{B}_\text{s}$). Spatial meshes corresponding to the coarsest refinement level for different enriched discretization are shown in Figure~\ref{fig:turek_zoneMesh_mx1}, while element and $\mathrm{dof}$s population data is tabulated in Table~\ref{tab:turek_mesh}.
\begin{table}[!ht]
\begin{center}
\begin{adjustbox}{max width=\textwidth}
\begin{tabular}{| c | c | c | c | c | c | c | c | c | c | c | c | c | c | c | c |}
  \hline
\multicolumn{3}{|c|}{} 		& \multicolumn{2}{ c |}{}	& \multicolumn{3}{ c |}{}	& \multicolumn{4}{ c |}{} 	& \multicolumn{4}{ c |}{} \\
\multicolumn{3}{|c|}{\textbf{Refinement}} 	& \multicolumn{2}{ c |}{\textbf{pure L1}}	& \multicolumn{3}{ c |}{\textbf{surface enrich}}& \multicolumn{4}{ c |}{\textbf{zone1 enrich}}	& \multicolumn{4}{ c |}{\textbf{zone2 enrich}} \\ 
\multicolumn{3}{|c|}{}	& \multicolumn{2}{ c |}{}	& \multicolumn{3}{ c |}{}& \multicolumn{4}{ c |}{}	& \multicolumn{4}{ c |}{} \\  \cline{4-16}
\multicolumn{3}{|c|}{} 	& \textbf{Elements} & $\mathrm{dofs}$ & \multicolumn{2}{c|}{\textbf{Elements}} & $\mathrm{dofs}$ & \multicolumn{3}{c|}{\textbf{Elements}} & $\mathrm{dofs}$ & \multicolumn{3}{c|}{\textbf{Elements}} & $\mathrm{dofs}$ \\ \cline{1-4} \cline{6-7} \cline{9-11} \cline{13-15}	
$m$ & $\bar{\mathrm{h}}^e \,(\text{m})$	&		& $\mathrm{L}1\mathrm{L}1$ & & $\mathrm{L}1\mathrm{L}1$ & $\mathrm{L}1\mathrm{S}2$ & & $\mathrm{L}1\mathrm{L}1$ & $\mathrm{L}1\mathrm{S}2$ & $\mathrm{S}2\mathrm{S}2$ & & $\mathrm{L}1\mathrm{L}1$ & $\mathrm{L}1\mathrm{S}2$ & $\mathrm{S}2\mathrm{S}2$ &  \parbox{0pt}{\rule{0pt}{1.2ex+\baselineskip}} \\
\hline
\multirow{2}{*}{$0$} & \multirow{2}{*}{$0.05125$}	& Fluid		& 816		& 1,800 		& 742		& 72		& 1,820 		& 742 		& 72 	& - 			& 1,820 		& 348		& 68	& 400 		& 2,096  \\
 &							& Solid		& 80		& 210 		& 38  		& 40		& 220		& - 			& - 		& 80		& 276		& - 			& - 		& 80 		& 276 \\
\hline
\multirow{2}{*}{$1$} & \multirow{2}{*}{$0.02563$}	& Fluid		& 3,264 		& 6,864 		& 3,114 		& 148 		& 6,884		& 3,114		& 148 	& - 			& 6,884		& 1,528 		& 136	& 1,600		& 7,416\\
&							& Solid		& 320 		& 738		& 234 		& 84		& 748		&- 			& - 		& 320 		& 860		& -			& - 		& 320 		& 860 \\
\hline
\multirow{2}{*}{$2$} & \multirow{2}{*}{$0.01281$}	& Fluid		& 13,056	& 13,392	& 12,754	& 300		& 26,804	& 12,754	& 300 	& - 			& 26,804	& 6,384		& 272	& 6,400   	& 27,848\\
&							& Solid		& 1,280		& 2,754		& 1,106		& 172		& 2,764		& - 			& - 		& 1,280		& 2,988		& - 			& - 		& 1,280 		&  2,988 \\
\hline
\multirow{2}{*}{$3$} & \multirow{2}{*}{$0.00641$}	& Fluid		& 52,224	& 105,792	& 51,618	& 604		& 105,812	& 51,618	& 604 	& - 			& 105,812	& 26,080	& 544	& 25,600  	& 107,880\\
&							& Solid		& 5,120		& 10,626	& 4,770		& 348		& 10,636	& - 			& - 		& 5,120		& 11,084	& - 			& - 		& 5,120 		& 11,084\\
\hline
\end{tabular}	
\end{adjustbox}
\end{center}
\caption{Flow past a circular cylinder with flexible tail: Mesh statistics for succesive spatial mesh refinement study.}
\label{tab:turek_mesh}
\end{table}

\begin{figure}[!ht]
	\centering
  	\includegraphics[width=1\textwidth]{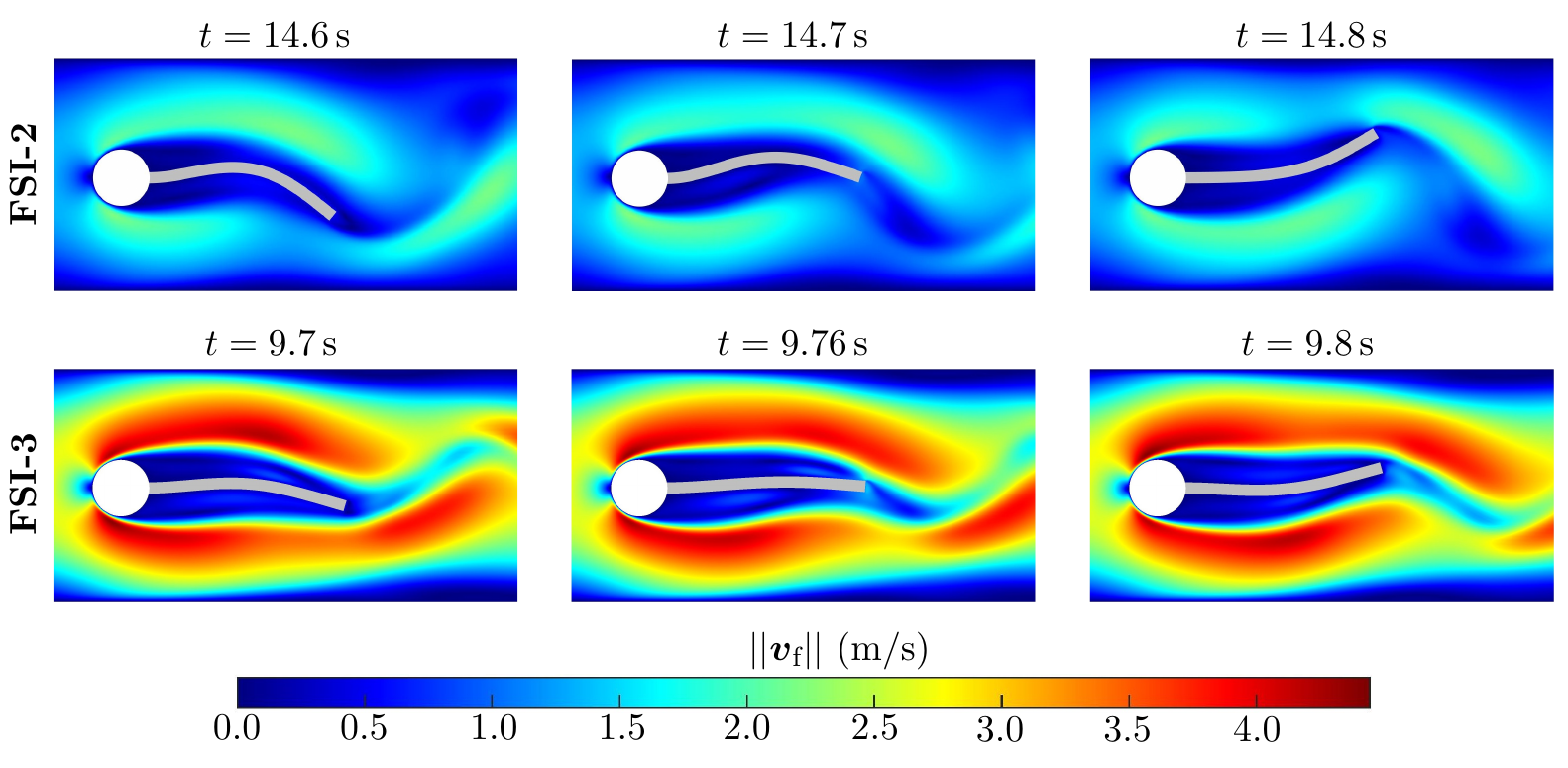}
	\caption{Flow past a circular cylinder with flexible tail: Contours of velocity magnitude at instants where $u_2$ of reference point A is minimum, zero and maximum.}
	\label{fig:turekFSI_contours}
\end{figure}

\begin{figure}[!ht]
	\centering
  	\includegraphics[width=1\textwidth]{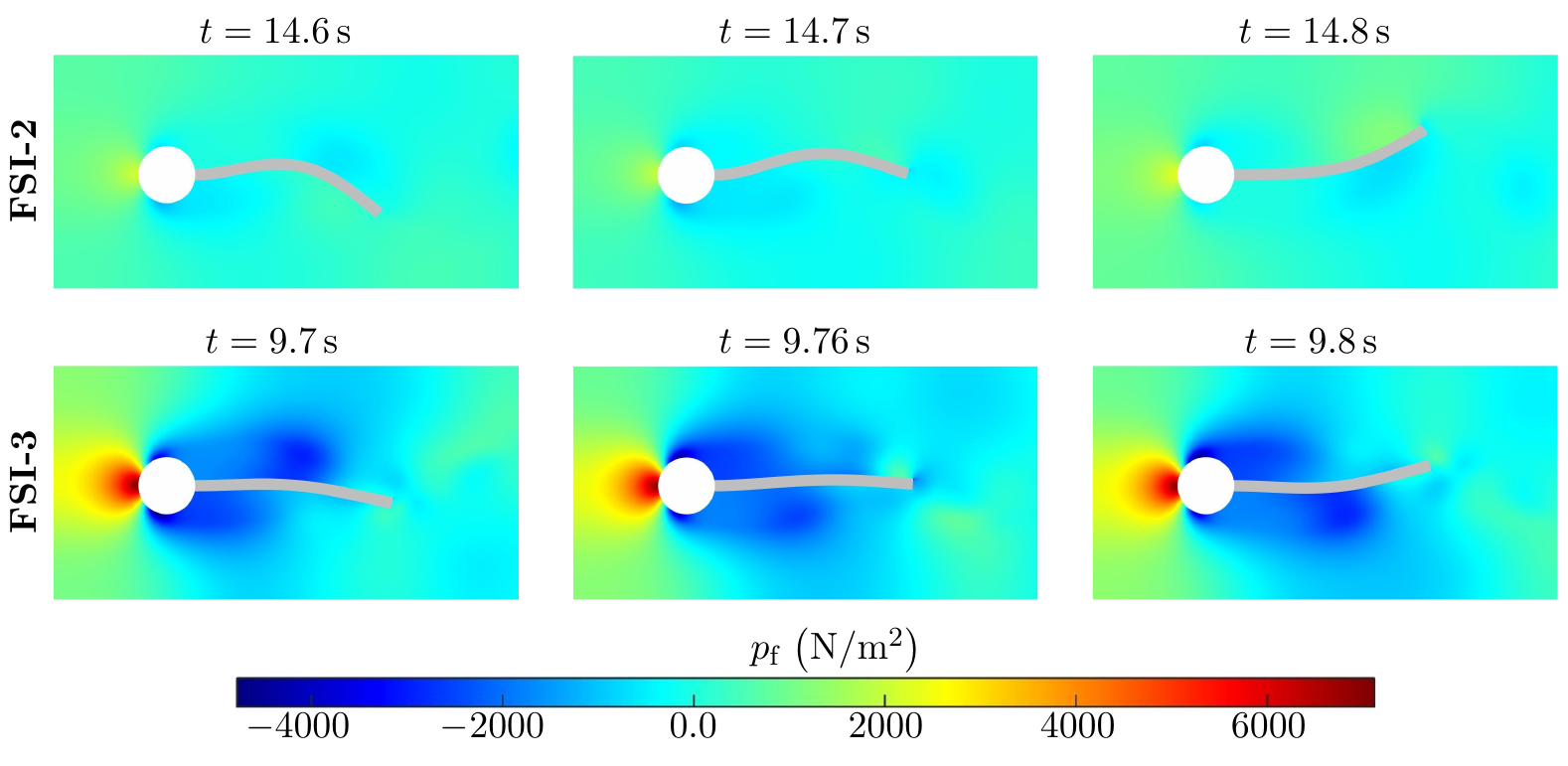}
	\caption{Flow past a circular cylinder with flexible tail: Contours of fluid pressure at instants where $u_2$ of reference point A is minimum, zero and maximum.}
	\label{fig:turekFSI_contours_press}
\end{figure}

\begin{figure}[h!]
	\centering
  	\includegraphics[width=1\textwidth]{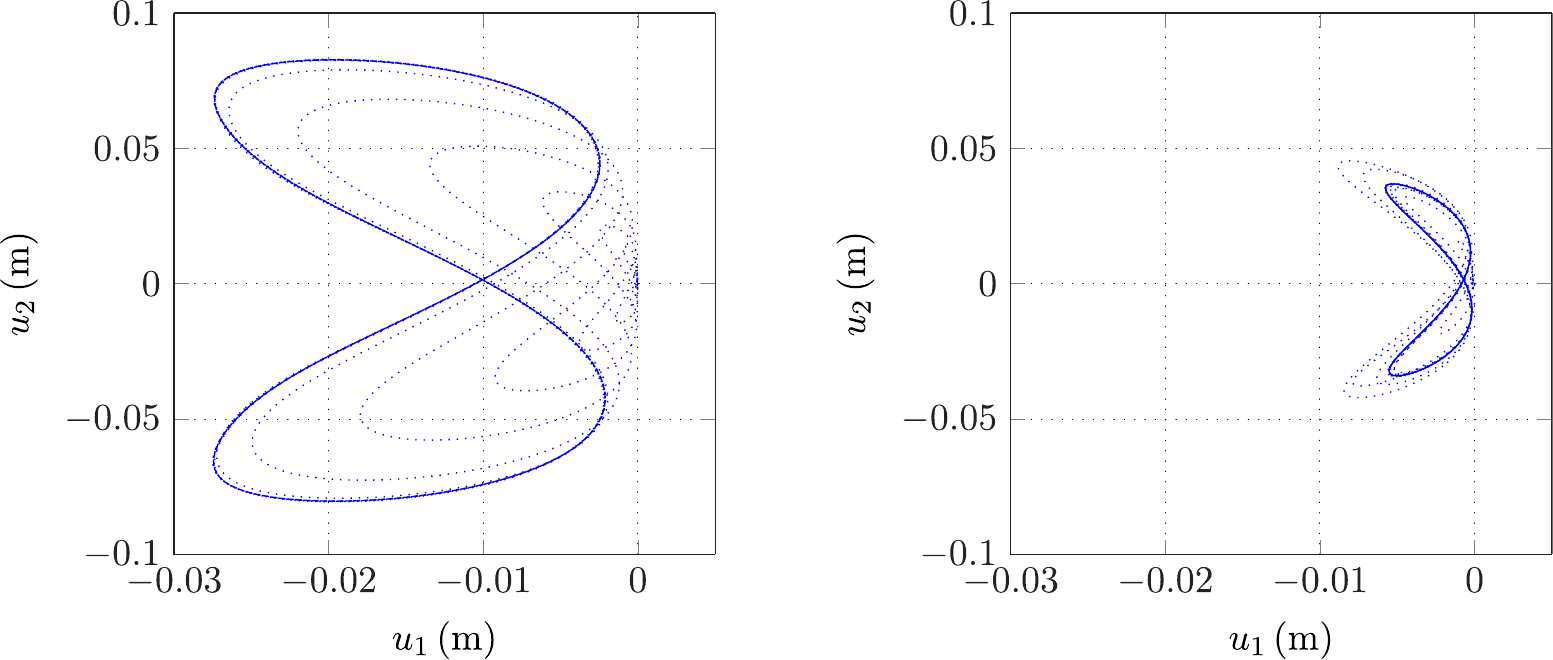}
	\caption{Flow past a circular cylinder with flexible tail: Evolution of the displacement of the reference point A for the FSI-2 and the FSI-3 benchmark case. A dotted blue line is used. Due to the periodic nature of the response, the dotted blue line appears to become a solid line due to proximate overlapping.}
	\label{fig:turekFSI_disp_evo} 
\end{figure}

Numerical simulations are performed over the interval $t\in\left[ 0,15\,\text{s}\right]$ for the FSI-2 test case and $t\in\left[ 0,10\,\text{s}\right]$ for the FSI-3 test case with a time step size of $0.001\,\text{s}$. The displacement of the reference point~A (see Figure~\ref{fig:turek_fsi_config}) is recorded over the entire simulation. Figure~\ref{fig:turekFSI_contours} shows contours of the velocity magnitude of the fluid at different deflection stages of the tail for the two investigated cases. Contours for fluid pressure at identical time-instances are shown in Figure~\ref{fig:turekFSI_contours_press}. Although the FSI-3 test case represents a higher Reynolds number flow (i.e., $Re=200$), the FSI-2 test case (with $Re=100$) is the more challenging one as the solid structure is softer and thus undergoes much larger deformations. The difference in the fluid-induced deformations in the flexible tail for the two test cases is also evident from Figure~\ref{fig:turekFSI_disp_evo}, where the evolution of the displacement of point A is plotted in terms of components $u_1$ and $u_2$.

\begin{figure}[h!]
	\centering
  	\includegraphics[width=1\textwidth]{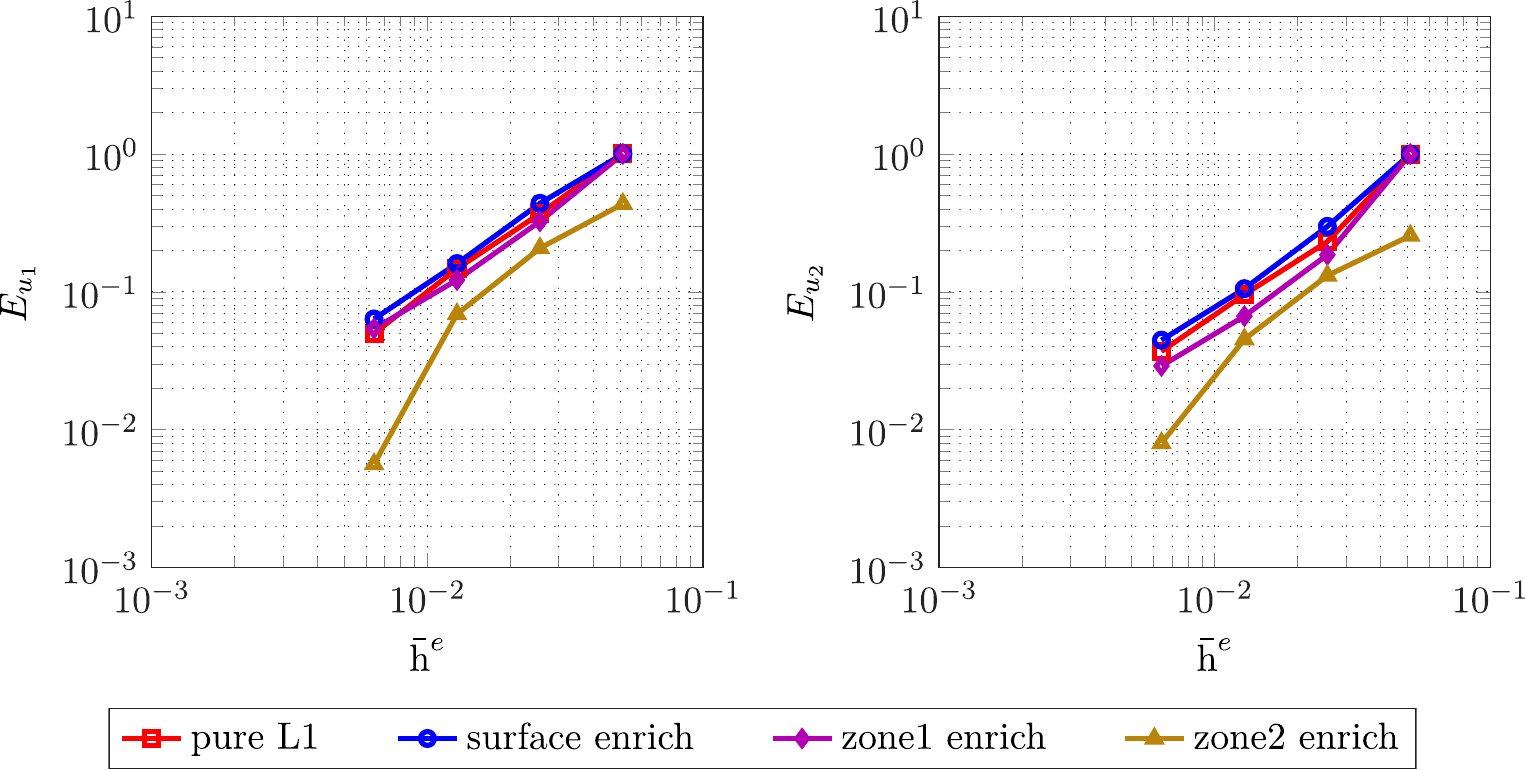}
	\caption{Flow past a circular cylinder with flexible tail: Mesh convergence behavior of the error norm in the Fourier coefficients, i.e., $E_{u_1} \text{ and } E_{u_2}$, for the FSI-2 benchmark case.}
	\label{fig:turekFSI2_err_fourier} 
\end{figure}

\begin{figure}[h!]
	\centering
  	\includegraphics[width=1\textwidth]{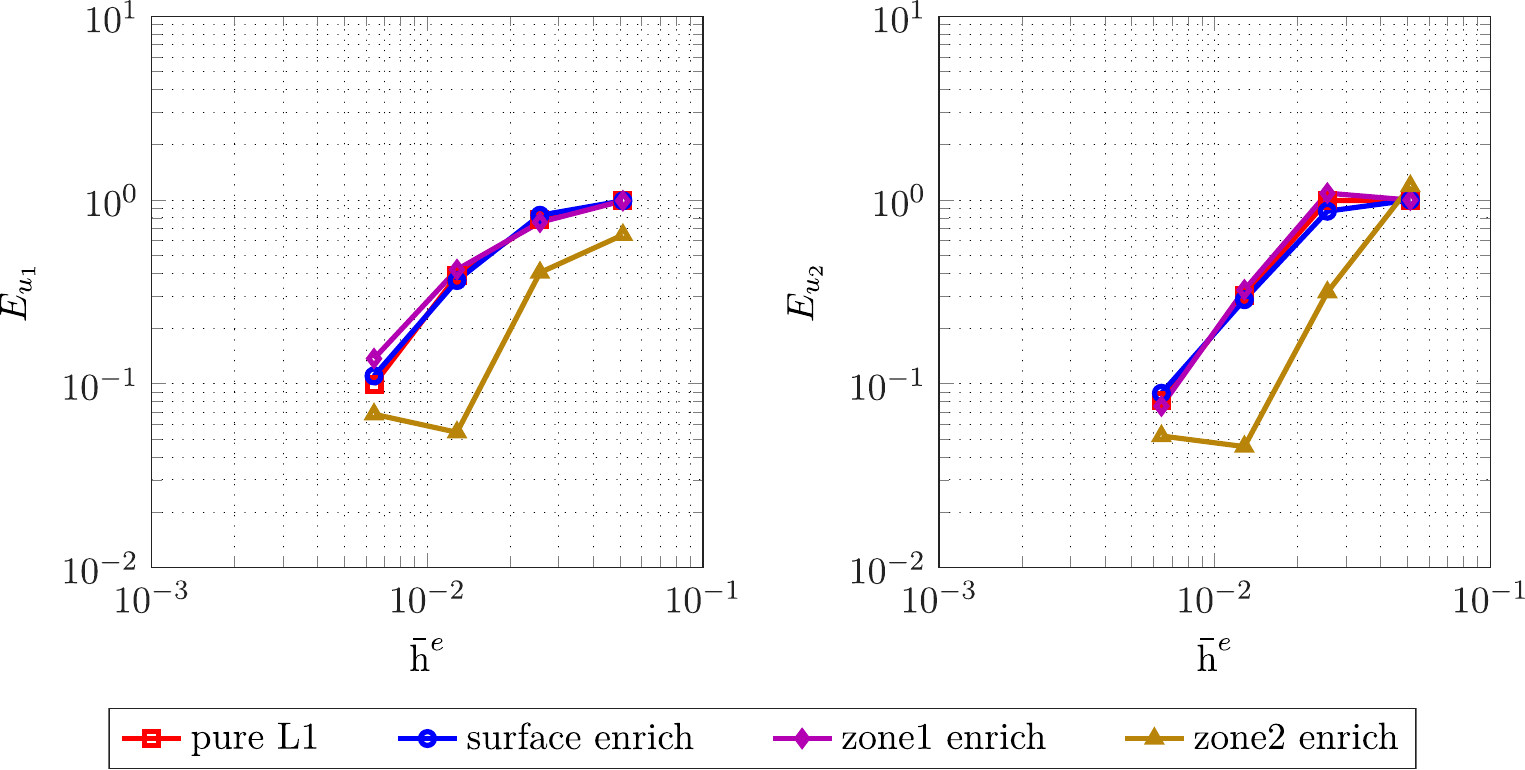}
	\caption{Flow past a circular cylinder with flexible tail: Mesh convergence behavior of the error norm in the Fourier coefficients, i.e., $E_{u_1} \text{ and } E_{u_2}$, for the FSI-3 benchmark case.}
	\label{fig:turekFSI3_err_fourier} 
\end{figure}

To gauge the effectiveness of different enrichment strategies used within this benchmark problem, we compare the error in the Fourier norm (see Eq.~\ref{eq:fourier_error_defn}) of the obtained numerical solutions with a reference solution. For this purpose, we use the benchmarking data available at~\cite{featflowFSI} as our reference solution, against which all the obtained numerical solutions are compared with. The response history of the horizontal $(u_1)$ and the vertical $(u_2)$ displacement of the reference point~A corresponding to the last two seconds of the simulation is used for this comparison. Within this interval, the displacement of the point~A becomes truly periodic. A Fourier series with $n=2$ offers a sound approximation, as any further increase in the number of Fourier coefficients $n$ yields identical approximation of the reference solution, and hence identical error plots. The results for this comparison are shown in Figure~\ref{fig:turekFSI2_err_fourier} for the FSI-2 test case and Figure~\ref{fig:turekFSI3_err_fourier} for the FSI-3 test case.

It is evident from the figures that the implemented scheme results in a consistent method. The consistent reduction in the Fourier error norms $(E_{u_1}$ and $E_{u_2})$ reflects that with each subsequent increment of the refinement level of the spatial mesh, the numerical solutions get closer to the reference solution reported in~\cite{featflowFSI}. This is particularly true for the FSI-2 case, as seen in Figure~\ref{fig:turekFSI2_err_fourier}. However, for the FSI-3 test case, the behavior of the error norms for the finer meshes (e.g., at $m=2,3$) reflect a minor difference between the reference solution and the converged solution from the obtained results. For both test cases, the zone2 enriched discretization yields results that are remarkably improved than all other enrichment cases. While all other enriched discretizations results in similar solutions (including the zone1 enriched with S2S2 elements for only $\mathcal{B}_\text{s}$), the zone2 enriched results are significantly more accurate and demonstrate higher order of convergence than others, particularly for the FSI-2 case. Hence for some applications, only analyzing the structure domain with IGA elements may not be enough and the inclusion of IGA zones for crucial fluid regions has the potential to offer significant accuracy gains. 
%
\section{Conclusion and outlook}			\label{sec-conc}
In this paper the efficacy of enriched finite element discretizations --- where isogeometric analysis is blended with classical Lagrangian finite elements --- is investigated in the context of fluid-structure interaction problems. In this regard, several different IGA enrichment strategies are considered. These include, among others, discretizations with IGA enriched surfaces and discretizations with IGA enriched volumetric zones. Results obtained from the studied numerical examples reveal interesting findings. For all the examples, similar solutions are obtained for the case of surface enriched discretizations (i.e., with IGA only at the surface of the fluid-solid interface) and classical Lagrangian finite elements discretizations. In some cases, classical Lagrangian finite elements even yield better accuracy than the surface enriched ones. However, with the addition of volumetric zones of IGA elements to the classical finite element mesh, the accuracy of the analysis increases remarkably. Analyzing only the solid domain with IGA analysis offers accuracy gains, however for some problems (as is the case in Section~\ref{sec:ex_turek}) these gains are not substantial. In this regard, enriching important regions of the much larger fluid domain (i.e., regions where flow gradients are expected to be large) offers a strategy that promises substantial accuracy gains. For all the studied examples, discretizations with IGA enriched elements in both the solid and the fluid domain offer the best quality numerical results without resorting to a full IGA analysis of the numerical problem.

IGA enriched finite elements offer a mechanism to enrich the classical finite element method with localized IGA. Such a treatment is extremely beneficial for FSI problems, where often large computational domains are employed to mitigate farfield boundary effects for the fluid domain. In this regard, careful IGA enrichment of only the crucial regions will result in improved numerical solutions with minimal increase in computational cost. Moreover, the strategy also provides a mechanism to analyze different interacting physical domains with different discretizations approaches in a conforming mesh setting (e.g., IGA for the solid and classical Lagrangian finite elements for the fluid).  

\bigskip
{\Large{\bf Acknowledgments}}
The financial support of the German Research Foundation (DFG) through the grants GSC 111 and SFB 1120 is gratefully acknowledged. 

\bigskip
%
\begin{appendices}
  \section{The finite element equations}		\label{appx:fe_eq}
In this section, we concisely present the implemented finite element formulation. In this regard, the derivation of the discrete residual vectors is first presented, which is followed by the important derivation of the tangent vectors. The formulation follows~{\cite{luginsland2017}}. 

  \subsection{The residual vectors}
To obtain the discrete finite element system discussed in Section~\ref{sec-continuum}, let us start with a finite element discretization of Eq.~(\ref{w_st_mom}) as suggested by Eq.~(\ref{eq:domain_fe}), i.e., 
\begin{align}
	&\bigcup_{e=1}^{n_{fe}} \left[ \,\, \int\limits_{\Omega^e} \rho_\text{f} \, \boldsymbol{w}^\text{h} \cdot \frac{\partial \boldsymbol{v}_\text{f}^\text{h}}{\partial t}  \bigg|_\chi \text{d}v + \int\limits_{\Omega^e} \!\rho_\text{f} \,\boldsymbol{w}^\text{h} \cdot \right.  \boldsymbol{c}^\text{h} \cdot \text{grad} \,\boldsymbol{v}^\text{h}_\text{f} \, \text{d}v   + \int\limits_{\Omega^e}\!\text{grad}\,\boldsymbol{w}^\text{h}: \boldsymbol{\sigma}_\text{f}^\text{h}  \, \text{d}v - \int\limits_{\Omega^e} \!\rho_\text{f} \, \boldsymbol{w}^\text{h} \cdot \boldsymbol{b}_\text{f} \,\text{d}v \Bigg. \, \Bigg]  \text{\hspace{1.0cm}} \nonumber \\
	 &\text{\hspace{5.0cm}}- \bigcup_{l=1}^{n_{fl}} \,\int\limits_{\Gamma^l} \! \boldsymbol{w}^\text{h} \cdot \bar{\boldsymbol{t}}_\text{f} \, \text{d}a + \sum_{e=1}^{n_{fe}} \, \int\limits_{\Omega^e} \! \tau_m \, \boldsymbol{c}^\text{h} \cdot \text{grad}\, \boldsymbol{w}^\text{h} \cdot \boldsymbol{\mathcal{R}}_m(\boldsymbol{v}_\text{f}^\text{h},p_\text{f}^\text{h})\, \text{d}v \nonumber \\ 
	 &\text{\hspace{8.7cm}}+ \sum_{e=1}^{n_{fe}} \,  \int\limits_{\Omega_e} \tau_c \,  (\,\text{div} \, \boldsymbol{w}^\text{h}\,) \, \mathcal{R}_c(\boldsymbol{v}_\text{f}^\text{h})\, \text{d}v= 0 \,. \label{eq:fe_fluid_eq1}
\end{align}
where $n_{fl}$ is the total number of surface elements in $\partial_t \mathcal{B}_\text{f}^\text{h}$ and $\Gamma^l$ is its $l^{th}$-element. Upon the substitution of approximation relations of Eqs.~(\ref{eq:disc_delu}-\ref{eq:disc_x}), we obtain
\begin{align}
	&\int\limits_{\Omega^e} \rho_\text{f} \, \boldsymbol{w}^\text{h} \cdot \frac{\partial \boldsymbol{v}_\text{f}^\text{h}}{\partial t}  \bigg|_\chi \text{d}v = \left(\textbf{w}^e\right)^\text{T} \int\limits_{\Omega^e} \rho_\text{f} \,\textbf{N}^\text{T} \, \textbf{N} \, \text{d}v \, \frac{\partial \textbf{v}^e_\text{f}}{\partial t}  \bigg|_\chi  = \left(\textbf{w}^e\right)^\text{T} \textbf{M}_\text{f}^e \, \frac{\partial \textbf{v}^e_\text{f}}{\partial t}  \bigg|_\chi \, , \\	
&\int\limits_{\Omega^e} \!\rho_\text{f} \,\boldsymbol{w}^\text{h} \cdot \boldsymbol{c}^\text{h} \cdot \text{grad} \,\boldsymbol{v}^\text{h}_\text{f} \, \text{d}v  = \left(\textbf{w}^e\right)^\text{T} \int\limits_{\Omega^e} \rho_\text{f} \, \textbf{N}^\text{T} \, \left( \boldsymbol{c}^\text{h} \cdot \text{grad}\, \boldsymbol{v}_\text{f}^\text{h}\right) \, \text{d}v = \left(\textbf{w}^e\right)^\text{T} \textbf{f}_\text{adv}^{\,e} (\textbf{v}^e_\text{f}) \, , \\ 
&\int\limits_{\Omega^e}\!\text{grad}\,\boldsymbol{w}^\text{h}: \boldsymbol{\sigma}_\text{f}^\text{h}  \, \text{d}v = \left(\textbf{w}^e\right)^\text{T} \int\limits_{\Omega^e} \textbf{B}^\text{T} \, \boldsymbol{\sigma}^\text{h}_\text{f} \, \text{d}v = \left(\textbf{w}^e\right)^\text{T} \, \textbf{f}_\text{fint}^{\,e} (\textbf{v}^e_\text{f}, \textbf{p}^e_\text{f}) \, , \label{eq:fl_int_elem}\\	
&\int\limits_{\Omega^e} \rho_\text{f} \, \boldsymbol{w}^\text{h} \cdot  \boldsymbol{b}_\text{f} \, \text{d}v =  \left(\textbf{w}^e\right)^\text{T} \int\limits_{\Omega^e} \rho_\text{f} \, \textbf{N}^\text{T} \, \boldsymbol{b}_\text{f} \, \text{d}v = \left( \textbf{w}^e\right)^\text{T} \, \textbf{f}^{\,e}_{\text{fextb}}\, , \\ 
&\int\limits_{\Gamma^l} \boldsymbol{w}^\text{h} \cdot \bar{\boldsymbol{t}}_\text{f}\, \text{d}a = \left( \textbf{w}^l\right)^\text{T} \int\limits_{\Gamma^l} \textbf{N}^\text{T}\, \bar{\boldsymbol{t}}_\text{f} \,\text{d}a= \left( \textbf{w}^l\right)^\text{T} \, \textbf{f}^{\,l}_{\text{fextt}}\, ,	
\end{align}
where $\boldsymbol{\sigma}$ now has a Voigt representation with the arrangement \mbox{$ \boldsymbol{\sigma} = [\sigma_{11}, \,\sigma_{22}, \,\sigma_{33}, \,\sigma_{23} , \, \sigma_{13}, \, \sigma_{12}]^\text{T}$ }, while $\textbf{B}$ is the strain rate-velocity matrix which in a three-dimensional setting takes the form
\begin{align}
	\textbf{B} &= \left[ \, \textbf{B}^1, \, \textbf{B}^2, \dots ,\, \textbf{B}^{n_n} \, \right]  \,, \\
	\text{such that\hspace{0.5cm}} \textbf{B}^\text{A} &= \left[ 
	\begin{array}{c c c c c c}
		N_{\text{A},1}	& 0					& 0					& 0					& N_{\text{A},3} 	& N_{\text{A},2} \\
		0	 			& N_{\text{A},2}	& 0					& N_{\text{A},3}	& 0 			  	& N_{\text{A},1}\\
		0 				& 0 				& N_{\text{A},3} 	& N_{\text{A},2}	& N_{\text{A},1}	& 0 
	\end{array}
\right]^\text{T}\,, \text{\hspace{1.75cm}}
\end{align} 
The discrete Cauchy stress tensor $\boldsymbol{\sigma}_\text{f}^\text{h}$, using the constitutive law of Eq.~(\ref{eq:fl_const_NS}), is expressed as
\begin{align}
	\boldsymbol{\sigma}_\text{f}^\text{h} = - \textbf{C}_p \, \tilde{\textbf{N}} \, \textbf{p}^e_\text{f} + 2 \, \mu_\text{f}\, \mathbb{I} \, \textbf{B} \, \textbf{v}_\text{f}^e \,,
\end{align} 
where $\textbf{C}_p = \left[\, 1, 1, 1, 0, 0, 0 \,\right]^\text{T}$ and $\mathbb{I}= (\delta_{ik}\delta_{jl} + \delta_{il} \delta_{jk})/2$. Let us now represent the terms involving the SUPG and the LSIC stabilization as,
\begin{align}
	 &\int\limits_{\Omega^e}  \tau_m \, \boldsymbol{c}^\text{h} \cdot \text{grad}\, \boldsymbol{w}^\text{h}  \! \cdot \!  \boldsymbol{\mathcal{R}}_m(\boldsymbol{v}_\text{f}^\text{h},p_\text{f}^\text{h})\, \text{d}v  = \left(\textbf{w}^e\right)^\text{T} \!\! \int\limits_{\Omega^e} \! \tau_m \, \textbf{B}_v^\text{T} \, \boldsymbol{\mathcal{R}}_m(\boldsymbol{v}_\text{f}^\text{h},p_\text{f}^\text{h}) \, \text{d}v = \left(\textbf{w}^e\right)^\text{T}  \textbf{f}^{\,e}_\text{SUPG}(\textbf{v}^e_\text{f},\textbf{p}^e_\text{f}) \,, \label{eq:app_SUPG_stab_disc}\\
	  &\int\limits_{\Omega_e} \tau_c \,  (\,\text{div} \, \boldsymbol{w}^\text{h}\,) \, \mathcal{R}_c(\boldsymbol{v}_\text{f}^\text{h})\, \text{d}v = \left(\textbf{w}^e\right)^\text{T} \int\limits_{\Omega_e} \tau_c \, \textbf{D} \, \mathcal{R}_c(\boldsymbol{v}^\text{h}_\text{f}) \, \text{d}v = \left(\textbf{w}^e\right)^\text{T} \textbf{f}^{\,e}_\text{LSIC}(\textbf{v}^e_\text{f}) \,,
\end{align}
where $\textbf{B}_v$ in Eq.~(\ref{eq:app_SUPG_stab_disc}) is given as, 
\begin{align}
 	\textbf{B}^\text{T}_v &= \left[ 
	\begin{array}{c}
	 \boldsymbol{c}^\text{h} \! \cdot \text{grad} (N_1) \, \textbf{I} \\ 
	 \boldsymbol{c}^\text{h} \! \cdot \text{grad} (N_2)\, \textbf{I} \\
	 \vdots \\
	 \boldsymbol{c}^\text{h} \! \cdot \text{grad} (N_{n_n}) \, \textbf{I} 
	\end{array}
	\right]  \, ,
\end{align}
while the finite element divergence operator $\textbf{D}$ is formulated as,
\begin{align}
\textbf{D} &= \left[ \, \textbf{G}^1, \, \textbf{G}^2, \, \dots , \textbf{G}^{n_n}\, \right] \,, \text{\hspace{2.2cm}} \\
	\text{with \hspace{1cm}}\textbf{G} &= \left[ \, \textbf{G}^1, \, \textbf{G}^2, \dots, \, \textbf{G}^{n_n} \, \right] \,,  \\
	\text{such that \hspace{0.5cm}}\textbf{G}^\text{A}&=\left[ \, N_{\text{A},1} \,, \, N_{\text{A},2} \,, \, N_{\text{A},3}\,  \right] \,.
\end{align}
The discrete residuals are evaluated through Eqs.~(\ref{eq:discrete_residual}-\ref{eq:discrete_residual_cont}). The divergence of $\boldsymbol{\sigma}_\text{f}^\text{h}$ can be obtained as
\begin{align}
	\text{div}\, \boldsymbol{\sigma}_\text{f}^\text{h} = -\textbf{G} \, \textbf{p}^e_\text{f} +  \textbf{C} \, \textbf{v}^e_\text{f} \,,
\end{align}
with 
\begin{align}
	\textbf{C} &= \mu_\text{f} \,\left[ \, \textbf{C}^1, \, \textbf{C}^2, \dots, \, \textbf{C}^{n_n} \, \right] \,, \text{\hspace{0.25cm}such that}  \\
 \textbf{C}^\text{A} \!&=\! \left[ \! \!
	\begin{array}{c c c}
		2N_{\text{A},11}\!+\!N_{\text{A},22}\!+\!N_{\text{A},33}	& N_{\text{A},12}													& N_{\text{A},13} \\
		N_{\text{A},21}												& N_{\text{A},11}\!+\!2N_{\text{A},22}\!+\!N_{\text{A},33}		& N_{\text{A},23} \\
		N_{\text{A},31}												& N_{\text{A},32} 														& N_{\text{A},11}\!+\!N_{\text{A},22}\!+\!2N_{\text{A},33}
	\end{array}
	\!\!\right ] \!.
\end{align}
We can now collect the individual terms representing Eq.~(\ref{eq:fe_fluid_eq1}) such that
\begin{align}
	\bigcup_{e=1}^{n_{fe}} \left( \textbf{w}^e\right)^\text{T} \left[ \, \textbf{M}_\text{f}^e \,\frac{\partial \textbf{v}_\text{f}^e}{\partial t} \bigg|_\chi + \textbf{f}_\text{adv}^{\,e} (\textbf{v}_\text{f}^e) + \textbf{f}_\text{fint}^{\,e} (\textbf{v}_\text{f}^e,\textbf{p}_\text{f}^e) + \textbf{f}_\text{SUPG}^{\,e} (\textbf{v}_\text{f}^e,\textbf{p}_\text{f}^e) + \textbf{f}_\text{LSIC}^{\,e} (\textbf{v}_\text{f}^e) - \textbf{f}_\text{fextb}^{\,e} \right] \nonumber \\
	 - \bigcup_{l=1}^{n_{fl}}\, ( \textbf{w}^l )^\text{T} \, \textbf{f}_\text{fextt}^{\,l} = 0 \,. \label{eq:fluid_fe_mom_elem}
\end{align}
The necessary ingredients for computing the element level representation of Eq.~(\ref{eq:fluid_fe_mom_elem}) have been defined. The element level contributions are then relocated to a global system using a finite element assembly operation, which ultimately results in
\begin{align}
		\left( \textbf{w}\right)^\text{T} \left [ \textbf{M}_\text{f}\, \frac{\partial \textbf{v}_\text{f}}{\partial t} \bigg|_\chi + \textbf{f}_\text{adv}(\textbf{v}_\text{f}) +\textbf{f}_\text{fint}(\textbf{v}_\text{f},\textbf{p}_\text{f}) + \textbf{f}_{\text{SUPG}}(\textbf{v}_\text{f},\textbf{p}_\text{f}) + \textbf{f}_\text{LSIC} (\textbf{v}_\text{f}) -  \textbf{f}_\text{fext} \right] = 0  \,,
\end{align}
where 
\begin{align}
	\textbf{f}_\text{fext} = \bigcup_{e=1}^{n_{fe}} \textbf{f}_\text{fextb}^{\,e} + \bigcup_{l=1}^{n_{fl}}\, \textbf{f}_\text{fextt}^{\,l} \,,
\end{align}
and $\textbf{w}$ contains the values of the test functions $\boldsymbol{w}$ for all the nodes in $\mathcal{B}_\text{f}^\text{h}$. Noting that the weak form holds for all admissible choice of test functions, we arrive at Eq.~(\ref{eq:fe_fl_mom_stab}), i.e.,
\begin{align}
	\textbf{R}_\text{m} \coloneqq \textbf{M}_\text{f}\, \frac{\partial \textbf{v}_\text{f}}{\partial t} \bigg|_\chi + \textbf{f}_\text{adv}(\textbf{v}_\text{f}) +\textbf{f}_\text{fint}(\textbf{v}_\text{f},\textbf{p}_\text{f}) + \textbf{f}_{\text{SUPG}}(\textbf{v}_\text{f},\textbf{p}_\text{f}) + \textbf{f}_\text{LSIC} (\textbf{v}_\text{f}) -  \textbf{f}_\text{fext}  = \textbf{0}  \,.
\end{align}
Adopting a similar procedure for Eq.~(\ref{w_st_mass}), the discrete finite element equation is of the form
\begin{align}
	\bigcup_{e=1}^{n_{fe}} \, \int\limits_{\Omega^e} q^\text{h} \, \text{div}\, \boldsymbol{v}_\text{f}^\text{h} \,\text{d}v + \sum_{e=1}^{n_{fe}} \, \int\limits_{\Omega^e}  \tau_m \,\, \text{grad}\,q^\text{h} \cdot \boldsymbol{\mathcal{R}}_m(\boldsymbol{v}_\text{f}^\text{h},p_\text{f}^\text{h})\,\text{d}v= 0 \,,
\end{align}
where
\begin{align}
	&\int\limits_{\Omega^e} q^\text{h} \, \text{div}\, \boldsymbol{v}_\text{f}^\text{h} \,\, \text{d}v =  \left(\textbf{q}^e\right)^\text{T} \int\limits_{\Omega^e} \tilde{\textbf{N}}^\text{T} \, \textbf{D} \, \text{d}v \, \, \textbf{v}^e_\text{f} = \left(\textbf{q}^e\right)^\text{T} \, \textbf{f}^{\,e}_\text{con}(\textbf{v}^e_\text{f}) \,, \\
	&\int\limits_{\Omega^e}  \tau_m \,\, \text{grad}\,q^\text{h} \! \cdot \! \boldsymbol{\mathcal{R}}_m (\boldsymbol{v}_\text{f}^\text{h},p_\text{f}^\text{h})\,\text{d}v = \left(\textbf{q}^e\right)^\text{T} \!\! \int\limits_{\Omega^e} \! \tau_p \, \textbf{G}^\text{T} \, \boldsymbol{\mathcal{R}}_m (\boldsymbol{v}_\text{f}^\text{h},p_\text{f}^\text{h}) \,\text{d}v = \left(\textbf{q}^e\right)^\text{T} \textbf{f}^{\,e}_\text{PSPG}(\textbf{v}^e_\text{f},\textbf{p}^e_\text{f})  \,,
\end{align}

The individual terms can be collected and then subsequently moved to a global system using a finite element assembly operator. Further noting that the global discrete system holds true for all admissible $\textbf{q}$, we arrive at Eq.~(\ref{eq:fe_fl_mass_stab}), i.e.,
\begin{align}
	\textbf{R}_\text{c} \coloneqq \textbf{f}_\text{con}(\textbf{v}_\text{f}) + \textbf{f}_{\text{PSPG}}(\textbf{v}_\text{f},\textbf{p}_\text{f})  = \textbf{0}  \,.
\end{align}
Applying a finite element discretization to Eq.~(\ref{eq:w_st_solid}), we obtain
\begin{align}
\bigcup_{e=1}^{n_{se}} \left[ \, \int\limits_{\Omega_\text{o}^e} \rho_{\text{s}_\text{o}} \delta \boldsymbol{u}^\text{h} \cdot\frac{D^2 \boldsymbol{u}_\text{s}^\text{h}}{Dt^2} \, \text{d}V + \int\limits_{\Omega_\text{o}^e} \text{Grad} \, \delta \boldsymbol{u}^\text{h} : \left( \boldsymbol{F} \boldsymbol{S}_\text{s}^\text{h}\right) \, \text{d}V - \int\limits_{\Omega_\text{o}^e} \rho_{\text{s}_\text{o}} \,\delta \boldsymbol{u}^\text{h} \cdot  \boldsymbol{b}_\text{s} \, \text{d}V  \, \right] \nonumber\\
 - \bigcup_{l=1}^{n_{sl}} \, \int\limits_{\Gamma_\text{o}^l} \delta \boldsymbol{u}^\text{h} \cdot \bar{\boldsymbol{t}}_\text{s}\, \text{d}A	= 0 \,. \label{eq:fe_solid_eq1}
 \end{align}
Here, $\Omega_\text{o}^e$ represents a unique finite element obtained from the discretization of the reference configuration. The ${l^{th}}$ surface element on the boundary $\partial_t \mathcal{B}_{\text{s}_\text{o}}^\text{h}$ is represented as $\Gamma_\text{o}^l$, while $n_{sl}$ represents the total number of surface elements on $\partial_t \mathcal{B}_{\text{s}_\text{o}}^\text{h}$. The individual terms of Eq.~(\ref{eq:fe_solid_eq1}) can be further expressed as
\begin{align}
	&\int\limits_{\Omega_\text{o}^e} \rho_{\text{s}_\text{o}} \delta \boldsymbol{u}^\text{h} \cdot\frac{D^2 \boldsymbol{u}_\text{s}^\text{h}}{Dt^2} \, \text{d}V = \left(\delta \textbf{u}^e\right)^\text{T}\int\limits_{\Omega_\text{o}^e} \rho_{\text{s}_\text{o}} \textbf{N}^\text{T} \, \textbf{N} \, \text{d}V	 \, \frac{D^2 \textbf{u}^e_\text{s}}{Dt^2} = \left(\delta \textbf{u}^e\right)^\text{T} \, \textbf{M}^e_\text{s} \, \frac{D^2 \textbf{u}^e_\text{s}}{Dt^2} \,, \label{eq:sol_iner_elem}\\
 &\int\limits_{\Omega^e_\text{o}} \text{Grad} \, \delta \boldsymbol{u}^\text{h} : \left( \boldsymbol{F} \boldsymbol{S}_\text{s}^\text{h}\right) \, \text{d}V =  \left(\delta \textbf{u}^e\right)^\text{T} \int\limits_{\Omega^e_\text{o}} \textbf{B}_\text{L}^\text{T} \, \boldsymbol{S}_s^\text{h}\, \text{d}V =  \left(\delta \textbf{u}^e\right)^\text{T} \, \textbf{f}^{\,e}_\text{sint}(\textbf{u}_\text{s}^e) \,, \\
&\int\limits_{\Omega^e_\text{o}} \rho_{\text{s}_\text{o}} \,\delta \boldsymbol{u}^\text{h} \cdot  \boldsymbol{b}_\text{s} \, \text{d}V =  \left(\delta \textbf{u}^e\right)^\text{T} \int\limits_{\Omega^e_\text{o}} \rho_{\text{s}_\text{o}} \, \textbf{N}^\text{T} \, \boldsymbol{b}_\text{s} \, \text{d}V = \left(\delta \textbf{u}^e\right)^\text{T} \, \textbf{f}^{\,e}_{\text{sextb}}\, , \\
&\int\limits_{\Gamma^l_\text{o}} \delta \boldsymbol{u}^\text{h} \cdot \bar{\boldsymbol{t}}_\text{s}\, \text{d}A = \left(\delta \textbf{u}^l\right)^\text{T} \, \int\limits_{\Gamma^l_\text{o}} \textbf{N}^\text{T}\, \bar{\boldsymbol{t}}_\text{s} \,\text{d}A= \left(\delta \textbf{u}^l\right)^\text{T} \, \textbf{f}^{\,l}_{\text{sextt}}\, . \label{eq:sol_extt_elem}
\end{align}
where $\textbf{B}_\text{L}$ is the strain-displacement matrix, which in $\mathbb{R}^{d}$ is taken as
\begin{align}
	\textbf{B}_\text{L} = \left[ \, \textbf{B}^1_\text{L}, \,\textbf{B}^2_\text{L}, \, \dots , \, \textbf{B}^{n_n}_\text{L}\, \right] \, ,
\end{align}
such that
\begin{align}
	\textbf{B}_\text{L} ^ \text{A} = \left[ 
	\begin{array}{c c c}
		F_{11} \, N_{\text{A},1}								&	F_{21} \, N_{\text{A},1}								& F_{31} \, N_{\text{A},1} \\
		F_{12} \, N_{\text{A},2}								&	F_{22} \, N_{\text{A},2}								& F_{32} \, N_{\text{A},2} \\
		F_{13} \, N_{\text{A},3}								&	F_{23} \, N_{\text{A},3}								& F_{33} \, N_{\text{A},3} \\
		F_{12} \, N_{\text{A},3}+F_{13} \, N_{\text{A},2}	& F_{22} \, N_{\text{A},3}+F_{23} \, N_{\text{A},2}	& F_{32} \, N_{\text{A},3}+F_{33} \, N_{\text{A},2}  \\
		F_{11} \, N_{\text{A},3}+F_{13} \, N_{\text{A},1}	& F_{21} \, N_{\text{A},3}+F_{23} \, N_{\text{A},1}	& F_{31} \, N_{\text{A},3}+F_{33} \, N_{\text{A},1} \\
F_{11} \, N_{\text{A},2}+F_{12} \, N_{\text{A},1}	& F_{21} \, N_{\text{A},2}+F_{22} \, N_{\text{A},1}	& F_{31} \, N_{\text{A},2}+F_{32} \, N_{\text{A},1} 
	\end{array}
\right] \,,
\end{align}
while the value of the second Piola-Kirchhoff stress tensor $\boldsymbol{S}_\text{s}^\text{h}$ can be obtained using Eq.~(\ref{eq:sol_cont_StVen}) with \mbox{$\boldsymbol{E}_\text{s} = ( \boldsymbol{F}^\text{T}\boldsymbol{F} - \textbf{I})/2$}. Subsequently performing the finite element assembly procedure, we obtain the global nodal force balance of the solid system, represented by Eq.~(\ref{eq:fe_s_mom_stab}), i.e.,
\begin{align}
  \textbf{R}_{\text{s}} \coloneqq \textbf{M}_\text{s} \, \frac{D^2 \textbf{u}_\text{s}}{Dt^2} + \textbf{f}_\text{sint}(\textbf{u}_\text{s}) - \textbf{f}_\text{sext}  = \boldsymbol{0} \,.
\end{align}
\subsection{The tangent matrices}
The nodal force balance system of Eqs.~(\ref{eq:fe_fl_mom_stab}-\ref{eq:fe_s_mom_stab}) consists of a system of nonlinear equations. Using a predictor-multicorrector algorithm, a Newton-Raphson approach is employed to linearize the nonlinear system, and subsequently advanced in time using the generalized-$\alpha$-method. A major component of this solution approach is the setup and solution of the linearized system of Eq.~(\ref{eq:residual_eq}). In-order to derive the terms associated with this system, let us expand Eq.~(\ref{eq:residual_eq}) in terms of its components, i.e.,  
\begin{align}
\left[ \def\arraystretch{2.2}
	\begin{array} {c c c}
\dfrac{\partial \textbf{R}_\text{m}^{n+1,k}}{\partial \dot{\textbf{v}}_\text{f}^{n+1}}	& \dfrac{\partial \textbf{R}_\text{m}^{n+1,k}}{\partial \textbf{p}_\text{f}^{n+1}} & \dfrac{\partial \textbf{R}_\text{m}^{n+1,k}}{\partial \ddot{\textbf{u}}_\text{s}^{n+1}}	 \\
\dfrac{\partial \textbf{R}_\text{c}^{n+1,k}}{\partial \dot{\textbf{v}}_\text{f}^{n+1}}	& \dfrac{\partial \textbf{R}_\text{c}^{n+1,k}}{\partial \textbf{p}_\text{f}^{n+1}} & \dfrac{\partial \textbf{R}_\text{c}^{n+1,k}}{\partial \ddot{\textbf{u}}_\text{s}^{n+1}}	  \\
\dfrac{\partial \textbf{R}_\text{s}^{n+1,k}}{\partial \dot{\textbf{v}}_\text{f}^{n+1}}	& \dfrac{\partial \textbf{R}_\text{s}^{n+1,k}}{\partial \textbf{p}_\text{f}^{n+1}}  & \dfrac{\partial \textbf{R}_\text{s}^{n+1,k}}{\partial \ddot{\textbf{u}}_\text{s}^{n+1}}	
	\end{array}
\right] \, \left\{ \def\arraystretch{2.2}
	\begin{array}{c}
	\Delta \dot{\textbf{v}}_\text{f}^{n+1}	\\
	\Delta \textbf{p}_\text{f}^{n+1}	\\
	\Delta \ddot{\textbf{u}}_\text{s}^{n+1}	
	\end{array}
\right\} = - \left\{ \def\arraystretch{2.2}
	\begin{array}{c}
		 \textbf{R}_\text{m}^{n+1,k} \\
		 \textbf{R}_\text{c}^{n+1,k} \\
		 \textbf{R}_\text{s}^{n+1,k} 
	\end{array}
\right\} \, ,		\label{eq:mat_residual_sys}
\end{align}
where each component is a finite element assembly of the corresponding element-level components. With reference to the notation used in Eq.~(\ref{eq:mon_fsi_orig}), we have
\begin{align}
	&\textbf{K}_\text{f} = \bigcup_{e=1}^{n_{fe}}\left[ \def\arraystretch{2.2}
	\begin{array}{c c}
		\dfrac{\partial \textbf{R}_\text{m}^e}{\partial \dot{\textbf{v}}_\text{f}^{n+1}}	& \dfrac{\partial \textbf{R}_\text{m}^e}{\partial \textbf{p}_\text{f}^{n+1}} \\
\dfrac{\partial \textbf{R}_\text{c}^e}{\partial \dot{\textbf{v}}_\text{f}^{n+1}}	& \dfrac{\partial \textbf{R}_\text{c}^e}{\partial \textbf{p}_\text{f}^{n+1}} 
	\end{array}
	\right]\, ,\text{\hspace{0.5cm}} \textbf{f}_\text{f} = \bigcup_{e=1}^{n_{fe}} \left\{ \! \def\arraystretch{2.2}
	\begin{array}{c}
		 \textbf{R}^{e}_{\text{m}} \\
		 \textbf{R}^{e}_{\text{c}}
	\end{array}
\!\right\} \,,\\
 	&\textbf{K}_\text{s} = \bigcup_{e=1}^{n_{se}}\dfrac{\partial \textbf{R}_\text{s}^e}{\partial \ddot{\textbf{u}}_\text{s}^{n+1}} \,, \text{\hspace{3.0cm}} \textbf{f}_\text{s} = \bigcup_{e=1}^{n_{se}} \, \textbf{R}^{e}_{\text{s}} \,,
\end{align}
where we have temporarily dropped the superscript $(n+1,k)$ for clarity. All subsequent representation of the residual and its tangent are understood to be evaluations at iteration $k$ for time-level $n+1$. Using Eqs.~(\ref{eq:fe_fl_mom_stab}-\ref{eq:fe_s_mom_stab}), it is it is evident that
\begin{align}
	\dfrac{\partial \textbf{R}_\text{m}}{\partial \ddot{\textbf{u}}_\text{s}^{n+1}} = \boldsymbol{0} \,, \text{\hspace{0.5cm}}\dfrac{\partial \textbf{R}_\text{c}}{\partial \ddot{\textbf{u}}_\text{s}^{n+1}}  = \boldsymbol{0} \,, \text{\hspace{0.5cm}} \dfrac{\partial \textbf{R}_\text{s}}{\partial \dot{\textbf{v}}_\text{f}^{n+1}} =  \boldsymbol{0}\text{\hspace{0.5cm}and\hspace{0.5cm}}\dfrac{\partial \textbf{R}_\text{s}}{\partial \textbf{p}_\text{f}^{n+1}} = \boldsymbol{0} \, .
\end{align}
Since the residual in Eq.~(\ref{eq:residual_eq}) is obtained using the field representation at the state $n+\alpha_m$ and $n+\alpha_f$, we express the corresponding tangent at the state $n+1$ in the following manner: 
\begin{align}
	\dfrac{\partial \textbf{R}_\text{m}}{\partial \dot{\textbf{v}}_\text{f}^{n+1}} &= \dfrac{\partial \textbf{R}_\text{m}}{\partial \dot{\textbf{v}}_\text{f}^{n+\alpha_m}} \, \frac{\partial \dot{\textbf{v}}_\text{f}^{n+\alpha_m}}{\partial \dot{\textbf{v}}_\text{f}^{n+1}} + \dfrac{\partial \textbf{R}_\text{m}}{\partial \textbf{v}_\text{f}^{n+\alpha_f}} \, \frac{\partial \textbf{v}_\text{f}^{n+\alpha_f}}{\partial \dot{\textbf{v}}_\text{f}^{n+\alpha_m}}\, \frac{\partial \dot{\textbf{v}}_\text{f}^{n+\alpha_m}}{\partial \dot{\textbf{v}}_\text{f}^{n+1}}\nonumber \\
	&= \alpha_m \, \dfrac{\partial \textbf{R}_\text{m}}{\partial \dot{\textbf{v}}_\text{f}^{n+\alpha_m}} + \alpha_f \, \gamma \, \Delta t\, \dfrac{\partial \textbf{R}_\text{m}}{\partial \textbf{v}_\text{f}^{n+\alpha_f}}  \, . \label{eq:tan_fl_alp1} 
\end{align}
Adopting the above given definition, we express the tangents associated with $\textbf{K}_\text{f}$ as
\begin{align}
	\dfrac{\partial \textbf{R}_\text{m}^e}{\partial \dot{\textbf{v}}_\text{f}^{n+1}} &= \alpha_m \left[ \,\, \int\limits_{\Omega^e} \rho_\text{f} \,\textbf{N}^\text{T} \, \textbf{N} \, \text{d}v + \int\limits_{\Omega^e} \tau_m \, \textbf{B}_v^\text{T} \,\rho_\text{f}\,  \textbf{N} \, \text{d}v  \right] + \alpha_f \gamma \Delta t \left[ \, \, \int\limits_{\Omega^e} \! \rho_\text{f} \, \textbf{N}^\text{T} \, \textbf{N}_v\,\text{d}v  \right. \nonumber \\
	&\text{\hspace{1.4cm}}\left.+ \! \int\limits_{\Omega^e} \!  \textbf{B}^\text{T} \mathbb{C} \, \textbf{B}\, \text{d}v + \! \int\limits_{\Omega^e}\! \tau_m \!\left( \,\textbf{B}^\text{T}_v \, \rho_\text{f} \,\textbf{N}_v - \textbf{B}_v^\text{T} \textbf{C} + \textbf{B}_f^\text{T} \textbf{N} \, \right) \text{d}v + \int\limits_{\Omega^e} \tau_c \, \textbf{D}^\text{T}  \, \textbf{D} \, \text{d}v \,  \right] , \\
	\dfrac{\partial \textbf{R}_\text{m}^e}{\partial \dot{\textbf{p}}_\text{f}^{n+1}}	 &= \int\limits_{\Omega^e} (-1) \, \textbf{D}^\text{T} \, \tilde{\textbf{N}}\, \text{d}v + \int\limits_{\Omega^e} \tau_m \, \textbf{B}_v^\text{T} \, \textbf{G} \, \text{d}v \,, \\
	\dfrac{\partial \textbf{R}_\text{c}^e}{\partial \dot{\textbf{v}}_\text{f}^{n+1}}	&= \alpha_m \int\limits_{\Omega^e} \tau_m \, \textbf{G}^\text{T} \,\rho_\text{f}\,  \textbf{N} \, \text{d}v + \alpha_f \gamma \Delta t \left[ \,\,\int\limits_{\Omega^e} \tilde{\textbf{N}}^\text{T} \textbf{D} \, \text{d}v +  \int\limits_{\Omega^e} \tau_m \, \textbf{G}^\text{T} \left( \rho_\text{f} \,\textbf{N}_v - \textbf{C} \right) \, \text{d}v \, \right]\,, \\
	\dfrac{\partial \textbf{R}_\text{c}^e}{\partial \dot{\textbf{p}}_\text{f}^{n+1}}	&= \int\limits_{\Omega^e} \tau_m \, \textbf{G}^\text{T} \, \textbf{G}\, \text{d}v \, ,
\end{align}
where $\mathbb{C} = 2 \, \mu_{\text{f}} \, \mathbb{I}$, while the matrices
\begin{align}
	\textbf{N}_v^\text{T} = \left[ 
	\begin{array}{c} 
		\boldsymbol{c}^\text{h}\cdot \text{grad}(N_1) \, \textbf{I}+ N_1 \, \text{grad}(\boldsymbol{v}_\text{f}^ \text{h}) \\
		\boldsymbol{c}^\text{h}\cdot \text{grad}(N_2) \, \textbf{I}+ N_2 \, \text{grad}(\boldsymbol{v}_\text{f}^ \text{h}) \\
		\vdots \\
		\boldsymbol{c}^\text{h}\cdot \text{grad}(N_{n_n}) \, \textbf{I}+ N_{n_n} \, \text{grad}(\boldsymbol{v}_\text{f}^ \text{h}) 
	\end{array}
	\right] 
	\text{\hspace{0.25cm}and\hspace{0.25cm}}\textbf{B}_f^\text{T} = \left[  
	\begin{array}{c}
		\boldsymbol{\mathcal{R}} \otimes \text{grad}(N_1) \\
		\boldsymbol{\mathcal{R}} \otimes \text{grad}(N_2) \\
		\vdots \\
		\boldsymbol{\mathcal{R}} \otimes \text{grad}(N_{n_n})
	\end{array}
	 \right] .
\end{align}
For the tangent matrices associated with the solid model, we note that using the chain-rule and the definitions from the generalized $\alpha$-method, we have
\begin{align}
	\dfrac{\partial \textbf{R}_\text{s}}{\partial \ddot{\textbf{u}}_\text{s}^{n+1}} = \alpha_m \, \dfrac{\partial \textbf{R}_\text{s}}{\partial \ddot{\textbf{u}}_\text{s}^{n+\alpha_m}} + \alpha_f \, \gamma \Delta t  \,\dfrac{\partial \textbf{R}_\text{s}}{\partial \dot{\textbf{u}}_\text{s}^{n+\alpha_f}} + \alpha_f \, \beta \, \Delta t^2 \dfrac{\partial \textbf{R}_\text{s}}{\partial \textbf{u}_\text{s}^{n+\alpha_f}} \,.  \label{eq:appen_res_sol}
\end{align} 
Hence the element-level tangent matrix associated with $\textbf{K}_\text{s}$ is expressed as
\begin{align}
	\dfrac{\partial \textbf{R}^e_\text{s}}{\partial \ddot{\textbf{u}}_\text{s}^{n+1}} = \alpha_m \int\limits_{\Omega_\text{o}^e} \rho_{\text{s}_\text{o}} \textbf{N}^\text{T} \, \textbf{N} \, \text{d}V + \alpha_f \, \beta \, \Delta t^2 \left[ \, \textbf{K}^e_\text{geo} + \textbf{K}^e_\text{mat} \, \right]\,,
\end{align}
where
\begin{align}
	\textbf{K}_\text{geo}^e = \left[  \def\arraystretch{1.2}
	\begin{array}{c c c c}
		\textbf{K}_\text{geo}^{11} 	& \textbf{K}_\text{geo}^{12}	& \dots 	& \textbf{K}_\text{geo}^{1n_n}	\\
		\textbf{K}_\text{geo}^{21} 	& \textbf{K}_\text{geo}^{22}	& \dots 	& \textbf{K}_\text{geo}^{2n_n} \\
		\vdots 								& \vdots 								& \ddots 	& \vdots \\
		\textbf{K}_\text{geo}^{n_n1} 	& \textbf{K}_\text{geo}^{n_n2}	& \dots 	& \textbf{K}_\text{geo}^{n_nn_n}
	\end{array}
	\right]	\text{\hspace{0.25cm}with\hspace{0.25cm}} \textbf{K}_\text{geo}^\text{AB} = \textbf{I} \int\limits_{\Omega^e_\text{o}} N_{\text{A},I} \, S_{IJ}  \, N_{\text{B},J} \, \text{d}V
\end{align}
is the geometrical stiffness contribution ($S_{IJ}$ is a component of $\boldsymbol{S}_\text{s}$), while
\begin{align}
	\textbf{K}_\text{mat}^e = \int\limits_{\Omega^e_\text{o}} \textbf{B}_\text{L}^\text{T} \,\mathbb{D} \, \textbf{B}_\text{L} \, \text{d}V
\end{align}
is the material stiffness with $\mathbb{D}$ being the fourth-order constitutive tensor, such that $\boldsymbol{S}_\text{s} = \mathbb{D} : \boldsymbol{E}_\text{s}$. For the considered Saint-Venant material model, the constitutive tensor is
\begin{align}
	\mathbb{D} =   \lambda_\text{s} \, \textbf{I} \otimes \textbf{I} + 2 \, \mu_\text{s} \, \mathbb{I}\, .
\end{align}

\end{appendices}

\bibliography{literature}

\begin{thebibliography}{88}
\providecommand{\natexlab}[1]{#1}
\providecommand{\url}[1]{\texttt{#1}}
\expandafter\ifx\csname urlstyle\endcsname\relax
  \providecommand{\doi}[1]{doi: #1}\else
  \providecommand{\doi}{doi: \begingroup \urlstyle{rm}\Url}\fi

\bibitem[fea(2016)]{featflowFSI}
Featflow {B}enchmark {S}uite.
\newblock
  \url{http://www.featflow.de/en/benchmarks/cfdbenchmarking/fsi_benchmark.html},
  July 2016.

\bibitem[Ahn et~al.(2010)Ahn, Branets, and Carey]{ahn2010}
H.~T. Ahn, L.~Branets, and G.~F. Carey.
\newblock Moving boundary simulations with dynamic mesh smoothing.
\newblock \emph{International Journal for Numerical Methods in Fluids},
  64:\penalty0 887--907, 2010.

\bibitem[Akkerman et~al.(2008)Akkerman, Bazilevs, Calo, Hughes, and
  Hulshoff]{akkerman2008}
I.~Akkerman, Y.~Bazilevs, V.~Calo, T.~Hughes, and S.~Hulshoff.
\newblock The role of continuity in residual-based variational multiscale
  modeling of turbulence.
\newblock \emph{Computational Mechanics}, 41\penalty0 (3):\penalty0 371--378,
  2008.

\bibitem[Bates and Watts(1988)]{batesbook}
D.~M. Bates and D.~G. Watts.
\newblock \emph{{N}onlinear {R}egression {A}nalysis and its {A}pplications}.
\newblock John Wiley \& Sons, 1st edition, 1988.
\newblock ISBN 978-0471816430.

\bibitem[Bazilevs et~al.(2007)Bazilevs, Calo, Cottrell, Hughes, Reali, and
  Scovazzi]{bazilevs2007vms}
Y.~Bazilevs, V.~M. Calo, J.~A. Cottrell, T.~J.~R. Hughes, A.~Reali, and
  G.~Scovazzi.
\newblock Variational multiscale residual-based turbulence modeling for large
  eddy simulation of incompressible flows.
\newblock \emph{Computer Methods in Applied Mechanics and Engineering},
  197\penalty0 (1-4):\penalty0 173--201, 2007.

\bibitem[Bazilevs et~al.(2008)Bazilevs, Calo, R., Hughes, and
  Zhang]{bazilevs2008}
Y.~Bazilevs, V.~M. Calo, T.~J. R., Hughes, and Y.~Zhang.
\newblock Isogeometric fluid-structure interaction: theory, algorithms, and
  computations.
\newblock \emph{Computational Mechanics}, 43\penalty0 (1):\penalty0 3--37,
  2008.

\bibitem[Bazilevs et~al.(2009)Bazilevs, Gohean, Hughes, Moser, and
  Zhang]{bazilevs2009}
Y.~Bazilevs, J.~Gohean, T.~Hughes, R.~Moser, and Y.~Zhang.
\newblock Patient-specific isogeometric fluid-structure interaction analysis of
  thoracic aortic blood flow due to implantation of the {J}arvik 2000 left
  ventricular assist device.
\newblock \emph{Computer Methods in Applied Mechanics and Engineering},
  198\penalty0 (45-46):\penalty0 3534--3550, 2009.
\newblock Models and Methods in Computational Vascular and Cardiovascular
  Mechanics.

\bibitem[Bazilevs et~al.(2010)Bazilevs, Michler, Calo, and
  Hughes]{bazilevs2010}
Y.~Bazilevs, C.~Michler, V.~M. Calo, and T.~J.~R. Hughes.
\newblock Isogeometric variational multiscale modeling of wall-bounded
  turbulent flows with weakly enforced boundary conditions on unstretched
  meshes.
\newblock \emph{Computer Methods in Applied Mechanics and Engineering},
  199\penalty0 (13-16):\penalty0 780--790, 2010.

\bibitem[Bazilevs et~al.(2011)Bazilevs, Hsu, Kiendl, W{\"u}chner, and
  Bletzinger]{bazilevs2011}
Y.~Bazilevs, M.-C. Hsu, J.~Kiendl, R.~W{\"u}chner, and K.-U. Bletzinger.
\newblock 3{D} simulation of wind turbine rotors at full scale. {P}art {II}:
  Fluid-structure interaction modeling with composite blades.
\newblock \emph{International Journal for Numerical Methods in Fluids},
  65\penalty0 (1-3):\penalty0 236--253, 2011.

\bibitem[Bazilevs et~al.(2012)Bazilevs, Hsu, and Scott]{bazilevs2012}
Y.~Bazilevs, M.-C. Hsu, and M.~A. Scott.
\newblock Isogeometric fluid-structure interaction analysis with emphasis on
  non-matching discretizations, and with application to wind turbines.
\newblock \emph{Computer Methods in Applied Mechanics and Engineering},
  249-252:\penalty0 28--41, 2012.
\newblock Higher Order Finite Element and Isogeometric Methods.

\bibitem[Bazilevs et~al.(2016)Bazilevs, Korobenko, Deng, and Yan]{bazilevs2016}
Y.~Bazilevs, A.~Korobenko, X.~Deng, and J.~Yan.
\newblock Fluid-structure interation modeling for fatigue-damage prediction in
  full-scale wind-turbine blades.
\newblock \emph{Journal of Applied Mechanics}, 83, 2016.

\bibitem[Belvins(1990)]{belvinsbook}
R.~D. Belvins.
\newblock \emph{{F}low-{I}nduced {V}ibration}.
\newblock Kreiger Publishing Company, 2nd edition, 1990.
\newblock ISBN 1-57524-183-8.

\bibitem[Bendiksen(1991)]{bendiksen1991}
O.~Bendiksen.
\newblock A new approach to computational aeroelasticity.
\newblock In \emph{AIAA Paper 91-0939-CP, 32nd Structures, Structural Dynamics
  and Materials Conference}, 1991.

\bibitem[Bhardwaj and Mittal(2012)]{bhardwaj2012}
R.~Bhardwaj and R.~Mittal.
\newblock Benchmarking a coupled immersed-boundary-finite-element solver for
  large-scale flow induced deformation.
\newblock \emph{AIAA Journal}, 50\penalty0 (7):\penalty0 1638--1642, 2012.

\bibitem[Brooks and Hughes(1982)]{brooks1982}
A.~N. Brooks and T.~J.~R. Hughes.
\newblock Streamline {U}pwind/{P}etrov--{G}alerkin formulations for convection
  dominated flows with particular emphasis on the incompressible
  {N}avier--{S}tokes equations.
\newblock \emph{Advances in Applied Mechanics}, 32:\penalty0 199--259, 1982.

\bibitem[Causin et~al.(2005)Causin, Gerbeau, and Nobile]{causin2005}
P.~Causin, J.~Gerbeau, and F.~Nobile.
\newblock Added-mass effect in the design of partitioned algorithms for
  fluid-structure problems.
\newblock \emph{Computer Methods in Applied Mechanics and Engineering},
  194\penalty0 (42-44):\penalty0 4506--4527, 2005.

\bibitem[Chivukula et~al.(2014)Chivukula, Mousel, Lu, and
  Vigmostad]{chivukula2014}
V.~Chivukula, J.~Mousel, J.~Lu, and S.~Vigmostad.
\newblock Micro-scale blood particulate dynamics using a non-uniform rational
  {B}-spline-based isogeometric analysis.
\newblock \emph{International Journal for Numerical Methods in Biomedical
  Engineering}, 30\penalty0 (12):\penalty0 1437--1459, 2014.

\bibitem[Chung and Hulbert(1993)]{chung1993}
J.~H. Chung and G.~M. Hulbert.
\newblock A time integration algorithm for structural dynamics with improved
  numerical dissipation: the generalized-$\alpha$ method.
\newblock \emph{Journal of Applied Mechanics}, 60:\penalty0 371--375, 1993.

\bibitem[Codina(2000)]{codina2000}
R.~Codina.
\newblock Stabilization of incompressibility and convection through orthogonal
  sub-scales in finite element methods.
\newblock \emph{Computer Methods in Applied Mechanics and Engineering},
  190\penalty0 (13-14):\penalty0 1579--1599, 2000.

\bibitem[Collier et~al.(2012)Collier, Pardo, Dalcin, Paszynski, and
  Calo]{collier2012}
N.~Collier, D.~Pardo, L.~Dalcin, M.~Paszynski, and V.~M. Calo.
\newblock The cost of continuity: A study of the performance of isogeometric
  finite elements using direct solvers.
\newblock \emph{Computer Methods in Applied Mechanics and Engineering},
  213-216:\penalty0 353--361, 2012.

\bibitem[Collier et~al.(2013)Collier, Dalcin, Pardo, and Calo]{collier2013}
N.~Collier, L.~Dalcin, D.~Pardo, and V.~M. Calo.
\newblock The cost of continuity: performance of iterative solvers on
  isogeometric finite element.
\newblock \emph{SIAM Journal on Scientific Computing}, 35\penalty0
  (2):\penalty0 A767--A784, 2013.

\bibitem[Corbett and Sauer(2014)]{corbett2014}
C.~J. Corbett and R.~A. Sauer.
\newblock {NURBS}-enriched contact finite elements.
\newblock \emph{Computer Methods in Applied Mechanics and Engineering},
  275:\penalty0 55--75, 2014.

\bibitem[Corbett and Sauer(2015)]{corbett2015}
C.~J. Corbett and R.~A. Sauer.
\newblock Three-dimensional isogeometrically enriched finite elements for
  frictional contact and mixed-mode debonding.
\newblock \emph{Computer Methods in Applied Mechanics and Engineering},
  284:\penalty0 781--806, 2015.
\newblock Isogeometric Analysis Special Issue.

\bibitem[Cottrell et~al.(2006)Cottrell, Reali, Bazilevs, and
  Hughes]{cottrell2006}
J.~A. Cottrell, A.~Reali, Y.~Bazilevs, and T.~J.~R. Hughes.
\newblock Isogeometric analysis of structural vibrations.
\newblock \emph{Computer Methods in Applied Mechanics and Engineering},
  195\penalty0 (41-43):\penalty0 5257--5296, 2006.
\newblock John H. Argyris Memorial Issue. Part {II}.

\bibitem[Cottrell et~al.(2009)Cottrell, Hughes, and Bazilevs]{cottrellbook}
J.~A. Cottrell, T.~J.~R. Hughes, and Y.~Bazilevs.
\newblock \emph{{I}sogeometric {A}nalysis: {T}oward {I}ntegration of {CAD} and
  {FEA}}.
\newblock John Wiley \& Sons, 2009.
\newblock ISBN 978-0-470-74873-2.

\bibitem[Dettmer and Peri{\'c}(2006)]{dettmer2006}
W.~Dettmer and D.~Peri{\'c}.
\newblock A computational framework for fluid-structure interaction: Finite
  element formulation and applications.
\newblock \emph{Computer Methods in Applied Mechanics and Engineering},
  195\penalty0 (41-43):\penalty0 5754--5779, 2006.

\bibitem[Donea and Huerta(2003)]{doneabook}
J.~Donea and A.~Huerta.
\newblock \emph{{F}inite {E}lement {M}ethods for {F}low {P}roblems}.
\newblock John Wiley \& Sons, 2003.
\newblock ISBN 0-471-49666-9.

\bibitem[Duong et~al.(2017)Duong, Roohbakhshan, and Sauer]{duong2016}
T.~X. Duong, F.~Roohbakhshan, and R.~A. Sauer.
\newblock A new rotation-free isogeometric thin shell formulation and a
  corresponding continuity constraint for patch boundaries.
\newblock \emph{Computer Methods in Applied Mechanics and Engineering},
  316:\penalty0 43 -- 83, 2017.
\newblock Special Issue on Isogeometric Analysis: Progress and Challenges.

\bibitem[{D\"utsch} et~al.(1998){D\"utsch}, Durst, Becker, and
  Lienhart]{duetsch1998}
H.~{D\"utsch}, F.~Durst, S.~Becker, and H.~Lienhart.
\newblock Low-{R}eynolds-number flow around an oscillating circular cylinder at
  low {K}eulegan-{C}arpenter numbers.
\newblock \emph{Journal of Fluid Mechanics}, 360:\penalty0 249--271, 1998.

\bibitem[Eken and Mehmet(2016)]{eken2016}
A.~Eken and M.~S. Mehmet.
\newblock A parallel monolithic algorithm for the numerical simulation of
  large-scale fluid structure interaction problems.
\newblock \emph{International Journal for Numerical Methods in Fluids},
  80\penalty0 (12):\penalty0 687--714, 2016.

\bibitem[Eken and Sahin(2017)]{eken2017}
A.~Eken and M.~Sahin.
\newblock A parallel monolithic approach for fluid-structure interaction in a
  cerebral aneurysm.
\newblock \emph{Computers \& Fluids}, 153:\penalty0 61--75, 2017.

\bibitem[Farhat(2004)]{farhat2004}
C.~Farhat.
\newblock \emph{CFD-Based Nonlinear Computational Aeroelasticity}.
\newblock John Wiley \& Sons, Ltd, 2004.
\newblock ISBN 9780470091357.

\bibitem[Farhat et~al.(1998)Farhat, Degand, Koobus, and Lesoinne]{farhat1998}
C.~Farhat, C.~Degand, B.~Koobus, and M.~Lesoinne.
\newblock Torsional springs for two-dimensional dynamic unstructured fluid
  meshes.
\newblock \emph{Computer Methods in Applied Mechanics and Engineering},
  163\penalty0 (1-4):\penalty0 231--245, 1998.

\bibitem[Felippa et~al.(2001)Felippa, Park, and Farhat]{felippa2001}
C.~A. Felippa, K.~C. Park, and C.~Farhat.
\newblock Partitioned analysis of coupled mechanical systems.
\newblock \emph{Computer Methods in Applied Mechanics and Engineering},
  190\penalty0 (24-25):\penalty0 3247--3270, 2001.
\newblock Advances in Computational Methods for Fluid-Structure Interaction.

\bibitem[Formaggia et~al.(2001)Formaggia, Gerbeau, Nobile, and
  Quarteroni]{formaggia2001}
L.~Formaggia, J.~Gerbeau, F.~Nobile, and A.~Quarteroni.
\newblock On the coupling of 3{D} and 1{D} {N}avier-{S}tokes equations for flow
  problems in compliant vessels.
\newblock \emph{Computer Methods in Applied Mechanics and Engineering},
  191\penalty0 (6-7):\penalty0 561--582, 2001.
\newblock Minisymposium on Methods for Flow Simulation and Modeling.

\bibitem[F{\"o}rster et~al.(2007)F{\"o}rster, Wall, and Ramm]{foerster2007}
C.~F{\"o}rster, W.~A. Wall, and E.~Ramm.
\newblock Artificial added mass instabilities in sequential staggered coupling
  of nonlinear structures and incompressible viscous flows.
\newblock \emph{Computer Methods in Applied Mechanics and Engineering},
  196\penalty0 (7):\penalty0 1278--1293, 2007.

\bibitem[Franca and Valentin(2000)]{franca2000}
L.~P. Franca and F.~Valentin.
\newblock On an improved unusual stabilized finite element method for the
  advective-reactive-diffusive equation.
\newblock \emph{Computer Methods in Applied Mechanics and Engineering},
  190:\penalty0 1785--1800, 2000.

\bibitem[Gamnitzer(2010)]{gamnitzer2010b}
P.~Gamnitzer.
\newblock \emph{Residual-based variational multiscale methods for turbulent
  flows and fluid-structure interaction}.
\newblock PhD thesis, Lehrstuhl f{\"u}r Numerische Mechanik, Technische
  Universit{\"a}t M{\"u}nchen, 2010.

\bibitem[Ge et~al.(2016)Ge, Guo, Yang, Sun, and Lu]{ge2016}
J.~Ge, B.~Guo, G.~Yang, Q.~Sun, and J.~Lu.
\newblock Blending isogeometric and {L}agrangian elements in three-dimensional
  analysis.
\newblock \emph{Finite Elements in Analysis and Design}, 112:\penalty0 50--63,
  2016.

\bibitem[Gee et~al.(2011)Gee, {K\"{u}ttler}, and Wall]{gee2011}
M.~W. Gee, U.~{K\"{u}ttler}, and W.~A. Wall.
\newblock Truly monolithic algebraic multigrid for fluid-structure interaction.
\newblock \emph{International Journal for Numerical Methods in Engineering},
  85\penalty0 (8):\penalty0 987--1016, 2011.

\bibitem[Gerbeau and Vidrascu(2003)]{gerbeau2003}
J.-F. Gerbeau and M.~Vidrascu.
\newblock A quasi-newton algorithm based on a reduced model for fluid-structure
  interaction problems in blood flows.
\newblock \emph{{ESAIM}: Mathematical Modelling and Numerical Analysis},
  37:\penalty0 631--647, 7 2003.

\bibitem[Hansbo and Szepessy(1990)]{hansbo1990}
P.~Hansbo and A.~Szepessy.
\newblock A velocity-pressure streamline diffusion finite element method for
  the incompressible {N}avier-{S}tokes equations.
\newblock \emph{Computer Methods in Applied Mechanics and Engineering},
  84\penalty0 (2):\penalty0 175--192, 1990.

\bibitem[Harari and Hughes(1992)]{harari1992}
I.~Harari and T.~J.~R. Hughes.
\newblock What are $c$ and $h$?: {I}nequalities for the analysis and design of
  finite element methods.
\newblock \emph{Computer Methods in Applied Mechanics and Engineering},
  97:\penalty0 157--192, 1992.

\bibitem[Harmel et~al.(2017)Harmel, Sauer, and Bommes]{harmel2016}
M.~Harmel, R.~A. Sauer, and D.~Bommes.
\newblock Volumetric mesh generation from {T}-spline surface representations.
\newblock \emph{Computer-Aided Design}, 82:\penalty0 13--28, 2017.
\newblock Isogeometric Design and Analysis.

\bibitem[Heil(2004)]{heil2004a}
M.~Heil.
\newblock An efficient solver for the fully coupled solution of
  large-displacement fluid-structure interaction problems.
\newblock \emph{Computer Methods in Applied Mechanics and Engineering},
  193\penalty0 (1-2):\penalty0 1--23, 2004.

\bibitem[Heil et~al.(2008)Heil, Hazel, and Boyle]{heil2008}
M.~Heil, A.~L. Hazel, and J.~Boyle.
\newblock Solvers for large-displacement fluid--structure interaction problems:
  segregated versus monolithic approaches.
\newblock \emph{Computational Mechanics}, 43\penalty0 (1):\penalty0 91--101,
  2008.

\bibitem[{H\"{u}bner} et~al.(2004){H\"{u}bner}, Walhorn, and
  Dinkler]{huebner2004}
B.~{H\"{u}bner}, E.~Walhorn, and D.~Dinkler.
\newblock A monolithic approach to fluid-structure interaction using space-time
  finite elements.
\newblock \emph{Computer Methods in Applied Mechanics and Engineering},
  193\penalty0 (23-26):\penalty0 2087--2104, 2004.

\bibitem[Hughes et~al.(1986)Hughes, Franca, and Balestra]{hughes1986b}
T.~J.~R. Hughes, L.~P. Franca, and M.~Balestra.
\newblock A new finite element formulation for computational fluid dynamics:
  {V}. {C}ircumventing the {B}abu{\v s}ka-{B}rezzi condition: a stable
  {P}etrov-{G}alerkin formulation of the {S}tokes problem accommodating
  equal-order interpolations.
\newblock \emph{Computer Methods in Applied Mechanics and Engineering},
  59\penalty0 (1):\penalty0 85--99, 1986.

\bibitem[Hughes et~al.(2005)Hughes, Cottrell, and Bazilevs]{hughes2005}
T.~J.~R. Hughes, J.~A. Cottrell, and Y.~Bazilevs.
\newblock {I}sogeometric {A}nalysis: {CAD}, finite elements, {NURBS}, exact
  geometry and mesh refinement.
\newblock \emph{Computer Methods in Applied Mechanics and Engineering},
  194:\penalty0 4135--4195, 2005.

\bibitem[Hughes et~al.(2008)Hughes, Reali, and Sangalli]{hughes2008}
T.~J.~R. Hughes, A.~Reali, and G.~Sangalli.
\newblock Duality and unified analysis of discrete approximations in structural
  dynamics and wave propagation: {C}omparison of p-method finite elements with
  k-method {NURBS}.
\newblock \emph{Computer Methods in Applied Mechanics and Engineering},
  197\penalty0 (49-50):\penalty0 4104--4124, 2008.

\bibitem[Jansen et~al.(1999)Jansen, Whiting, and Hulbert]{jansen1999}
K.~E. Jansen, C.~H. Whiting, and G.~M. Hulbert.
\newblock A generalized-$\alpha$ method for integrating the filtered
  {N}avier--{S}tokes equations with a stabilized finite element method.
\newblock \emph{Computer Methods in Applied Mechanics and Engineering},
  190:\penalty0 305--319, 1999.

\bibitem[Johnson and Tezduyar(1994)]{johnson1994}
A.~A. Johnson and T.~E. Tezduyar.
\newblock Mesh update strategies in parallel finite element computations of
  flow problems with moving boundaries and interfaces.
\newblock \emph{Computer Methods in Applied Mechanics and Engineering},
  119\penalty0 (1):\penalty0 73--94, 1994.

\bibitem[Johnson and Tezduyar(1999)]{johnson1999}
A.~A. Johnson and T.~E. Tezduyar.
\newblock Advanced mesh generation and update methods for {3D} flow
  simulations.
\newblock \emph{Computational Mechanics}, 23\penalty0 (2):\penalty0 130--143,
  1999.

\bibitem[Kassiotis et~al.(2011)Kassiotis, Ibrahimbegovic, Niekamp, and
  Matthies]{kassiotis2011}
C.~Kassiotis, A.~Ibrahimbegovic, R.~Niekamp, and H.~G. Matthies.
\newblock Nonlinear fluid-structure interaction problem. {P}art {I}: implicit
  partitioned algorithm, nonlinear stability proof and validation examples.
\newblock \emph{Computational Mechanics}, 47\penalty0 (3):\penalty0 305--323,
  2011.

\bibitem[Kiendl et~al.(2009)Kiendl, Bletzinger, Linhard, and
  W{\"u}chner]{kiendl2009}
J.~Kiendl, K.-U. Bletzinger, J.~Linhard, and R.~W{\"u}chner.
\newblock Isogeometric shell analysis with {K}irchhoff--{L}ove elements.
\newblock \emph{Computer Methods in Applied Mechanics and Engineering},
  198\penalty0 (49):\penalty0 3902--3914, 2009.

\bibitem[{Kl\"{o}ppel} et~al.(2011){Kl\"{o}ppel}, Popp, {K\"{u}ttler}, and
  Wall]{kloeppel2011}
T.~{Kl\"{o}ppel}, A.~Popp, U.~{K\"{u}ttler}, and W.~A. Wall.
\newblock Fluid-structure interaction for non-conforming interfaces based on a
  dual mortar formulation.
\newblock \emph{Computer Methods in Applied Mechanics and Engineering},
  200\penalty0 (45-46):\penalty0 3111--3126, 2011.

\bibitem[Kollmannsberger et~al.(2009)Kollmannsberger, Geller, D\"{u}ster,
  T\"{o}lke, Sorger, Krafczyk, and Rank]{kollmannsberger2009}
S.~Kollmannsberger, S.~Geller, A.~D\"{u}ster, J.~T\"{o}lke, C.~Sorger,
  M.~Krafczyk, and E.~Rank.
\newblock Fixed-grid fluid-structure interaction in two dimensions based on a
  partitioned lattice boltzmann and p-{FEM} approach.
\newblock \emph{International Journal for Numerical Methods in Engineering},
  79\penalty0 (7):\penalty0 817--845, 2009.

\bibitem[Kuhl et~al.(2003)Kuhl, Hulshoff, and {de Borst}]{kuhl2003}
E.~Kuhl, S.~Hulshoff, and R.~{de Borst}.
\newblock An arbitrary {L}agrangian {E}ulerian finite-element approach for
  fluid-structure interaction phenomena.
\newblock \emph{International Journal for Numerical Methods in Engineering},
  57\penalty0 (1):\penalty0 117--142, 2003.

\bibitem[{K\"uttler} and Wall(2008)]{kuettler2008}
U.~{K\"uttler} and W.~A. Wall.
\newblock Fixed-point fluid-structure interaction solvers with dynamic
  relaxation.
\newblock \emph{Computational Mechanics}, 43\penalty0 (1):\penalty0 61--72,
  2008.

\bibitem[{K\"uttler} et~al.(2006){K\"uttler}, {F\"orster}, and
  Wall]{kuettler2006}
U.~{K\"uttler}, C.~{F\"orster}, and W.~A. Wall.
\newblock A solution for the incompressibility dilemma in partitioned
  fluid-structure interaction with pure {D}irichlet fluid domains.
\newblock \emph{Computational Mechanics}, 38\penalty0 (4-5):\penalty0 417--429,
  2006.

\bibitem[{K\"{u}ttler} et~al.(2010){K\"{u}ttler}, Gee, {F\"{o}rster},
  Comerford, and Wall]{kuettler2010}
U.~{K\"{u}ttler}, M.~W. Gee, C.~{F\"{o}rster}, A.~Comerford, and W.~A. Wall.
\newblock Coupling strategies for biomedical fluid-structure interaction
  problems.
\newblock \emph{International Journal for Numerical Methods in Biomedical
  Engineering}, 26\penalty0 (3-4):\penalty0 305--321, 2010.

\bibitem[Lee and You(2013)]{lee2013}
J.~Lee and D.~You.
\newblock Study of vortex-shedding-induced vibration of a flexible splitter
  plate behind a cylinder.
\newblock \emph{Physics of Fluids}, 25\penalty0 (11), 2013.

\bibitem[{L\"ohner} and Yang(1996)]{loehner1996}
R.~{L\"ohner} and C.~Yang.
\newblock Improved {ALE} mesh velocities for moving bodies.
\newblock \emph{Communications in Numerical Methods in Engineering},
  12\penalty0 (10):\penalty0 599--608, 1996.

\bibitem[Lu et~al.(2013)Lu, Yang, and Ge]{lu2013}
J.~Lu, G.~Yang, and J.~Ge.
\newblock Blending {NURBS} and {L}agrangian representations in isogeometric
  analysis.
\newblock \emph{Computer Methods in Applied Mechanics and Engineering},
  257:\penalty0 117--125, 2013.

\bibitem[Malan and Oxtoby(2013)]{malan2013}
A.~Malan and O.~Oxtoby.
\newblock An accelerated, fully-coupled, parallel 3d hybrid finite-volume
  fluid-structure interaction scheme.
\newblock \emph{Computer Methods in Applied Mechanics and Engineering},
  253:\penalty0 426--438, 2013.

\bibitem[Matthies et~al.(2006)Matthies, Niekamp, and Steindorf]{matthies2006}
H.~G. Matthies, R.~Niekamp, and J.~Steindorf.
\newblock Algorithms for strong coupling procedures.
\newblock \emph{Computer Methods in Applied Mechanics and Engineering},
  195\penalty0 (17-18):\penalty0 2028--2049, 2006.

\bibitem[Mayr et~al.(2015)Mayr, {Kl\"{o}ppel}, Wall, and Gee]{mayr2015}
M.~Mayr, T.~{Kl\"{o}ppel}, W.~A. Wall, and M.~W. Gee.
\newblock A temporal consistent monolithic approach to fluid-structure
  interaction enabling single field predictors.
\newblock \emph{SIAM Journal on Scientific Computing}, 37\penalty0
  (1):\penalty0 B30--B59, 2015.

\bibitem[Mok et~al.(2001)Mok, Wall, and Ramm]{mok2001b}
D.~P. Mok, W.~A. Wall, and E.~Ramm.
\newblock {A}ccelerated iterative substructuring schemes for instationary
  fluid-structure interaction.
\newblock In \emph{{P}roceedings of {F}irst {MIT} {C}onference on
  {C}omputational {F}luids and {S}olid {M}echanics}, volume~2, pages
  1325--1328, {A}msterdam, {L}ondon, 2001. Elsevier.

\bibitem[Morganti et~al.(2015)Morganti, Auricchio, Benson, Gambarin, Hartmann,
  Hughes, and Reali]{morganti2015}
S.~Morganti, F.~Auricchio, D.~J. Benson, F.~I. Gambarin, S.~Hartmann, T.~J.~R.
  Hughes, and A.~Reali.
\newblock Patient-specific isogeometric structural analysis of aortic valve
  closure.
\newblock \emph{Computer Methods in Applied Mechanics and Engineering},
  284:\penalty0 508--520, 2015.
\newblock Isogeometric Analysis Special Issue.

\bibitem[Motlagh et~al.(2013)Motlagh, Ahn, Hughes, and Calo]{motlagh2013}
Y.~G. Motlagh, H.~T. Ahn, T.~J.~R. Hughes, and V.~M. Calo.
\newblock Simulation of laminar and turbulent concentric pipe flows with the
  isogeometric multiscale method.
\newblock \emph{Computers \& Fluids}, 71:\penalty0 146--155, 2013.

\bibitem[Ohayon(2004)]{ohayon2004}
R.~Ohayon.
\newblock Fluid-structure interaction problems.
\newblock In \emph{Encyclopedia of Computational Mechanics}. John Wiley \&
  Sons, Ltd, 2004.
\newblock ISBN 9780470091357.

\bibitem[Piegel and Tiller(1997)]{piegelbook}
L.~Piegel and W.~Tiller.
\newblock \emph{The {NURBS} {B}ook}.
\newblock Springer, 1997.
\newblock ISBN 978-3-642-59223-2.

\bibitem[Piperno et~al.(1995)Piperno, Farhat, and Larrouturou]{piperno1995}
S.~Piperno, C.~Farhat, and B.~Larrouturou.
\newblock Partitioned procedures for the transient solution of coupled
  aeroelastic problems {P}art {I}: {M}odel problem, theory and two-dimensional
  application.
\newblock \emph{Computer Methods in Applied Mechanics and Engineering},
  124\penalty0 (1):\penalty0 79--112, 1995.

\bibitem[Rasool et~al.(2016)Rasool, Corbett, and Sauer]{rasool2016}
R.~Rasool, C.~J. Corbett, and R.~A. Sauer.
\newblock A strategy to interface isogeometric analysis with lagrangian finite
  elements -- {A}pplication to incompressible flow problems.
\newblock \emph{Computers \& Fluids}, 127:\penalty0 182--193, 2016.

\bibitem[Rugonyi and Bathe(2001)]{rugonyi2001}
S.~Rugonyi and K.~Bathe.
\newblock On finite element analysis of fluid flows fully coupled with
  structural interactions.
\newblock \emph{CMES - Computer Modeling in Engineering and Sciences},
  2\penalty0 (2):\penalty0 195--212, 2001.

\bibitem[Sauer(2011)]{roger2011}
R.~A. Sauer.
\newblock Enriched contact finite elements for stable peeling computations.
\newblock \emph{International Journal for Numerical Methods in Engineering},
  87:\penalty0 593--616, 2011.

\bibitem[Sauer(2013)]{roger2013}
R.~A. Sauer.
\newblock Local finite element enrichment strategies for 2{D} contact
  computations and a corresponding postprocessing scheme.
\newblock \emph{Computational Mechanics}, 52:\penalty0 301--319, 2013.

\bibitem[{Sauer} and {Luginsland}(2017)]{luginsland2017}
R.~A. {Sauer} and T.~{Luginsland}.
\newblock {A monolithic fluid-structure interaction formulation for solid and
  liquid membranes including free-surface contact}.
\newblock \emph{ArXiv e-prints}, Oct. 2017.

\bibitem[Sauer et~al.(2014)Sauer, Duong, and Corbett]{sauer2014}
R.~A. Sauer, T.~X. Duong, and C.~J. Corbett.
\newblock A computational formulation for constrained solid and liquid
  membranes considering isogeometric finite elements.
\newblock \emph{Computer Methods in Applied Mechanics and Engineering},
  271:\penalty0 48--68, 2014.

\bibitem[Tezduyar and Osawa(2000)]{tezduyar2000}
T.~E. Tezduyar and Y.~Osawa.
\newblock Finite element stabilization parameters computed from element
  matrices and vectors.
\newblock \emph{Computer Methods in Applied Mechanics and Engineering},
  190\penalty0 (3-4):\penalty0 411--430, 2000.

\bibitem[Tezduyar et~al.(2006)Tezduyar, Sathe, Keedy, and Stein]{tezduyar2006}
T.~E. Tezduyar, S.~Sathe, R.~Keedy, and K.~Stein.
\newblock Space-time finite element techniques for computation of
  fluid-structure interactions.
\newblock \emph{Computer Methods in Applied Mechanics and Engineering},
  195\penalty0 (17-18):\penalty0 2002--2027, 2006.

\bibitem[Turek and Hron(2006)]{turek2006}
S.~Turek and J.~Hron.
\newblock Proposal for numerical benchmarking of fluid-structure interaction
  between an elastic object and laminar incompressible flow.
\newblock In H.-J. Bungartz and M.~Sch\"{a}fer, editors, \emph{Fluid-Structure
  Interaction}, volume~53 of \emph{Lecture Notes in Computational Science and
  Engineering}, pages 371--385. Springer Berlin Heidelberg, 2006.
\newblock ISBN 978-3-540-34595-4.

\bibitem[{van Zuijlen} et~al.(2007){van Zuijlen}, Bosscher, and
  Bijl]{vanzuijlen2007}
A.~H. {van Zuijlen}, S.~Bosscher, and H.~Bijl.
\newblock Two level algorithms for partitioned fluid-structure interaction
  computations.
\newblock \emph{Computer Methods in Applied Mechanics and Engineering},
  196\penalty0 (8):\penalty0 1458--1470, 2007.

\bibitem[Walhorn et~al.(2005)Walhorn, {K\"{o}lke}, {H\"{u}bner}, and
  Dinkler]{walhorn2005}
E.~Walhorn, A.~{K\"{o}lke}, B.~{H\"{u}bner}, and D.~Dinkler.
\newblock Fluid-structure coupling within a monolithic model involving free
  surface flows.
\newblock \emph{Computers \& Structures}, 83\penalty0 (25-26):\penalty0
  2100--2111, 2005.

\bibitem[Wall(1999)]{wall1999}
W.~A. Wall.
\newblock \emph{Fluid-{S}truktur-{I}nteraktion mit stabilisierten {F}initen
  {E}lementen}.
\newblock PhD thesis, Institute of Structural Mechanics, Universit{\"a}t
  Stuttgart, 1999.

\bibitem[Wang et~al.(2004)Wang, Nogami, Dasari, and Lin]{wang2004}
J.~G. Wang, T.~Nogami, G.~R. Dasari, and P.~Z. Lin.
\newblock A weak coupling algorithm for seabed-wave interaction analysis.
\newblock \emph{Computer Methods in Applied Mechanics and Engineering},
  193\penalty0 (36-38):\penalty0 3935--3956, 2004.

\bibitem[Yigit et~al.(2008)Yigit, {Sch\"afer}, and Heck]{yigit2008}
S.~Yigit, M.~{Sch\"afer}, and M.~Heck.
\newblock Grid movement techniques and their influence on laminar
  fluid-structure interaction computations.
\newblock \emph{Journal of Fluids and Structures}, 24\penalty0 (6):\penalty0
  819--832, 2008.

\bibitem[Zhang et~al.(2007)Zhang, Bazilevs, Goswami, Bajaj, and
  Hughes]{zhang2007}
Y.~Zhang, Y.~Bazilevs, S.~Goswami, C.~L. Bajaj, and T.~J.~R. Hughes.
\newblock Patient-specific vascular {NURBS} modeling for isogeometric analysis
  of blood flow.
\newblock \emph{Computer Methods in Applied Mechanics and Engineering},
  196:\penalty0 2943--2959, 2007.

\end{thebibliography}

\end{document}